\documentclass[iop]{emulateapj}
\pdfoutput=1
\usepackage{epstopdf}
\usepackage{lineno}
\usepackage[usenames,dvipsnames,svgnames,table]{xcolor}  
\usepackage{ulem}  

\slugcomment{Published in ApJ 796, 53 - revised 2014 Dec 02}
\shorttitle{The Variable Sky}
\shortauthors{Ridgway et al.}

\begin{document}

\title{The Variable Sky of Deep Synoptic Surveys\footnote{Revised 2014 Nov 11}}
\author{Stephen T. Ridgway, Thomas Matheson, Kenneth J. Mighell, Knut A. Olsen}
\affil{National Optical Astronomy Observatory, Tucson, AZ 85725, USA}
\and
\author{Steve B. Howell}
\affil{NASA Ames Research Center, P.O. Box 1, M/S 244-30, Moffett Field, CA 94035, USA}
\email{ridgway@noao.edu}

\begin{abstract}
The discovery of variable and transient sources is an essential product of synoptic surveys.   The alert stream will require filtering for personalized criteria -- a process managed by a functionality commonly described as a Broker.  In order to understand quantitatively the magnitude of the alert generation and Broker tasks, we have undertaken an analysis of the most numerous types of variable targets in the sky -- Galactic stars, QSOs, AGNs and asteroids.   It is found that the Large Synoptic Survey Telescope will be capable of discovering $\sim 10^5$ high latitude ($|b| > 20^\circ$) variable stars per night at the beginning of the survey.  (The corresponding number for $|b| < 20^\circ$ is orders of magnitude larger, but subject to caveats concerning extinction and crowding.)  However, the number of new discoveries may well drop below 100/night within  2 years.  The same analysis applied to GAIA clarifies the complementarity of the GAIA and LSST surveys.   Discovery of active galactic nuclei (AGNs) and Quasi Stellar Objects (QSOs) are each predicted to begin at $\sim 3000$ per night, and decrease by 50$\times$ over 4 years.  SNe are expected at $\sim 1100$/night, and after several survey years will dominate the new variable discovery rate. LSST asteroid discoveries will start at $> 10^5$ per night, and if orbital determination has a 50\% success rate per epoch, will drop below 1000/night within 2 years.  
\end{abstract}


\section{Introduction}
The Large Synoptic Survey Telescope project \citep[LSST, ][]{ivez2008} is being designed to carry out a 10--year survey of the available southern sky, with $\sim$800 visits per field and an expected single visit depth {\it r}~$\simeq24.5$ (5$\sigma$).  The survey will detect unprecedented numbers of variable/transient targets, including variable stars, active galactic nuclei (AGNs), supernovae (SNe), and variable and main belt asteroids (MBAs).  The LSST Project is committed to identifying variable targets and rapidly publishing information packets on them with a time lag not longer than 60 seconds.  The total number of these detections and the rate at which they occur will have a significant impact on the processing power required to support the LSST pipeline and feed subsequent follow-up resources.  The LSST Science Requirements Document, v5.2.3 \citep[henceforth SRD]{lsst2011} specifies a minimum alert rate capability of $10^3$ per visit, and a stretch-goal rate of $10^5$, with visits on a $\simeq$40 second cadence.  It has not been clear how these specifications compare to likely peak variable detection rates.

While the production of alerts completes the LSST near-real-time response to variability detection, it signals just the beginning of the scientific response.  From the alert stream, it will be necessary to apply filters, some simple and some complex, to select the targets for particular programs.  For those requiring rapid follow-up, the filtering process must respond on a $\sim$minute time scale to avoid degrading the timeliness of the LSST alert stream.  This filtering process is sometimes referred to as a Broker \citep{2012PASP..124.1175B} - e.g., a Skyalert\footnote{http://www.skyalert.org/} event stream.  We understand a Broker to refer to a process that applies one or more selection criteria, possibly including correlation with the target history if any, other LSST data including nearby objects, other surveys, and very likely including among its services a probabilistic classification of many or all targets \citep{rich2012}.  Naively, the required Broker processing power should scale with the LSST alert production rate, though we will question this below.

There have been estimates of alert rates in the past, but they have not been detailed or documented and have not always been consistent.  The objective of this study is to organize basic data and a rationale needed to determine order of magnitude (goal 50\%) estimates for discovery rates of variable targets in deep synoptic surveys.  In most of the paper, LSST will be taken as a model, guide, and example for observing parameters.  Variable Galactic stars will be examined in most detail, with briefer discussion of AGNs, QSOs, SNe and MBAs. There are of course other synoptic surveys. In section \ref{GAIA} the methodology will be applied to GAIA as a second example.

\section{Variable stars}

The most numerous variable/transient targets in the LSST catalog will be stars, active galaxies, and asteroids.  Simple estimates suggest that any one of these could dominate actual detections, hence all should be considered.  Each of these requires a different approach.  

Initially, we address the number of Galactic variable stars that will be detected and generate alerts in the LSST survey.  It is known that at the level of a few percent variation, a few percent of stars are variable \citep{howe2008}.  This suggests very large variable object counts.  However, even if more exactly formulated, such a single data-point does not suffice to determine the actual detection rates within an order of magnitude.  

There have been several studies for prediction of variable star numbers in surveys, e.g., \citet{maud1987}, \citet{eyer2000}, which typically have taken a bottom-up approach of considering each major variable type in turn, reviewing the status of information on the occurrence of these types, and extrapolating to the number that will be detected in a future survey.  These suffer from uncertain completeness.   Estimating the probability of variability can be problematic.  Most large surveys give some attention to identifying variable targets.  But in order to estimate probability, it is necessary to know the non-variable count for each stellar type, which involves determining the properties of the non-variable objects.  Furthermore, we need data that goes deep and wide enough to produce good statistical characterization of the targets that will be the most numerous variable stars in the deep surveys.  Existing surveys provide large and important databases of variability.  An example is the OGLE-II catalog of 200,000 candidate variables \citep{wozn2002}.  Published studies of such data sets have typically been focussed, especially toward detecting specific types of periodic variables rather than a high level description of the incidence of variability.  

Two elements are needed to estimate the detection rates: a (possibly simulated) catalog of the objects, and a statistical basis for the probability of variability of each type of object as a function of variability amplitude (which we will call a variability probability distribution function, or VPDF) for at least the most abundant variable types.    
Our approach to predicting variable star detections combines knowledge of the variability of stars by spectral type, with a Galactic model.  As will be explained below, this requires photometry suitable for estimating effective temperatures, $T_{\rm eff}$, (and to a lesser extent luminosity) in combination with well-populated photometric time series.  Thus, for example, we considered the overlap of the Sloan Digital Sky Survey \citep{gunn2006} for multicolor photometry and the Catalina Sky Survey \citep{drak2009} for time series.  This remains a promising avenue.

We have chosen to work with data from the Kepler mission.  This data has obvious strengths and conveniences owing to the high target count, the photometric quality, and the pre-survey characterization effort.  It has some minor drawbacks discussed below.  Our approach is to characterize the stars primarily by temperature as determined in the Kepler target pre-selection studies, and we find simple statistics of variability from their photometric time series.    
In order to apply the Kepler-derived statistics to a region of the sky, we generate star catalogs for a grid of Galactic coordinates with the Besan\c{c}on on-line Galaxy model. These catalogs provide complete descriptions of each simulated star. 

\newpage

\section{The Besan\c{c}on Galaxy Model}

The Besan\c{c}on Galaxy model is described by \citet{robin2003}, and a web-based interface to its use is available\footnote{\url{http://model.obs-Besancon.fr}}.  For an input Galactic coordinate pair (or range) the model computes a simulation of the star population in the designated sky area.  The simulation is realized as a catalog of stars, with both fundamental parameters (age, mass, luminosity, effective temperature) as well as predicted observed parameters (brightness in selected filters).  We use only the $T_{\rm eff}$  and magnitudes.  The Galaxy model includes values for typical interstellar extinction, and should be directly comparable to observed stellar populations, although in the Galactic plane, where extinction is complex and clumpy, it is represented by a smooth model. (An improved representation of extinction for the Besan\c{c}on model is under development - A. Robin, 2013, private communication).

We have created a Besan\c{c}on model for the Kepler field in order to explore whether or not the model population distribution is consistent with the color distribution of the Kepler target sample (not required, but reassuring), and also as a guide to the luminosity content of the Kepler target set.  The model was computed for an area of 20 square degrees, covering Galactic coordinates b= 75.3-77.3$^\circ$ and l = 8.5 - 18.5$^\circ$, and thus representing a fair sample of the Kepler field.  With a faint limit imposed of {\it R} = 18 (UBV system), the catalog returned 209,742 stars.  This synthesis was done after the March 6, 2013 update in the Besan\c{c}on late M dwarf model.

\begin{figure}
\includegraphics[width=3.4in]{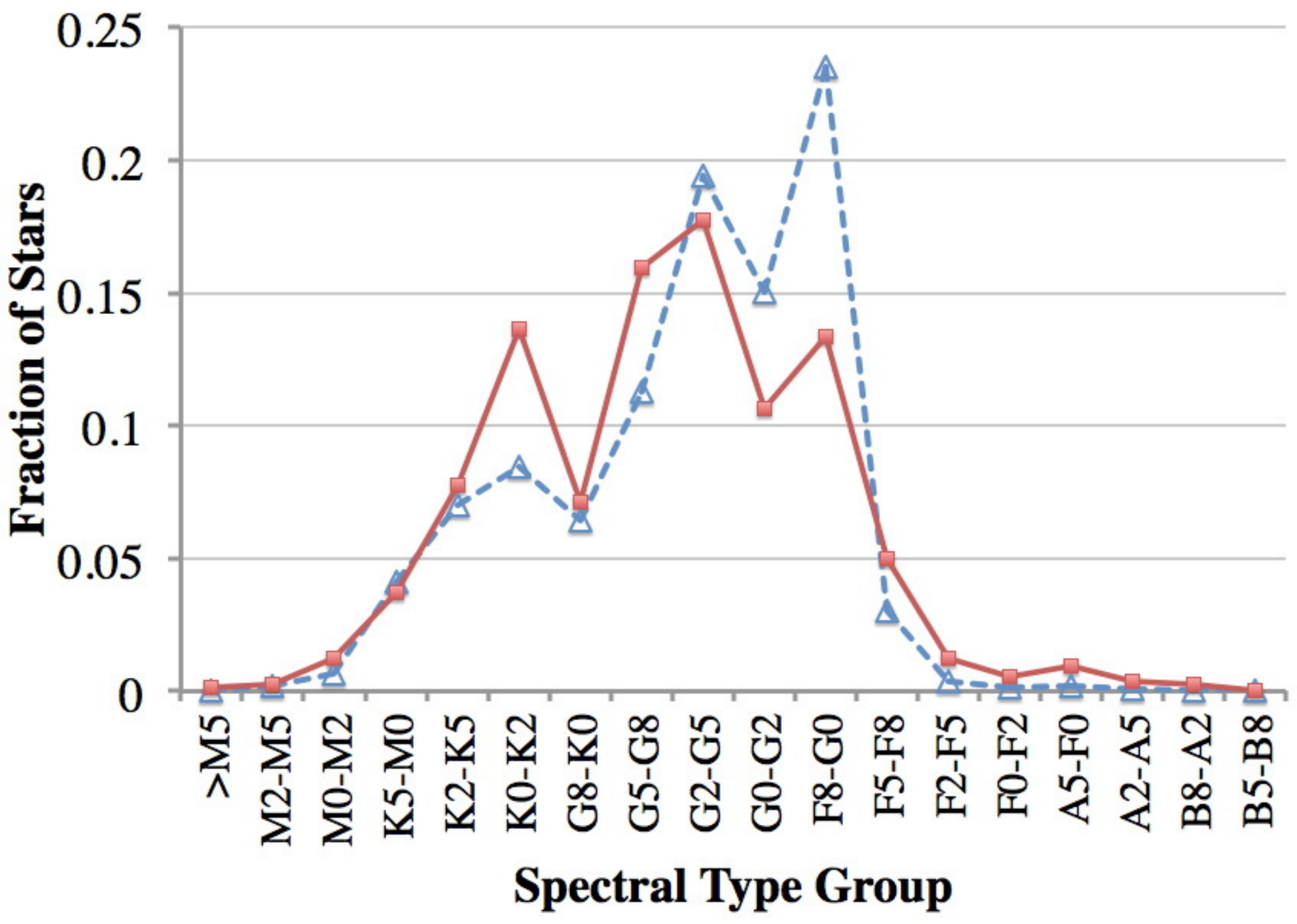}

\caption{Distribution of main sequence stars R $ \le 18$ in the Besan\c{c}on model for the Kepler field (solid line), and of all stars in Kepler Quarter 13 (dashed line).  Kepler target selection was designed to enhance the proportion of solar-type stars, as shown by the peak near G0.}
\label{Kepler-Besancon}
\end{figure}

Both the Besan\c{c}on catalog and the Kepler catalog include $T_{\rm eff}$ values for each object, and these are used as a primary basis for associating Kepler variability statistics with Besan\c{c}on catalog stars.   We have adopted the spectral type ranges and $T_{\rm eff}$  boundary values from Spectral Type - $T_{\rm eff}$  tables of \citet{cox2000}, and listed here in Table \ref{Kepler-Besancon}.  Figure 1 shows the distribution of the field contents by spectral type for the 210,000 Besan\c{c}on model stars sampling the Kepler field.  

\begin{table*}[width=6.8in]
 \begin{center}
 \caption{Spectral type groups used for analysis of variabiilty\label{AnalyticalApproximation}}
 \begin{tabular}{cccccccccc}
 \tableline\tableline
 Spectral Type & Tmax\tablenotemark{a} &  $\log g$ cut  & $K_p$  &  a  &  b &  c  &  d  &   Min valid & Max valid   \\
Range               &       &                          &           &    &        &       &     &   mmag & mmag   \\
 \tableline
 M5        & 3170 & $  ...     $   & 13.0 &   ...   &    ...    &  0.004  &  1.35  & 1  &  1000 \\
 M2-M5 & 3520 & $<3.5$  & 13.0 &   ...   &    ...    &   0.06   &   2.10    & 1  &  1000 \\
              &            & $>3.5$  & 13.0 &   ...   &   ...     &  0.12    &   1.30    &  1  &  100  \\
 M0-M2& 3840  & $<3.5$  & 13.0&   ...    &    ...    &  0.005  &  2.50    &   1  &  100  \\
              &            & $>3.5$  & 13.0 &   ...   &     ...   &  0.12     & 2.00    &    1  &  100 \\
 K5-M0 & 4410 & $<3.5$  &  13.0 &   ...   &    ...    &  0.15     & 1.80    &   1  &  100  \\
              &            & $>3.5$ &  13.0 &  0.15       &  -1.40 & ... &      ...      &  1  &  100 \\
 K2-K5 & 4830 & $<3.75$ &  13.5 & -0.50        &  -1.10&  ...  &     ...       & 1  &  200 \\
              &           & $>3.75$ &  13.5  &  ...  &   ...     & 0.40        & 1.50     &  1  & 200  \\
 K0-K2 & 5150 & $<4.0$    &  14.0  &  -0.75 &  -1.20  &    &           &   0.5  &  200 \\
              &           & $>4.0$  &     14.0 &  ... &   ...      & 0.55     &    1.50   &  0.5  & 200 \\
 G8-K0 & 5310  &$<4.1$ & 14.0  &  -0.60 &  -0.90  &  ...  &   ...    &       0.5  &  200 \\
              &             & $ >4.1$ & 14.0 &  ...  &   ...   &      0.55 &  1.50   &   0.5  &  200 \\
 G5-G8 & 5560 & $<4.2$ & 14.0  &  -0.10 &  -1.15  &  ...  &    ...   &       0.5  &  200 \\
              &             & $ >4.2$ & 14.0 &  -0.20  & -1.40     &  ...  &    ...     &   0.5  &  200 \\
 G2-G5 & 5790 & $<4.3$ & 14.0  &  -0.65 &  -0.90  &  ...  &   ...    &       0.2  &  150 \\
              &             & $ >4.3$ & 14.0 &     ...   &      ...    &  1.50   &   1.45      &   0.2  &  150 \\
 G0-G2 & 5940 & $<4.3$ & 14.0  &  -0.75 &  -0.95  &  ...  &   ...    &       0.2  &  200 \\
              &             & $ >4.3$ & 14.0 &  -0.65  &  -1.20 & ...  &    ...    &   0.2  &  200 \\
 F8-G0 & 6250 & $<4.3$ & 14.0  &  -0.75 &  -0.85  &  ...  &    ...   &       0.2  &  150 \\
              &             & $ >4.3$ & 14.0 &  -0.75  &  -1.20 & ...  &    ...    &   0.2  &  150 \\
 F5-F8 & 6650 &$<4.25$ & 14.0  &  -0.70 &  -0.80  &  ...  &    ...   &       0.2  &  150 \\
              &             & $ >4.25$ & 14.0 &  -0.75  &  -1.00 & ...  &    ...    &   0.2  &  150 \\
 F2-F5 & 7000 & $<4.25$ & 14.0  &  -0.50 &  -0.70  &  ...  &   ...    &       0.2  &  150 \\
              &             & $ >4.25$ & 14.0 &  -0.60  &  -0.90 &  ... &   ...     &   0.2  &  150 \\
 F0-F2 & 7300 & $<3.7$ &   ...  &    ...   &     ...    &  0.90  &   1.2    &       0.1  &  150 \\
              &             & $ >3.7$ &  ...  &     ...    &    ...     &  0.90 &   1.0     &   0.1  &  150 \\
 A5-F0 & 8180 &  ...  &  15.0   &    ...   &    ...     &  0.90  &   0.95    &       0.1  &  200 \\
 A2-A5 & 9000 &   ...  &  ...	   &    -0.7   &   -0.7      &    ...   &    ...      &       0.1  &  200 \\
 B8-A2 &11400  &  ... &  	...   &    -0.55   &   -0.4      &   ...    &    ...      &     0.1 &  7 \\
             &              &  ...  &  	...   &    0.5   &   -1.7      &   ...    &    ...      &        7  &  50 \\
 B5-B8 & 15200 &  ... &  	...   &    ...       &      ...    &   0.15    &     1.8     &     0.1 &  10 \\
 \tableline
 \end{tabular}
 \tablecomments{Spectral type ranges are defined by the $T_{max}$ limits noted for each range.  The $\log g$ cut is an empirical criterion, with no independent significance, based on published Kepler $\log g$, for selecting a clean set of main sequence stars.  The Kepler magnitude $K_p$ is a cut in brightness that was found convenient in eliminating higher luminosity stars from the main sequence set. The $a,b,c,d$ parameters are for the functional representations described in section \ref{representing}, and the following columns give the approximate range of usefulness of the empirical functions.}
 \tablenotetext{1}{The maximum $T_{\rm eff}$  assigned to the spectral type range}
 \end{center}
 \end{table*}

\newpage

\section{The Kepler Variable Star Data Sets}
\label{Kepler-section}

For our purposes, it is important to have good statistics for variability near the level of survey photometry calibration limits, which for LSST is to be $\simeq$5 mmag (SRD design specification). This quality of photometry is not common in large ground-based surveys, but has been reported by Pan-STARRS \citep{schl2013}.  For a data source we have chosen to go with the Kepler mission which monitors 170,000 stars brighter than R$\simeq$16, with a photometric precision $\simeq$80 ppm.  The Kepler targets are intentionally drawn from a large range of main sequence types.  Giants are also represented, though not  proportionately.  

Kepler benefits from a pre-survey study \citep{bata2010} that provided measures of $T_{\rm eff}$  and surface gravity, log(g), and the former are useful as they allow us to associate the Kepler stars to our Galaxy model stars.  Kepler has undergone a careful selection process associated with maximizing its planet detection performance.  This is only relevant to us if it tends to bias the variability statistics for any stellar type.  We will return to this point later, but it proves to be a small effect.

Kepler data was acquired in 3-month intervals, as constrained by regular reorienting of the space craft to maintain illumination of the solar panels.  Data first became publicly available with the release of Q1, the first quarter Kepler data set\footnote{\url{http://archive.stsci.edu/kepler/}} in 2009.  Soon after, \citet{cia2011} published a summary of variability statistics for the Kepler stars.  This study was not completely sufficient for our purposes, and Ciardi et al.(2013, private communication) recommended that we use their results cautiously and that a new analysis of the statistics of variability was merited when a better Kepler pipeline became available.   We have followed that advice.

Since that time, the Kepler team has made much progress in the photometry pipeline, as described in data release notes \citep{mast2013}.  Although the team takes care to point out that the bulk data processing does not produce the best photometry product for individual stars, it is now very effective at achieving good data quality across the breadth of the survey.  We have undertaken a new study of stellar variability with Kepler Quarter 13 (Q13) as described in Kepler Release 13 Notes  \citep{bar2012}, based on the SOC Pipeline 8.0.  

Each star data file contains a number of quality flags, and deleting all object datasets with any flag setting other than nominal, as well as a few non-standard FITS files, leaves a total of 155,370 stars, with the so-called long cadence sampling of $\sim30$ minutes, over a total interval of 93.4 days, with two gaps of less than 1 day each, as described in the Kepler Data Handbook  \citep{dawg2012}.

Using the Kepler-supplied values of $T_{\rm eff}$, the Kepler stars are shown in Figure \ref{Kepler-Besancon}, for comparison with the Besan\c{c}on model.  The comparison shows the good correspondence in distributions between the Kepler target list and the simulated Galaxy, and reflects some of the intended selection effects in the Kepler catalog, to enhance coverage of sun-like stars, cooler dwarfs and hot stars.  

The majority of Galactic stars detected in any deep survey will be dwarfs.  The Kepler sample is likewise dominated by dwarfs, with an estimated $\simeq 5$ \% contribution of giants.   In our application, luminosity is important primarily in avoiding contamination of the dwarf statistics with giants. Following \citet{mann2012} we have used apparent magnitude as a secondary indicator of luminosity.  By examining $\log g - M_k$ distributions, several cleaned dwarf data sets were generated for each spectral group, with differing $\log g$ and $M_k$ cuts.  This established and confirmed consistent characterization for clean  luminosity data sets.  The isolation of higher luminosity groups was less secure, especially for spectral types earlier then G.  Therefore we accept a lower degree of confidence in our evaluation of giant variability and statistics. This is of little importance to this study, but should be noted if the results are used for other purposes, and we do report results for luminosity dependence where the data justifies it (all spectral types except A and B).

\begin{figure}
\includegraphics[width=3.4in]{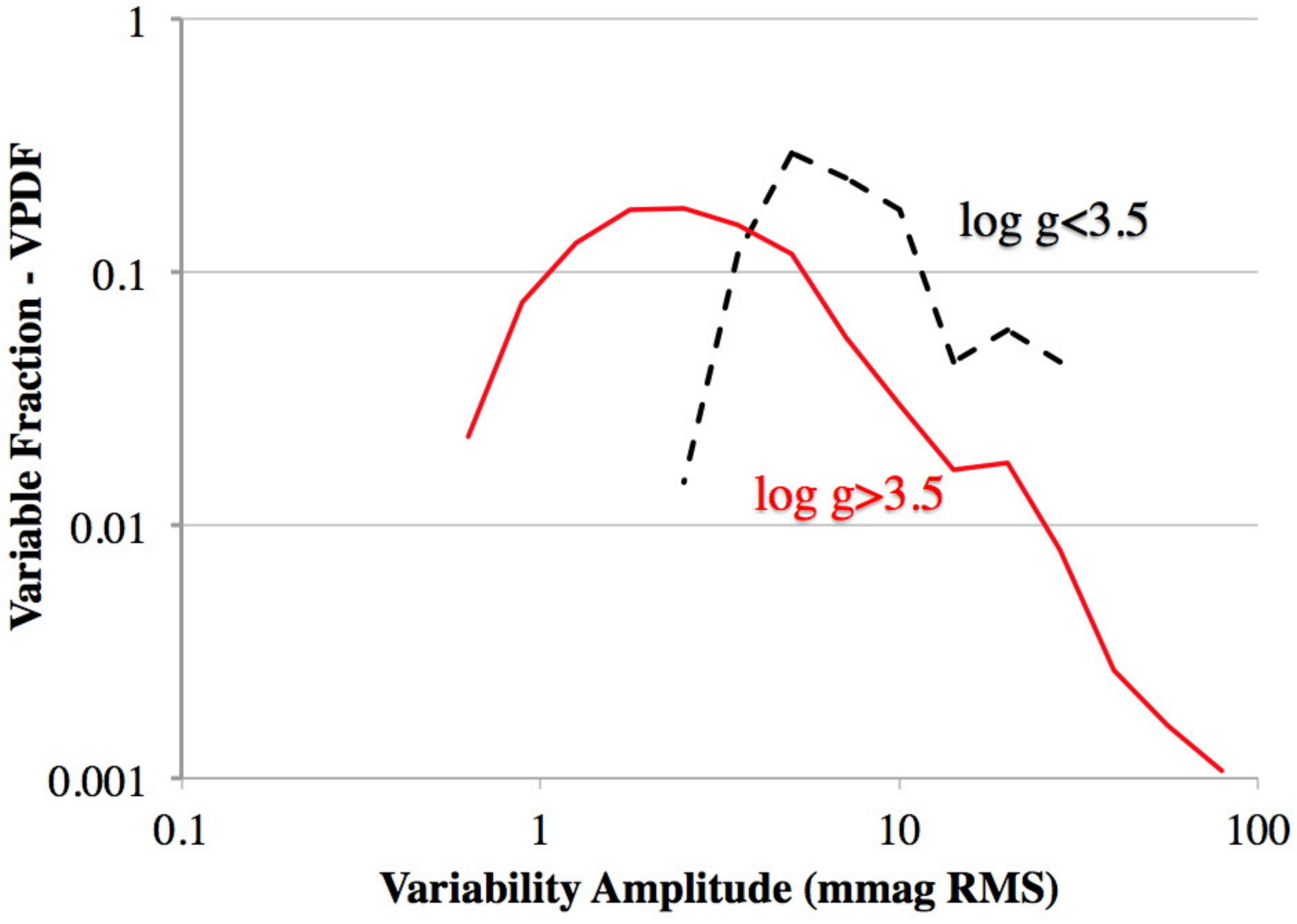}
\caption{Variability probability distribution functions (VPDFs) describing the probability that a Kepler M0 - M2 star will have rms brightness amplitude variability of {\it RMS}. Stars with Kepler $\log g > 3.5$, are represented by a solid line, and for $\log g < 3.5$ by a dashed line. 
\label{M-M2-VPDF}
}
\end{figure}

Figure \ref{M-M2-VPDF} shows an example of the kind of variability distribution found, and also one of the cases of clearly defined luminosity dependence.  The M0 -- M2 stars are separated into two groups, $\log g$ greater than and less than 3.5.  There are more than 2000 stars in this spectral group, and the luminosity dependence is distinct.  

\section{Representing the variability fraction}\label{representing}

Figure \ref{M-M2-VPDF} for M0-M2 stars is typical of all the VPDFs in several respects.  These empirical VPDFs predominantly have a power-law relationship to the rms variability amplitude in the range $\simeq$1-100 mmag.   It shows a turndown for very low amplitude variability essentially due to running out of stars that are less variable.  It also shows strong, spectral-group dependent, systematic deviations from a simple analytical description.  Therefore we have not tried to work with a functional representation.

\begin{figure}
\includegraphics[width=3.5in]{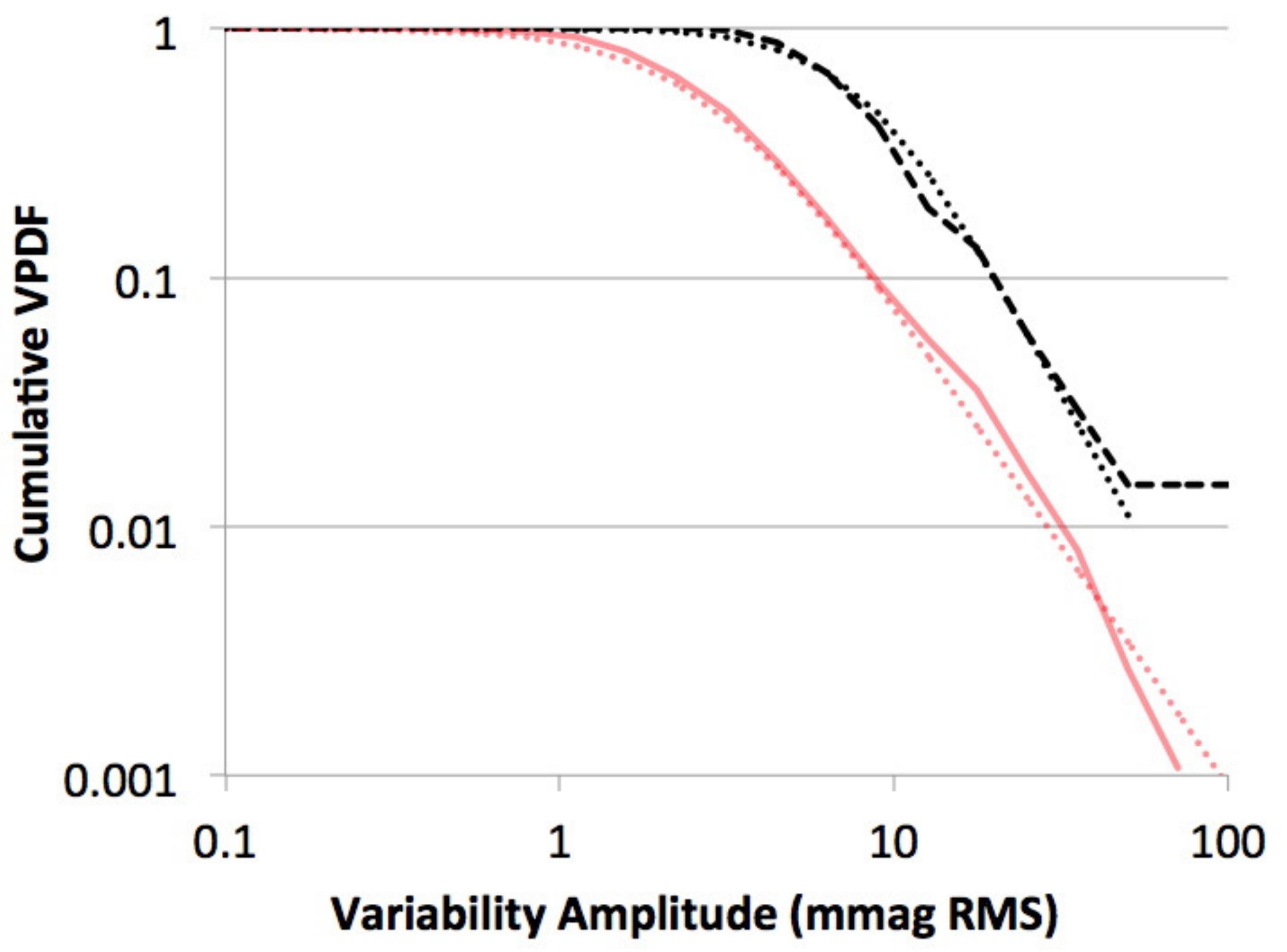}    
\caption{Cumulative VPDF that a Kepler M0 -- M2 star will have variability of amplitude $>RMS$ (derived from the data in Figure \ref{M-M2-VPDF}. Stars with Kepler $\log g > 3.5$, are represented by a solid line, and for $\log g < 3.5$ by a dashed line.  The dotted lines represent empirical approximations to the distributions with the functional representation $f(\sigma) = 1 / (1 + c\sigma^d)$.  \label{M0-M2-Cumulative}}
\end{figure}

For an application in counting detectable variables, it is convenient to make use of a cumulative VPDF, from high to low variability, giving the probability that a target will be more variable than a given rms amplitude.  Figure \ref{M0-M2-Cumulative} presents the cumulative distributions for Figure \ref{M-M2-VPDF}.  Not surprisingly, the integrated distribution shows less scatter  than the original data, but this is more a numerical effect than a reduction in noise, as the data in Figure \ref{M-M2-VPDF} has excellent signal-to-noise ratio.  It appears that the cumulative VPDFs might have a simple analytical  representation, and indeed a rather good fit is obtained with a function of the form $f(\sigma_{var}) = 1 / (1 + c\sigma_{var}^d)$, as shown in the figure.  In some cases a function $f(\sigma_{var}) = a + \sigma_{var}^b)$is more appropriate.  In Table 1 we give the parameters of empirical fits.  However, it is not clear that the fitted functions are ever a better description than the actual data, so we have chosen to work with the numerical cumulative VPDFs. Most of the cumulative VPDFs approach unity at or near 1 mmag rms.  This study does not examine closely the behavior much below this level. The cumulative VPDFs for all spectral type groups are shown in the Appendix, in Figures 18-26.

\begin{figure}
\includegraphics[width=3.5in]{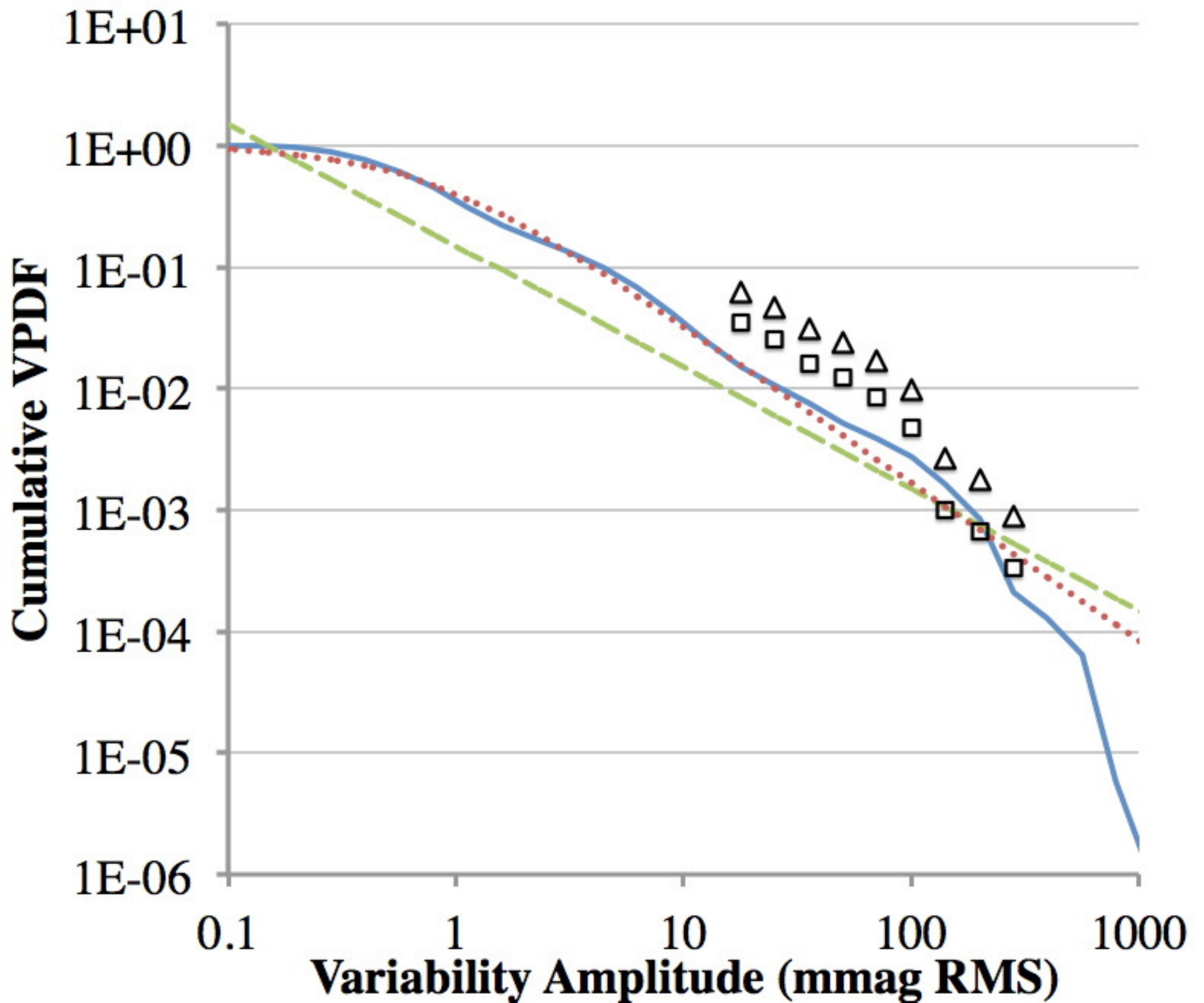}
\caption{Solid line: summed cumulative VPDFs of all represented spectral types with the weights of the Besan\c{c}on model population for the Kepler field. The fit in the range 1-100 mmag (dotted line) has $a = 2.2$ as described in section \ref{Kepler-section}.  The dashed line shows an analagous result from \citet{ever2002}, also representing a mixed population, but based on fewer stars and a restricted range of spectral types.  The open squares represent a cumulative VPDF derived from \cite{hube2006} for their group 01, and the open triangles similarly for their group 19 (further discussion in our section \ref{representative}).  \label{VariabilityBesanconDistribution}}
\end{figure}

It is reasonable to wonder why the cumulative VPDFs tend to relatively uniform distributions, since the underlying physics undoubtedly involves different processes at different amplitudes.  The VPDFs are possibly revealing a log-normal distribution \citep{limp2001} that can commonly characterize statistics of quantities that are predominantly small but cannot be negative (in this case, variability amplitude).  

\section{How Representative of the Sky is the Kepler Field?}
\label{representative}

The Kepler field, at galactic coordinates (76.32, +13.5)  samples regions near but mostly above the plane, and at approximately the solar distance from the galactic center.  It is reasonable to wonder whether or not this sample population, to $R \sim 18$, will be representative of a deeper survey and in other directions.  The \citet{ever2002} and and \citet{tonr2005}  surveys of variability studied fields at (167, +12) and (99, +60), to V $\sim 19.5$,  and showed a similar result for field stars, though based on a much smaller sample and omitting the cooler part of the population. 
The VPDF-equivalent representation of their data appears in Figure \ref{VariabilityBesanconDistribution}.  The sense of the difference is that the previous work did not include a Galactic mix of cool dwarfs, which are among the more variable stars, so it is understandable that the Besan\c{c}on mix shows somewhat more variability.  The \cite{groo2003}  Faint Sky Variability Survey studied 78 fields at mid and high galactic latitudes over a large range of longitudes, to V $\sim  24$ magnitude.  In the \cite{hube2006} analysis of this data set, data related to the VPDF is presented in figures 13 and 14 of that paper.  This information is in terms of peak to peak variation detected with $\sim 13$ epochs, that we convert to an rms estimate by dividing by 3.  The inferred cumulative mean VPDFs for magnitudes 18.15 - 19.25, in their regions 01 and 19,  are shown in Figure  \ref{VariabilityBesanconDistribution} as open symbols.  The paper reports on fainter stars, for which variability fraction is systematically lower with fainter magnitudes, due to a natural sensitivity cutoff in variable detection.

This comparison between surveys is somewhat problematic, as the comparison studies have different cadences and magnitude ranges.  Future surveys will probably detect significant population dependencies, but we conclude that these other studies do not offer a preferred or distinctly different characterization, and are qualitatively and to some extent quantitatively supportive.

\section{The RMS Representation of Variability Amplitude}
\label{Fsigma}
In section \ref{decline}, we will be interested in estimating how many photometric measurements will be required to determine that a target is variable.  This could be Monte Carlo modeled with the Kepler data, but it would be a great simplification if the estimate could be computed directly from the $\sigma_{var}$ distribution associated with a $T_{\rm eff}$  (and where appropriate a luminosity).
For a single visit to a target, what is the probability, $f_{\sigma}$, that it will be undergoing a brightness excursion (positive or negative) greater than the $\sigma_{var}$?  For stochastic variability governed by Gaussian statistics, the probability will be $f_{\sigma} \sim0.317$.  For sinusoidal variability, the probability is $f_{\sigma} =0.5$.  This range should reasonably span the ÒcontinuouslyÓ variable stars.  Transient variable types such as eclipsing binaries and flare stars can have low values of $f_{\sigma}$.  

In order to evaluate the utility of the rms value in characterizing the Kepler stars, we examined each photometric time series for both the rms value, and for the amount of time spent with excursion amplitudes up to $100 \sigma_{var}$.   Counting all Q13 observations for the set of Kepler stars, we found $f_{\sigma} = 0.314$, 
showing that Kepler variability statistics are at least somewhat noise-like (perhaps dominated by the more numerous stars with low amplitude variations).

 The statistics for all measurements of all stars are shown in Table \ref{ObservedDistribution}.  This shows that the actual distribution of brightness values in units of the rms follows a Gaussian distribution rather closely.  The deviations are a fraction of a percent reduction in the distribution for 1-4  $\sigma_{var}$ and a compensating increase in the range $>4 \sigma_{var}$. The relative deviations from Gaussian are small but with a large data set, the number of ``excess" data points $>4 \sigma_{var}$   is $\sim 400,000$!

\begin{table}
\begin{center}
\caption{Observed distribution of variable excursions from the mean compared to a Gaussian probability distribution. Each measured excursion in flux is converted to units of $\sigma_{var}$ for the particular star, and the measures are binned as shown in units of $\sigma_{var}$, showing a small excess of high amplitude excursions.\label{ObservedDistribution}}
\begin{tabular}{cccr}
\tableline\tableline
Excursion  & Observed &  Gaussian& Excess   \\
range ($\sigma_{var}$) & fraction & fraction  & fraction \\
\tableline
0.0 -- 0.2 & 0.15914 & 0.15852 & 0.00062 \\
0.2 -- 0.4 & 0.15259 & 0.15232 & 0.00027  \\
0.4 -- 0.6 &0.14087 & 0.14066 & 0.00021  \\
0.6 -- 0.8	& 0.12558 & 0.12478 & 0.00080  \\
0.8 -- 1.0 & 0.10761 & 0.10640 & 0.00121 \\
1.0 -- 2.0 & 0.27168 & 0.27182 & -0.00014 \\
2.0 -- 4.0 & 0.04180 & 0.04544 & -0.00364  \\
4.0 -- $\inf$ & 0.00070 & 0.00006 & 0.00063  \\
\tableline
\end{tabular}
\tablecomments{Based on 620M measurements of 155178 stars.}
\end{center}
\end{table}

\begin{figure}
\includegraphics[width=3.5in]{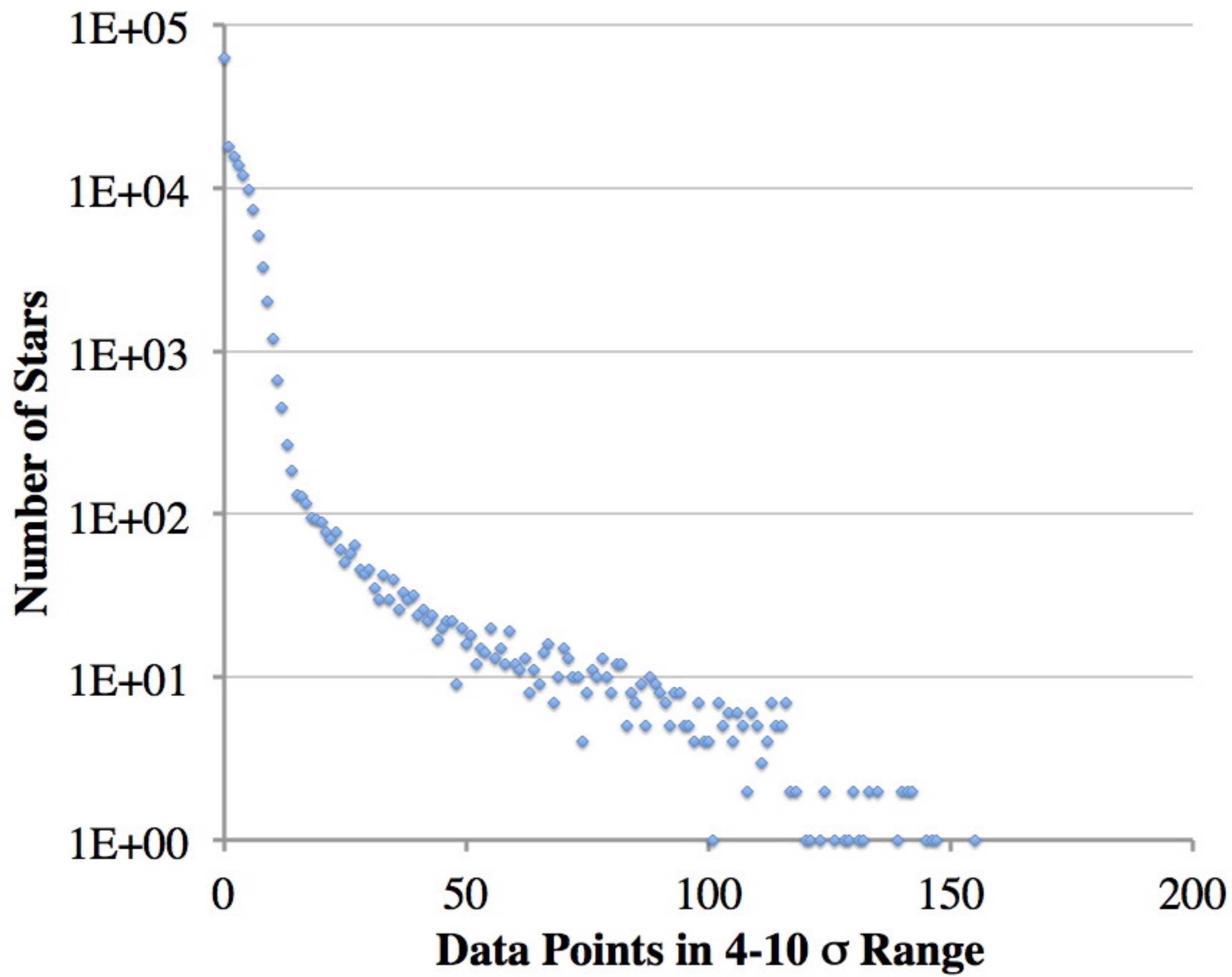}
\caption{Histogram of the numbers of stars binned according to the numbers of data points in the 4 -- 10 $\sigma_{var}$ range, showing additional detail on the same stars in Table \ref{ObservedDistribution}.  The number of data points in the 4 -- 10 $\sigma$ range can be compared to the total number of measurements per star, $\simeq$ 4000) and the normal distribution of Table \ref{ObservedDistribution}. \label{Fig:ObservedDistribution}}
\end{figure}

To better understand the nature of these excursions, the stars were ordered by the number of measurements that fell in the range 4-10 $\sigma_{var}$, which, for the $\simeq$4000 data points per star, should be typically 0-1 per star for a normal distribution.  The observed numbers are shown in Figure \ref{Fig:ObservedDistribution}.   From the left side of the graph, the first data point represents the 63,000 stars that have no large-excursion observations.  The next 2--3 data points account for as many total large-excursions as expected for Gaussian statistics.  Loosely speaking, the rest of the tail corresponds to the stars with an excess of large-excursion measurements.

A visual inspection was carried out for a sample of 70 stars, in groups of 10, to explore the different regimes of Figure \ref{Fig:ObservedDistribution}.  At the left end (small 4--10 $\sigma_{var}$ excess),  the high excursion points were contributed by one or more of three major brightness variation types.

\begin{itemize}
  \item Most Kepler  time series have small numbers of (usually low amplitude) single-point glitches. 
  \item A significant fraction of time series have small discontinuities at one or more of the breaks where the spacecraft was rotated and repointed.
  \item In some time series, the slow instrumental drifts characteristic of Kepler appear to have been corrected inexactly, leaving one or more spurious features.
\end{itemize}
These ``events" are important for peak-finding searches, but are generally not important for the kind of statistical measurements considered here, which are dominated by the $\sim4000$ well-behaved measurements for each star rather than by the few outliers.  

The same three effects appear in the light curves with a larger number of excursions, but in this group eclipsing binaries are increasingly common.    In the examined sample with more than 50 large excursion data points, 100\% were eclipsing binaries, with a range of periods and amplitudes.   Not surprisingly, these spend a relatively smaller fraction of their time at peak photometric excursion.  These mostly eclipsing binaries have $f_\sigma  \sim $ 0.07 -- 0.1 for the high-excursion tail of the distribution, and of course for individual rare objects the fraction could be arbitrarily small.  From this sample we deduce evidence for the signature of at least $\sim1000$ eclipsing binaries in Q13, consistent with \citet{slaw2011} who have identified as eclipsing binaries 2165 Kepler stars, or 1.4\% (integrated over multiple data releases, with overlapping but larger target list).  

In summary, there are only small and qualitatively understandable deviations from Gaussian statistics.  For purposes of a high-level characterization of variability we will adopt brightness rms and $f_{\sigma} = 0.314$ as the parameters for estimation of variability detectability.

At this point, the tools described can be used to compare the statistics derived here for Q13 with those available to \citet{cia2011} in their analysis of Q1.   Ciardi et al. (private communication) provided us with the Q1 data file used for their paper, which included a somewhat different set of stars and additional selection criteria that we did not apply.  We compared the variability distributions between the two quarters.  The summed distribution functions (i.e., weighting all spectral type groups equally) are in good agreement, show no systematic difference and in particular no variability amplitude-dependent difference above 1 mmag.  The Kepler photometric catalog quality improved significantly between Q1 and Q13, without changing the basic result for frequency of variability.  

\section{Limitations of the Kepler Sample}

At the faint limit of the Kepler survey, the S/N per data point is reduced, and this could potentially bias the results toward over-estimating the number of variables among the faint stars.  We have compared spectral type subsets grouped by brightness and confirmed that there is no strong brightness dependence at the 1--2  mmag level.

The Kepler survey target list \citep{bata2010} constitutes ~150,000 stars selected from ~0.5 million with  $m_K< 16$.  The selection process included explicit inclusion of the full range of temperature classes, M to O, suppression of higher luminosity stars in favor of main sequence stars, and a consideration of the estimated observability of a low-mass planet for each potential target.  These biases may be obviated by not relying on the relative counts between spectral types.  Within each spectral class, it is more important to consider what selection factors may impact variability.    \citet{bata2010} note that all then-known eclipsing binaries in the Kepler field ($>600$) were included in either the primary or ancillary target lists.  Therefore the \citet{slaw2011} eclipsing binary count of 1.4\% in the Kepler target list may include an enhancement from the fraction in the host population.  This likely error will be accepted here.

From examination of individual Kepler light curves, there appears in some cases a possible ambiguity between slow drifts of the instrument and slow evolution of the stellar brightness -- a point discussed in the Kepler Data Handbook \citep{dawg2012}.    The frequency of occurrence of changes over intervals longer than $\simeq 20$ days may be diluted.

The Kepler survey provides limited information about the frequency of stochastic transients such as flares in dwarfs, or mass transfer events in cataclysmic binaries.  These are discussed separately below.

\section{Estimating Numbers of Detectable Variable Stars in Synoptic Surveys}

The approach used to determine the count of detectable variables is as follows: (1) for representative areas on the sky, generate a model catalog of all starsÑthis catalog gives for each
star an apparent magnitude and fundamental parameters; (2)
step through the list of stars; (3)  for the assumed survey technology,
determine the fractional variation detectable for each star;  (4) from the VPDF function for that spectral type, find the probability that the star is more variable than the detection limit; (5) sum the probabilities over the catalog in useful ways.

In order to narrow the range of possible survey and analysis strategies, we assume a simple survey model, consisting of a sequence of measurements, with identification of variables by differential photometry between two or among many epochs.  We consider two cases of variable detection: rapid detection, as required for urgent alert generation (discussed here), and post-processing detection (section \ref{post}).  For clarity, the discussion is referenced to the S/N achieved in a single epoch, represented by its inverse, $\sigma_{phot}$, whether derived from single or multiple images, with the added assumption that the S/N in a stacked multi-epoch image will improve as $\sqrt{n}$.

LSST reports discuss a 5-$\sigma$ criterion for detections \citep{ivez2008}.  The development of criteria to ensure a desired level of statistical significance is complex.   In sophisticated difference imaging simulations and source detection simulations, \citet{beck2013} predict  that the 5-$\sigma$ cut would allow $\sim$3--13 false positives per sensor, or $\sim$500--2200 per field per visit.  In view of this number, Becker et al.  suggest perhaps increasing the threshold to 5.5-$\sigma$.  Indeed, a stringent criterion is needed to limit contamination of precious blank-field transient alerts.  However, only a tiny fraction of 5--$\sigma$ sky-photon/detection noise events will coincide with a known star position, thus a 5-$\sigma$ cut may be far more severe than necessary for identification of variables from among already catalogued stars. 

This example emphasizes that the significance threshold is not fundamental to a survey.  Furthermore it may be adjustable for various purposes and differently for different data streams.  In the following we will focus on significance of detection of variability only and not extend the discussion to the significance criteria for catalog inclusion.

\subsection{Photometric Detection and Calibration Noise}

The brightness-dependent component of photometric noise is the information typically offered by instrument ``exposure calculators" that determine the detection rms noise, $\sigma_{det}$, including detector noise, sky photon noise and possibly observing process factors.

In variability detection, we should also consider error in photometry calibration, $\sigma_{cal}$.  All-sky calibration closure better than 0.005 has been reported \citep{schl2013} and is increasingly the expectation for surveys.  An all-sky calibration of this quality may only be achieved post-facto months or years after data acquisition.  On the other hand, relative photometry within a single field-of-view is less subject to systematics, and $\sigma_{cal}$ = 5 mmag can be a conservative limit for such differential photometry.    

 For the total measurement noise, we use the RSS (root of the sum of squares) of the detection noise and calibration noise,
$\sigma_{phot} = (\sigma_{det}^2 + \sigma_{cal}^2)^{1/2}$.  This $\sigma_{phot}$ represents the single-epoch measurement noise.  The following discussion will be referenced to $\sigma_{phot}$.

\subsection{Procedure}

For application to a survey of the sky, we will use the Besan\c{c}on Galaxy model to compute a star catalog for representative fields.  If the number of stars in a field greatly exceeds $10^6$,  the model will be used to generate a catalog for a representative sub-area, and the results scaled to the full area. 

For each star in the model catalog, the apparent brightness of the star in the selected filter will be used with the measurement model to predict the single-epoch measurement error, $\sigma_{phot}$.   The threshold criterion, $t$ in units of $\sigma_{phot}$, will determine the minimum level of variability detectable for that star.  The Besan\c{c}on temperature will be used to select the appropriate Kepler VPDF.  The threshold, $t \sigma_{phot}$,  will be used to read from the cumulative VPDF the probability that the star has a detectable variability.  The fractional probabilities per star will be simply summed with binning appropriate to track across the catalog variable star counts that may be sorted by spectral type and/or apparent magnitude.

\subsection{Rapid Detection and Alerts\label{rapid}}

A common objective of synoptic surveys is the rapid identification of transient targets, in some cases requiring timely follow-up, hence the generation of alerts.  A low probability of false positives will greatly enhance the value of the alerts.  

For rapid identification, we assume simple comparison of one measurement at the current epoch with one measurement from a previous epoch.  We expect that  a few percent of stars will have detectable variability.  Each star will be observed multiple times, so if the probability of a false positive indication for variability is $\gamma$, then the integrated probability after $n$ epochs will be $\sim n\gamma$ (for small probability).  Requiring that the false attribution of variability be less than 0.0001 of the number of true detected variables, a clearly severe criterion, requires
$\gamma \leq 10^{-4}/n$.  
For a survey duration of order $10^2$--$10^3$ epochs, this implies a statistical threshold of 3.4--4.0 $\sigma_{phot}$.  Though much different from $5.0\;\sigma_{phot}$ as a statistical property (2+ orders of magnitude), it is only 20\% different as a variability amplitude, hence perhaps not importantly different, so in the following we will use a threshold of $5.0 \;\sigma_{phot}$ for rapid detection.

\subsection{Variability Detection by Post-processing}\label{post}

For a survey with many epochs, variability will be detectable in the full data set to a lower fractional level than possible with one visit pair.  We note from Table \ref{ObservedDistribution} that stars with rms variability $\sigma_{var}$ will typically spend fraction $f_{\sigma} = 0.314$ of their time at excursions exceeding amplitude $\sigma_{var}$, and ~0.683 of their time at smaller excursions.  For a survey with $n$ epochs , assume that one can optimally sort the data into bins greater than and less than $\sigma_{var}$.  Then for a per epoch noise of $\nu = 1/\sigma_{phot}$, the noise in the mean of the two bins will be improved by $\sqrt{1/0.314n}$ and $\sqrt{1/0.686n}$, respectively.  The noise in the comparison of two epochs will be $\sqrt{2}\nu$, and the noise in the comparison of $n$ epochs binned will be $\sim2.15\nu/\sqrt{n}$ (for $n> > 2$).  This shows that a survey of 100 -- 1000 epochs may reach variability levels $\sim$5--15$\times$ lower than the alert detection process.    Combining this and section \ref{rapid}, a useful threshold range for variable detection by post-processing of 100--1000 epochs would be 1 or 0.3 $\sigma_{phot}$, respectively. Of course the availability of other information or priors may allow further reducing the limits and further enlarging the set of detectable variables, so these thresholds probably imply lower bounds to the totals.

\subsection{The Declining Rate of Detection in a Survey\label{decline}}

It can be anticipated that the number of detectable variable stars will be large.  If the survey commences with no knowledge of the particular targets, the rate of discovery of variables should start out high and then taper off as the survey continues and an increasing fraction of the variables become known variables.  While this may not reduce the required alert rate, if all new variations are announced, the existence of previous detections and alerts can greatly reduce the alert generation and filtering tasks.  

The rate of variable detection can be cast with reference to the probability, $f_{\sigma}$, that a star when observed at one epoch will be undergoing a brightness excursion greater than the selected threshold.  We will assume a simple model in which all targets have only two distinguishable states, with frequency of occurrence of $f_\sigma$ and (1 - $f_\sigma$).  A variable will be detected when the star switches from one state to the other, assumed random.  Then at the $n$th epoch, the number of targets that have not yet been observed to have switched states for the first time will be $N_0 [f_\sigma^n + (1-f_\sigma)^n]$, where $N_0$ is the total number of detectable variables and $n$ is the number of epochs.  There are two terms here, one for beginning in the low state and switching to the high state, and {\it vice versa} - that are about equally important in this case.  This is a two-process exponential decay problem and the expression for the number of yet undetected variables can also be written $N = N_0 [f_\sigma e^{-\lambda _1 n} + (1-f_\sigma) e^{-\lambda _2 n]}$, where $\lambda _1 = \ln{f_\sigma}$ and $\lambda _2 = \ln(1-f_\sigma)$.

The exponential drives very rapidly toward zero, so for most types of continuous variability, the discovery rate is strongly front-loaded and decreases to ``noise" within a few tens of epochs.  

As used here the term ``epoch" refers to the characteristic time over which {\it observations} of the brightness of the type of variable considered decorrelate, so it is the larger of the intrinsic decorrelation time of the target and of the revisit interval.  Values will be introduced below in the discussion of discovery rates for specific surveys and target types.

\subsection{Discussion of False Positives}

Controlling the number of false positives in a survey can have multiple motivations and criteria, and the latter can differ across the data products, and can evolve with time.

In a mature survey, the rate of false detections should hopefully be relatively stable. The rate of real discoveries of variability will start out higher than pessimistic false positive rates, but after a few tens of epochs, should fall below some estimates of false positive rates \citep{beck2013}. This will be true for virtually any threshold, for example in the range 3--6$\, \sigma_{phot}$, hence the actual value assigned is more a matter of convenience than a fundamental decision.  For example, the threshold could be set low initially, as long as the ratio of false to real detections was small, or it could be set high initially to suppress both false and lower-significance real detections and reduce the volume of the alert stream.  Late in the survey, the threshold could be set high to reduce contamination of increasingly rare real discoveries, or it could be set low to minimize rejection of possibly real events, which would be most detectable at this stage.  An interesting approach possibly attractive to operations planning would be to vary the threshold so that the rate of new discovery alerts  remained relatively stable through the survey.

For our investigation, we will explore thresholds of 1 and $5\,\sigma_{phot}$.  A value in the vicinity of 5 will probably be most suitable for alert generation, and a value in the vicinity of 1 for post-processing detections at the end of a many-epoch  survey.

\section{Application of Besan\c{c}on Models and Kepler Statistics to Future Surveys}

It would be attractive to apply this model to past surveys, but we have been discouraged from doing so owing to the lack of published information that can be directly compared to total variable count predictions.  

For future surveys, the following results depend not only on hardware not yet tested on the sky, but assumptions about survey execution and analysis.   Therefore, the results should be understood as illustrative, hopefully a fair representation of expected survey products, but subject to adjustment for different assumptions.

\subsection{The Large Synoptic Survey Telescope}

Quantifying the LSST alert rate and Broker tasks is a major motivation for this study.  The LSST \citep{ivez2008} will survey the southern sky available from its site on Cerro Pach\'{o}n repeatedly in 6 filters for 10 years.  The actual cadence is not yet fixed, and hence the sky coverage per night and the distribution among filters is uncertain, so we have chosen to consider the variable statistics on a per-field-visit basis, rounding the LSST field size up from the design $\sim9.6$ to 10.0 square degrees for convenience of scaling.  The LSST photometric errors are constrained by a number of design specifications \citep{lsst2011}, but of course the delivered performance is not known and depends on observing conditions.  We have used an LSST ``exposure calculator", version 4.3 \citep{gee2008}, that implements equations (4) and (5) from \cite{ivez2008}, calculating S/N for point sources including sky subtraction. The Kepler bandpass has a response centroid of 643 nm.  We have used the LSST {\it r} bandpass for comparison (response centroid 620 nm).  It is understood that LSST will have a saturating bright limit - we have adopted $r = 15$, but the total counts are not significantly dependent on this cut.

LSST is expected to reach all-sky photometry calibration of $\sigma_{cal} \sim$ 5 mmag.  Though it will sometimes operate under conditions that are not photometric in the classical sense of low and stable extinction, the large field and great depth of LSST imagery will allow strong constraints with differential photometry relative to nearby stars.  Pending better analysis, or until it is resolved by on-sky experience, we adopt $\sigma_{cal}$ = 5 mmag as the rms calibration noise for a single visit, and we will require $5\, \sigma_{phot}$ significance for variability detection and alert generation.

\begin{figure*}[ht]
\centering
\includegraphics[width=6.8in]{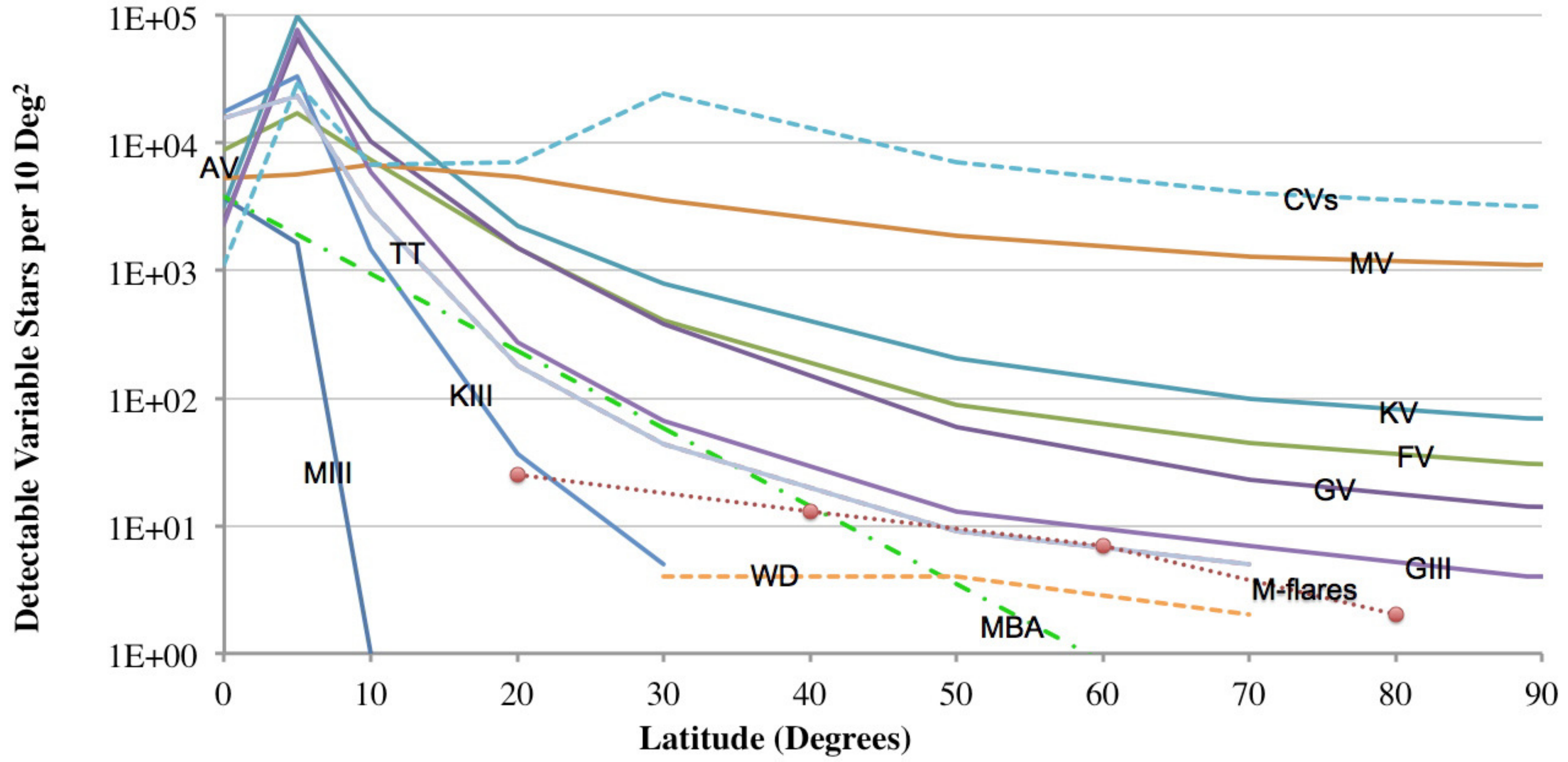}
\caption{Number of detectable variables per field of 10 square degrees, per spectral type, for an LSST-type survey, with detection in the  {\it r}-band at $5\,\sigma_{phot}$ significance.  Stars brighter than $r=15$ are excluded.  The ordinate follows a great circle that for the stars starts at the Galactic center, to the south Galactic pole (at 90 degrees). Solid lines show main sequence plus G and K giants, TTauri (TT) and white dwarf (WD).  These are total TT and WD counts, not considering level of variability.   The predicted dip at the Galactic equator is due to higher extinction in the plane.  The text symbol {\it MIII} reprents a single point for the M giants, which have a non-zero density only in the galactic plane (where extinction brings them into the $r > 15$ brightness range).   The text {\it AV} represents a similar single point on the plane for A dwarfs. Additional material is included to suggest rates for other targets.  Dashed line - estimated {\it upper limit} to the possible number of cataclysmic variables (CV, section \ref{CV}), without regard to possible detectability - the observable number is surely much lower.   The dash-dot line shows the predicted distribution of detectable main belt asteroids (MBA, section \ref{MBA}), for which the ordinate shows ecliptic latitude. The dotted line connects 4 circles that show the predicted number of detectable M-dwarf flare events (amplitude greater than 0.1 magnitudes in the {\it u}--band) {\it per visit} \citep{hilt2011} - most of these stars will be included in the dwarf data sets at the top of the figure. } \label{GreatCircle}
\end{figure*}

Figure \ref{GreatCircle} shows the expected numbers of detectable variable stars at the 5-$\sigma$ threshold, per 10 square degrees, as a function of sky position, along a great circle from the Galactic center to the south Galactic pole.  The total counts are dominated by the dwarfs, with the M dwarfs most numerous except near the Galactic plane, where K and G dwarfs are competitive due to a combination of star column density and line-of-sight extinction. The cataclysmic variables (CV) will be discussed in section \ref{CV}, and the main belt asteroids (MBA) in section \ref{MBA}

\begin{figure*}[hb]
\centering
\includegraphics[width=6.8in]{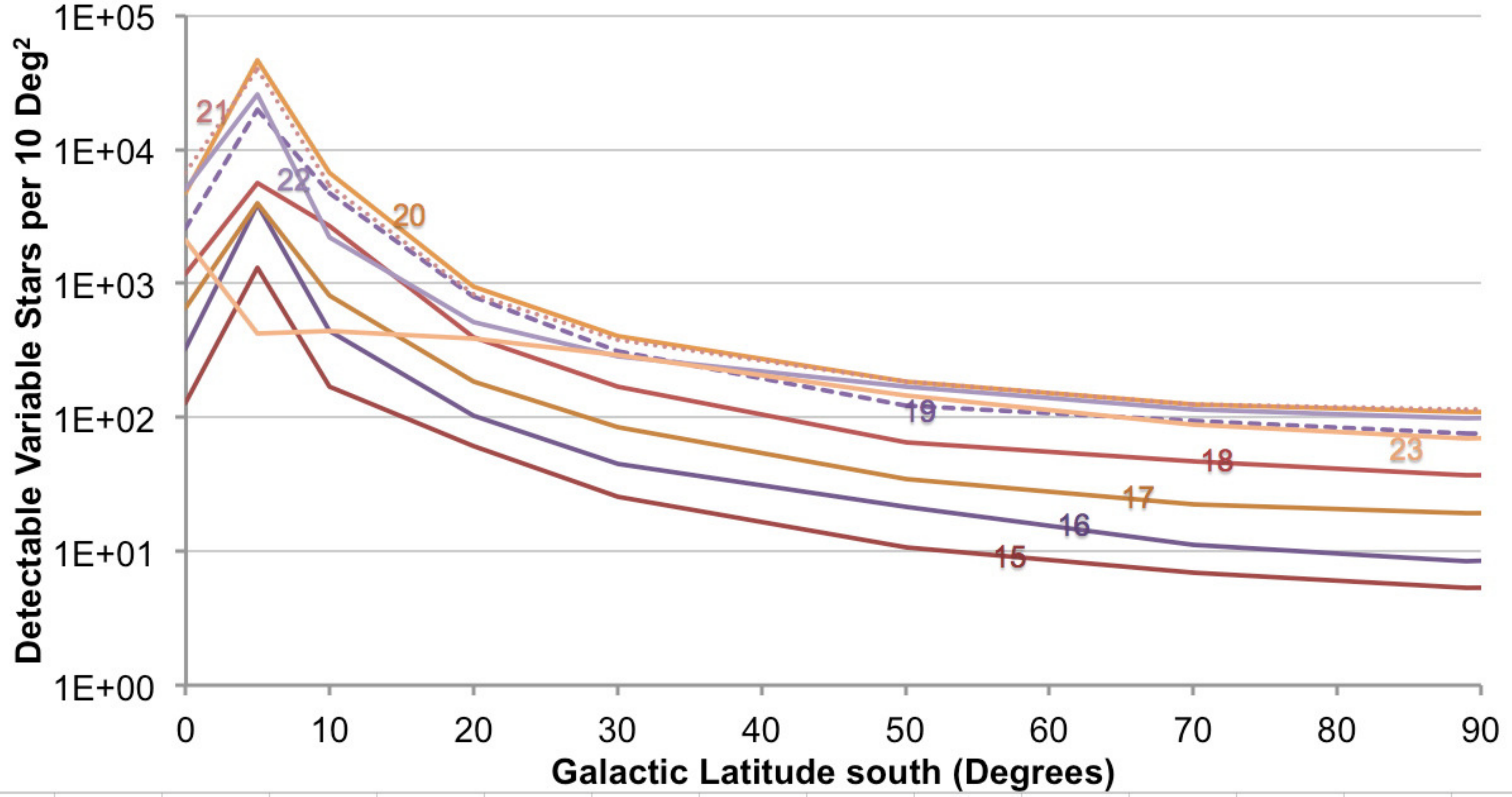}
\caption{Number of detectable variable stars per field of 10 square degrees, per 0.5 magnitude bin in apparent {\it r} brightness, for an LSST-type survey  at $5\,\sigma_{phot}$ significance.  Some curves (the fractional magnitude bins) have been deleted to avoid crowding.  The labels on the curves correspond to the center of the 0.5 magnitude bin.   \label{ByMags}}
\end{figure*}

Figure \ref{ByMags} illustrates how the number of detectable variables depends on the brightness range considered.  Each curve represents the total stars from Figure \ref{GreatCircle} sorted by 0.5 magnitude bins.   The majority of detectable variables at high latitude will be brighter than $r\simeq 21$. Near the Galactic center the peak brightness for the number of detectable variables shifts over the range 20--21 and will be direction dependent anywhere near the plane.

\begin{figure}
\includegraphics[width=3.5in]{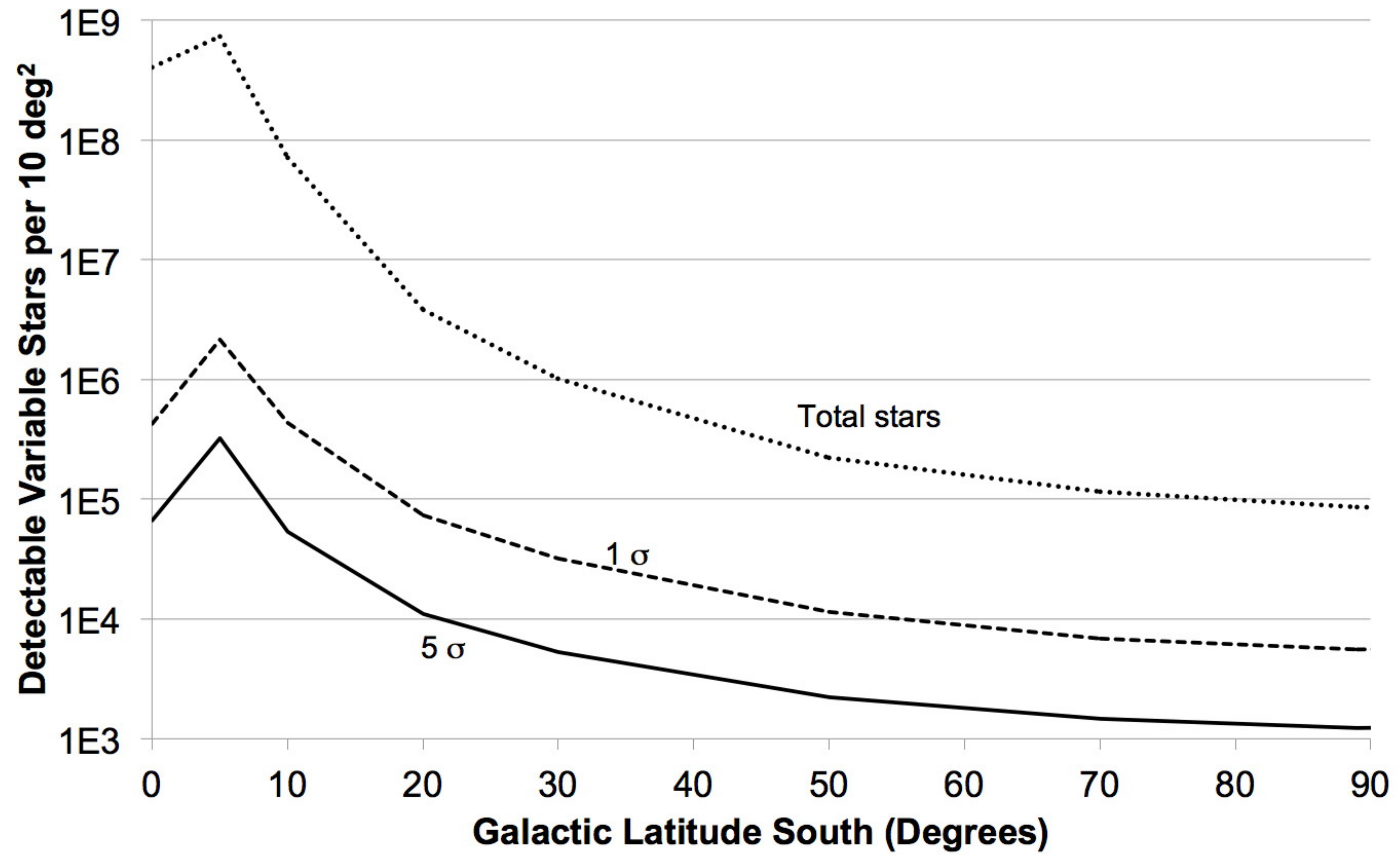}
\caption{Density of stars, per field of 10 square degrees, detectable by LSST in  {\it r}-band.  Dotted curve - the total number of stars in the range g= 15 to 24.5 detectable in one visit at 5--$\sigma$ significance.  Solid curve - the numbers of stars whose variability is discoverable, for threshold: 5--$\sigma$.  Dashed curve: the total number of stars detectable at 1--$\sigma$ at the end of the survey.  Assumption: $\sigma_{cal} = 5$ mmag.
 \label{TotalStars}}
\end{figure}

Figure \ref{TotalStars} shows how the total number of detectable variables depends on the detection threshold.   The $5\,\sigma_{phot}$ solid curve effectively sums the information in Figures \ref{GreatCircle} and \ref{ByMags} and shows the number of detectable variables with that criterion, which should approximate the number of stars that can alert for variability.  The dashed curve for $1\,\sigma_{phot}$ shows a likely number when end-survey post-processing for variable detection is considered.  The dotted line shows for comparison the total number of stars (variable {\it and} non-variable) detectable in the range {\it r}= 15 to 24.5, using a $5\,\sigma_{phot}$ single visit detection threshold.  This is to clarify that the $5\,\sigma_{phot}$ threshold applies somewhat differently to detection of non-variable stars, where it is the total stellar flux that must be detectable, and to discovery of variability, where it is the variation that must be detectable.

\begin{figure}
\includegraphics[width=3.5in]{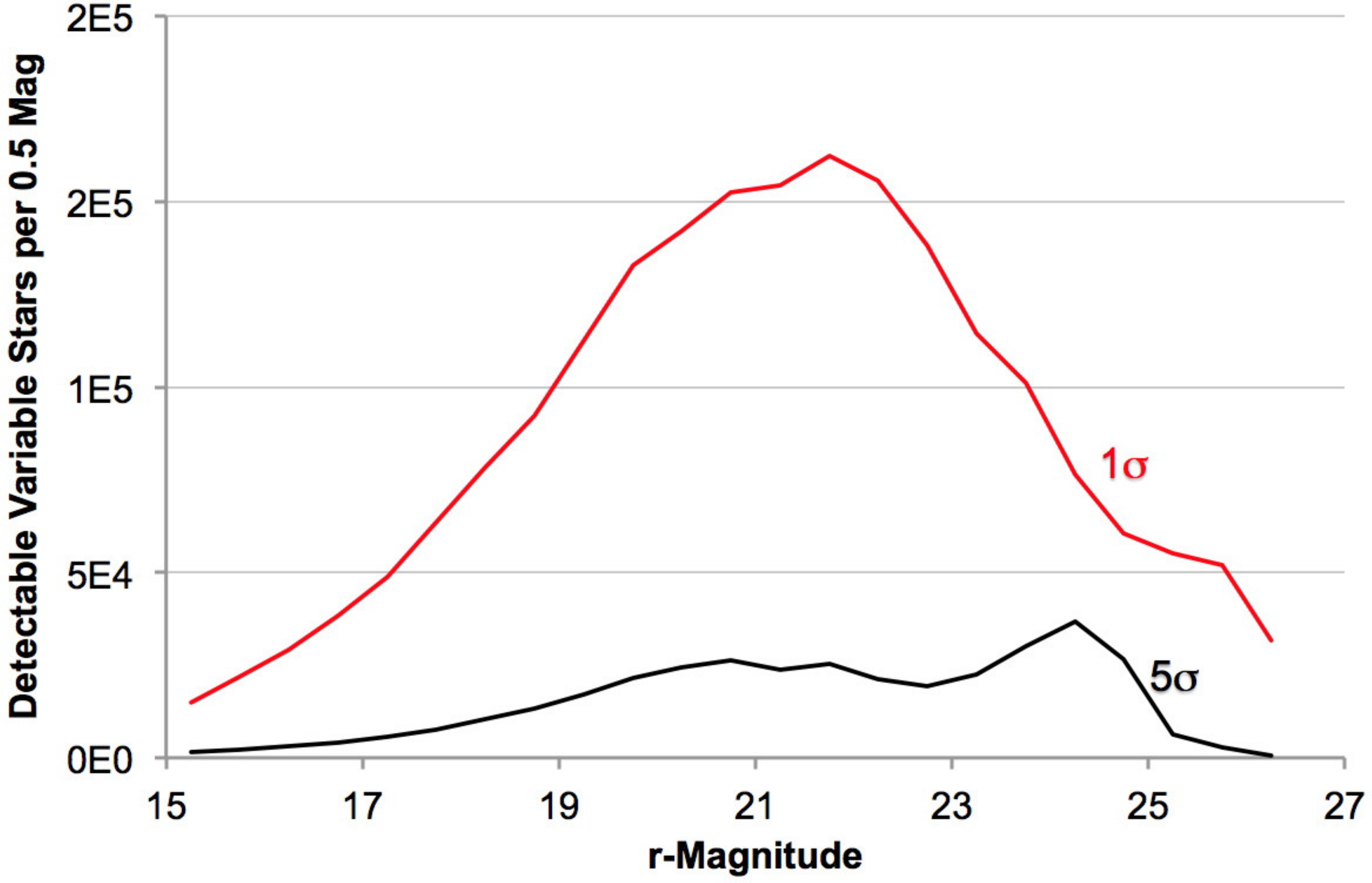}
\caption{Total number of LSST detectable variables integrated over high latitudes ($b =-20$ to $-90^\circ$), per bin of 0.5 magnitudes in the  {\it r}-band, for a threshold of $5 \,\sigma_{phot}$ (solid line, appropriate to alert generation) and 1$\sigma_{phot}$ (dotted line, appropriate to end-of-survey post processing).  Assumption: $\sigma_{cal} = 5$ mmag.
 \label{LSST-ByMags}}
\end{figure}

Since the total numbers across the sky are dominated by the Galactic plane, we consider next the plane and high latitude regions (taken here as $|b| > 20^\circ$) separately.  A grid of models sampling from the plane to high latitudes has been used, with appropriate weighting, to find the distribution of detectable variables across the sky.   The predicted total counts of detectable variables in the south high latitude region are   $4\times 10^6$ ($5\;\sigma_{phot}$), and $3\times10^7$ ($1\:\sigma_{phot}$).  Figure \ref{LSST-ByMags} shows the integrated total detectable variables for $|b| \ge20^\circ$, plotted as a histogram in 0.5 mag bins.  The totals are modest, even for a threshold of $1\,\sigma_{phot}$.  The distributions confirm that the most numerous variables will be detected 2-3 magnitudes brighter than the LSST single-visit detection threshold.

In the plane of the Galaxy the focus will be on Galactic targets, there will be interest in detecting large numbers of variable targets of all types, and compiling complete samples may be useful.  We have studied the Galactic plane with  samples on a grid.    An appropriately weighted sum for the Galactic plane ($|b| \leq 20^\circ$) gives counts of $2\times 10^7$ and $2\times 10^{8}$ for 5$\sigma_{phot}$ and 1$\sigma_{phot}$, respectively, always subject to caveats about extinction and crowding.  Again, most of these variables will not be close to the LSST single visit faint detection limit.

\placefigure{GreatCircle}

\placefigure{ByMags}

\subsection{GAIA}
\label{GAIA}

The capability of the GAIA mission to discover variable stars has been studied with the same tools described above.
The GAIA mission \citep{perr2001} will deliver  time-resolved photometry for large numbers of Galactic variables, with $\simeq70$ observations per target over an interval of 5 years.  The mission will make photometric measurements in two passbands, designated blue and red \citep{brui2005}.  The performance in terms of detection error per visit (i.e. taking into account the information obtained from multiple detectors during the progress of a single scan) has been characterized in terms of the GAIA G-magnitude, for stars in the range {\it G}= 12-20, as $\sigma_{det} = 10^{-3}(0.02076 z^2 + 2.7224z + 0.004352)^{1/2}$, where $z = 10^{0.4(G-15)}$ \citep{esa2014}.  Note that the GAIA {\it G} bandpass is very similar to the Kepler bandpass, and hence may be compared  to LSST {\it r}.

\begin{figure}
\includegraphics[width=3.5in]{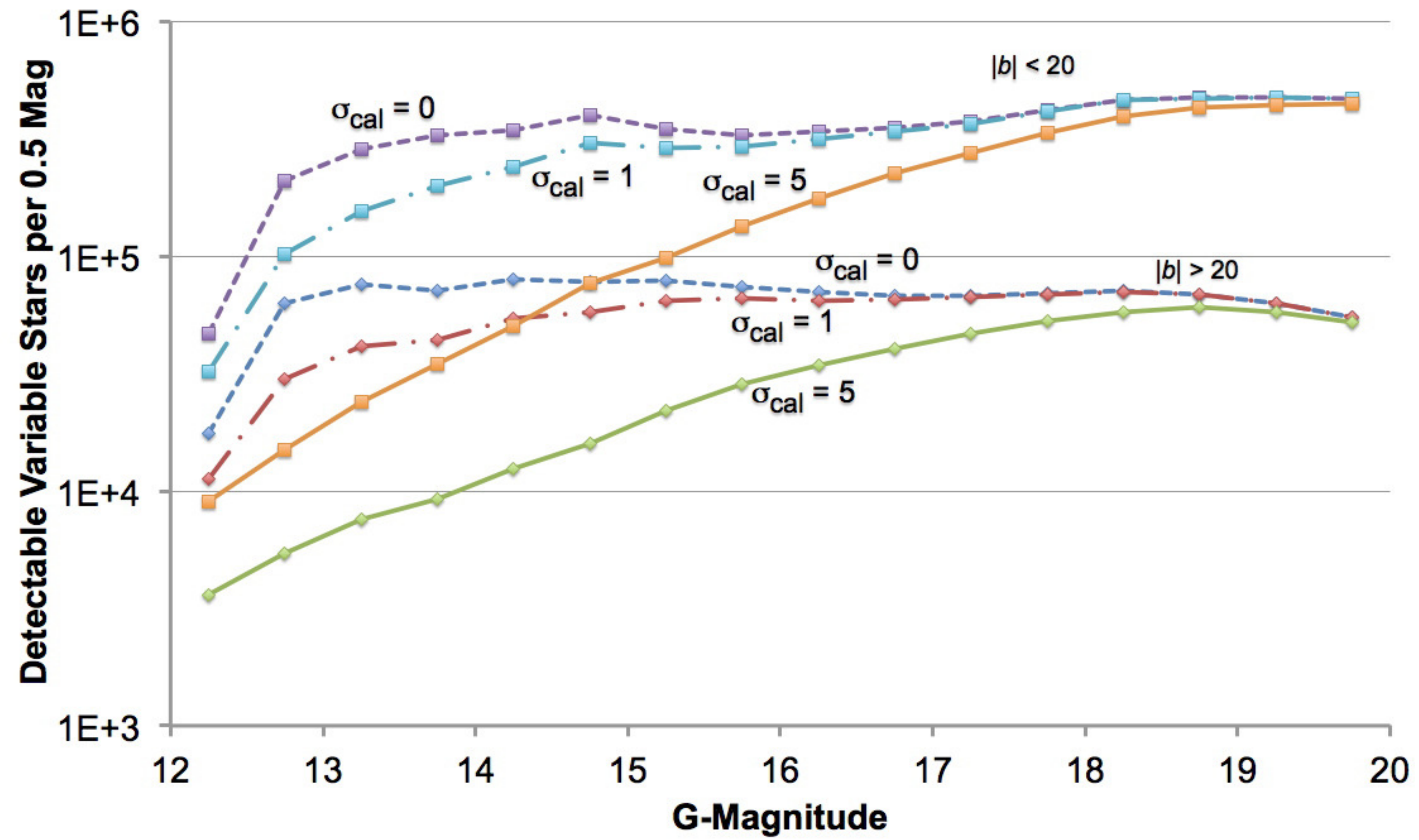}
\caption{Distribution of GAIA detectable variables  per 0.5 {\it G}--magnitude bin for calibration noise $\sigma_{cal}$ = 0 mmag (dashed line), 1 mmag (dash-dot line), and 5 mmag (solid line), for a detection threshold of $5\,\sigma_{phot}$.  The lower count group (triangle symbols) corresponds to the integrated high latitude sky ($|b| > 20^\circ$), and the higher count group (squares) corresponds to the integrated low latitude sky ($|b|\leq 20^\circ)$. The photometric calibration error will not be a major factor in the count of detectable variables. See section \ref{GAIA} for details.  
The integrated totals of the $\sigma = 1$ curves correspond to the entries {\it GAIA-S} in Table \ref{GAIA-LSST-counts}.}
 \label{GAIA-ByMags}
\end{figure}

The photometry calibration error for GAIA will naturally be less than for LSST, but it is a little unclear what value to use, particularly for the rapid processing and alert generation \citep{jord2010}.  We have experimented with calibration noise levels  of 0, 1, and 5 mmag, and show these in Figure \ref{GAIA-ByMags}. The GAIA variability detection limit at faint magnitudes is surely limited by photon noise rather than calibration, and the lower (and zero) calibration error assumptions simply increase the number of very bright and only slightly variable targets identified.  

Quantitatively, we find that for 1 mmag calibration error, the number of detectable variables will be  $9\times 10^5$ at high latitude and $5\times 10^6$ in the plane.  The numbers with 0 mmag calibration error are larger by $\simeq$20\%, and the numbers for a photometric noise of 5 mmag are $\simeq$35\% smaller, so the total range is $<2\times$. In the following we will take as representative a GAIA photometric error of 1 mmag. The large star count at low-latitude of course depends on GAIA's ability to deal with crowded fields, given the drift scan and data compression strategy.  In this context it is worth noting that GAIA operations has an optional Modified Scanning Law for crowded regions, that the angular resolution of GAIA is very good, and that scans will eventually be combined to form a high-resolution image of the vicinity of cataloged targets.

As with LSST, the detectability of variability for the complete survey dataset will be greater than for the single-epoch determination and we estimate that by considering a threshold noise level of $1\,\sigma_{phot}$.

A numerical experiment shows that for a 2-times increase in variability amplitude, the expected number of detectable GAIA variables increases by a factor 2.1 (3.1) for latitudes $ >20$ or $<20$.

\subsection{GAIA as an LSST precursor}

LSST and GAIA have similar dwell times per epoch, $\simeq30$  seconds for LSST and $\simeq40$ seconds for GAIA.  LSST is a large telescope and GAIA is a small one, but since study of low-amplitude variables is strongly photometry performance-limited, GAIA's space platform regains a lot of ground with respect to LSST.  At GAIA's operational cutoff of $G=20$, GAIA single-scan signal-to-noise is only a factor of 2.5 lower than LSST's single-visit S/N in {\it r}.  GAIA will precede LSST by a long enough interval that it is reasonable to expect the fully post-processed GAIA catalog to be available when LSST records its first data.  Therefore we will consider estimated GAIA end-of-survey performance, with 1 mmag photometric error and at the $1\,\sigma_{phot}$ (single-scan noise) threshold. This will be compared to the LSST early survey performance, assuming 5 mmag photometric performance, and a $5\,\sigma_{phot}$ (single-visit noise) threshold.

\begin{table}
\begin{center}
\caption{Predicted total detectable variable star counts for GAIA and for LSST, by brightness and latitude.   \label{GAIA-LSST-counts}}
\begin{tabular}{clccc}
\tableline\tableline
Sky region & Survey & {\it G} $\simeq r = 15-20$ &  $= 20-25$ & Total\\
\tableline

$\left| b \right|> 20$    & GAIA-E  & 5e+6 & ...     & 5e+6\\  
$\left| b \right|> 20$    & GAIA-S  & 9e+5 & ...     & 9e+5\\  
$\left| b \right|> 20$    & LSST-S  & 1e+6 & 4e+6& 5e+6\\ 
$\left| b \right| \leq 20$& GAIA-E & 2e+7 & ...     & 2e+7\\  
$\left| b \right| \leq 20$& GAIA-S & 5e+6 & ...     & 5e+6\\ 
$\left| b \right| \leq 20$ & LSST-S& 9e+6 & 2e+7& 3e+7\\  
\tableline
\end{tabular}
\tablecomments{E = detectable by end of survey analysis, S = detectable in a single scan or visit}
\end{center}
\end{table}

\begin{figure}
\includegraphics[width=3.5in]{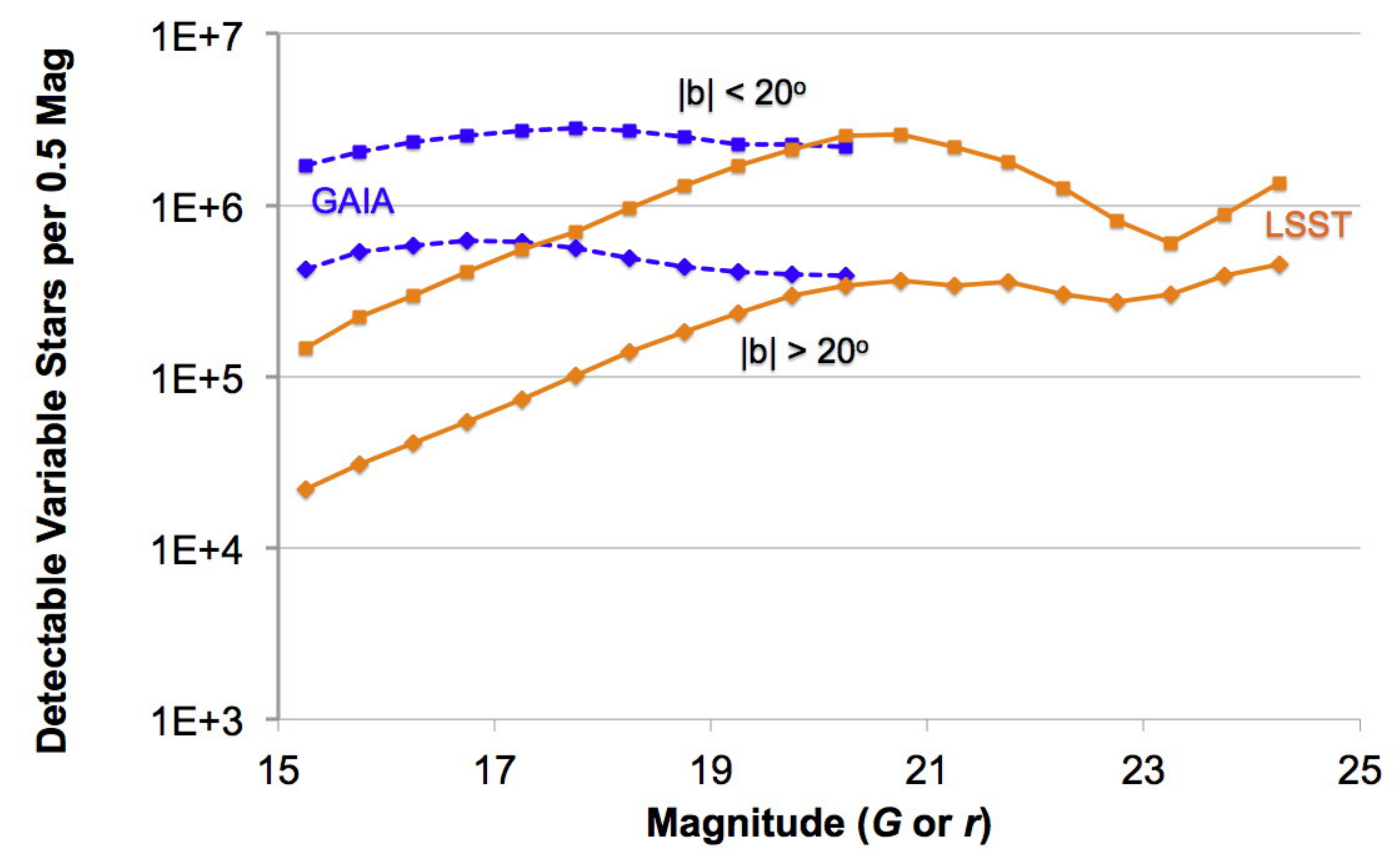}
\caption{GAIA as a precursor to LSST.  Distribution of detectable variables  per 0.5 {\it G}- or {\it r}--magnitude bin, comparing the single-epoch sensitivity for LSST, $5\sigma_{phot}$, to the GAIA performance for $1\sigma_{phot}$, approximating the performance for the end-of-mission processing of the full survey data set - hence the target counts are higher here than in Figure \ref{GAIA-ByMags}.  The curves stopping at {\it G}=20 (dashed lines) represent GAIA performance and the curves extending to {\it r} =25 (solid lines) represent LSST.  The lower count groups (diamond symbols) correspond to the integrated high latitude sky ($|b|> 20^\circ$), and the higher count groups (squares) correspond to the integrated low latitude sky ($|b|\leq 20^\circ)$. The complex structure for the Galactic plane arises due to interplay of the stellar distribution model and the extinction model.
The integrated totals of the GAIA curves correspond to the {\it -E} entries in Table \ref{GAIA-LSST-counts}, and the integrated totals for LSST correspond to the {\it -S} entries.}
 \label{GAIA-LSST-precursor}
\end{figure}

Figure \ref{GAIA-LSST-precursor} shows the expected number of detectable variable stars for GAIA and for LSST in the sky region of overlap.  Thanks to its expected high photometric quality, the GAIA survey will detect lower variability amplitudes for stars brighter than {\it G}=20.  LSST will reach fainter stars, showing a satisfying level of complementarity.  The integrated totals are shown in Table \ref{GAIA-LSST-counts}.  

\section{The Alert Rate for LSST and GAIA}

\begin{figure}
\includegraphics[width=3.5in]{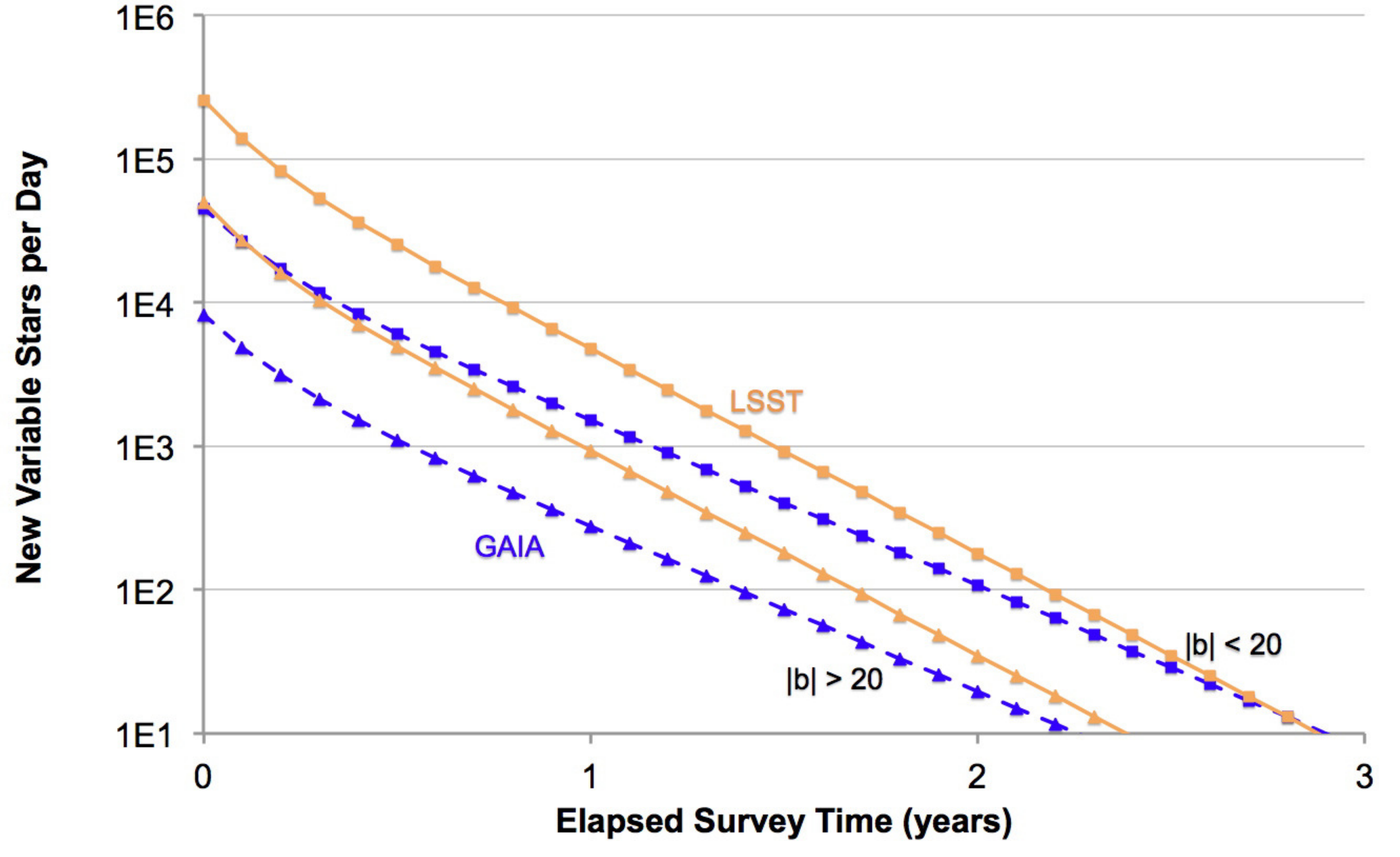}
\caption{Stellar variable discovery rates for GAIA and LSST.  The predicted nightly discovery rates for GAIA and LSST in the first years of their mission/surveys, for the full sky available to each survey, assuming LSST detections only in the  {\it r}--band.  LSST is represented by connected lines and GAIA by symbols only.   The square symbols represent integrated totals for $|b| < 20^\circ$, and the diamonds for $|b| \leq 20^\circ$.  The LSST discovery rate is shown as slightly higher than GAIA (steeper descent), but could actually be somewhat slower depending on the actual cadence adopted.
}
 \label{LSST-GAIA-Compare-Rates}
\end{figure}

Both GAIA and LSST plan to provide the community with timely alerts for newly identified targets of interest.  Particularly early in the survey,  it will sometimes be difficult to distinguish novel from mundane targets.  We expect that Galactic stars will be a major contributor to the count of variable/transient targets, and we have now assembled the information needed to predict the discovery rates for such variable stars. Our approach is to use the tools and results described above, especially in section \ref{decline}.  The initial total pool of targets to be discovered for LSST is the sum of the low and high latitude columns in Table \ref{GAIA-LSST-counts}.  The corresponding total for GAIA is Given by the GAIA-S entries in the Table. The predicted discovery rates for GAIA and LSST are shown in Figure \ref{LSST-GAIA-Compare-Rates}.  This is for a required $5\,\sigma_{phot}$ detection in all cases, and the photometric $1\,\sigma_{phot}$ noise is 1 mmag for GAIA and 5 mmag for LSST.  

The time scale is determined by the rate of measurement of decorrelated epochs of variability. In the case of GAIA, the $\sim70$ visits will be obtained in a complex and precisely known pattern over approximately 5 years.  For LSST, $\sim 180$ {\it r}--filter visits will occur  in a yearly modulated but not yet known complex pattern over 10 years.  For both surveys, there will be close pairing in time for some or most observations, reducing by $\simeq 2\times$ the number of independent epochs.   With this correction, we estimate typical values of 52 days between well-separated  {\it G}-band observations for GAIA and 42 for LSST {\it r}--band visits.  These are long enough to exceed the variability time scale of most stars in our study, and so we use these numbers as the interval between measurement epochs.

Figure \ref{LSST-GAIA-Compare-Rates} shows the variable star ``alert problem"  for GAIA and LSST, initially severe for the first year of the survey, nominally dropping to modest numbers within a few years.  The time scales are similar for the two surveys.   Of course the numbers are imprecise, but to the extent that an exponential relation applies, the qualitative conclusion is inescapable - most variable stars will be identified relatively early in the surveys, and need not overwhelm the search for rarer targets subsequently.  (Note that GAIA and LSST report different planned policies, with LSST attempting a more comprehensive alert broadcast stream and GAIA more selective.)

Figure \ref{LSST-GAIA-Compare-Rates} carries an important message for the processing power required to handle alert production.  In the case of LSST, the planned program is to emit an alert every time a target is observed to deviate from a reference image or value.  Variable stars will be the object of repeated alerts (in many cases, alerts will be produced on virtually every observation).  From the number of variable stars, this clearly implies a lot of alerts.  However, after a short break-in period, the majority of stars generating alerts will already have a previously generated data package, the only additional information will be a few added data points, and derivative characterizations can be formulated in incremental algorithms for rapid updating.

\begin{figure}
\includegraphics[width=3.5in]{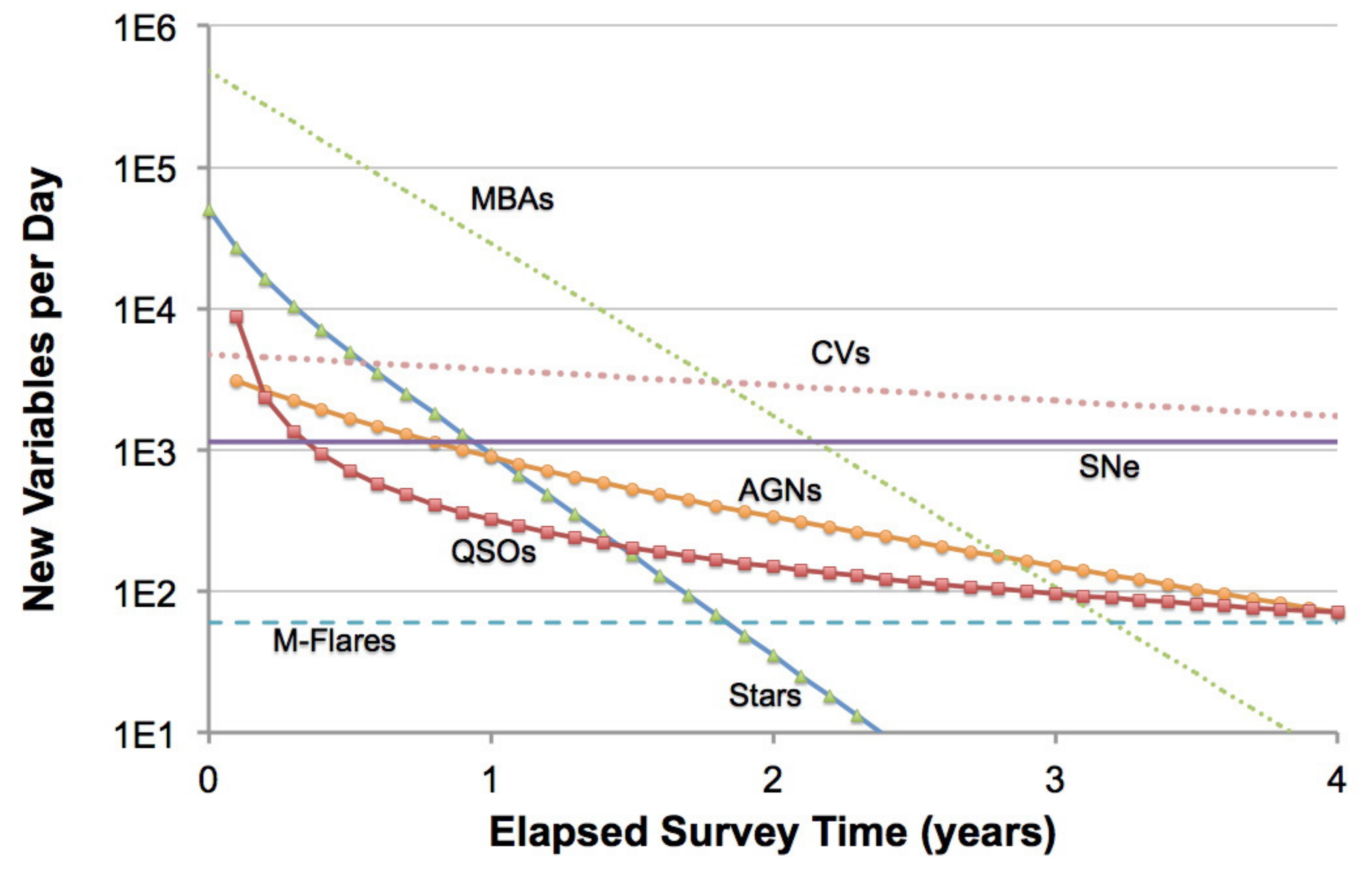}
\caption{Prediction of the LSST alert rates (i.e., discoveries based on 60 second analysis of new visits) for variable targets in high latitudes, $|b|>20^\circ$.  A solid line represents variable stars from the combination of Kepler statistics and the Besan\c{c}on model.  The solid curve labeled QSOs shows a predicted discovery rate based on the QSO variability structure function.  The solid curve labeled AGNs corresponds to the predicted variable AGN discovery rate assuming a characteristic variability time of 0.5 years.  
(The differences between AGNs and QSOs are due to different data availability and models, and may not be significant.)
The  dashed CV curve gives an estimated extreme upper limit to the CV discovery rate for an assumed burst interval of 2 years, and a discovery probability of 0.5 for each burst.  The dotted line is a prediction for the nightly rate of observation of main belt asteroids (MBA) that have not yet been adequately characterized, with numerous assumptions described in section \ref{MBA}, including a characterization success rate of 50\% per data set.
The dashed curve shows a probable conservative upper limit to the frequency of observed flares from M-dwarfs that are not detectable in the quiescent state \citep{hilt2011}.  The solid line for SNe represents all SN types, of which approximately 1/3 will be of type Ia.
}
 \label{Rate-of-Discovery}
\end{figure}

In the following sections we will gather information on detection rates for other types of variables, and these will be collected in Figure  \ref{Rate-of-Discovery}.  This figure shows numbers only for the LSST high latitude sky.  It is not directly applicable to GAIA because the detection limits and cadence are different, but a similar comparative analysis of GAIA will lead to similar qualitative conclusions.  The LSST high latitude detection rate for stellar variables from Figure \ref{LSST-GAIA-Compare-Rates} appears in Figure \ref{Rate-of-Discovery} as a single line labeled ``Stars".  The additional information in Figure \ref{Rate-of-Discovery} will be discussed below.

\section{Cataclysmic Variables}
\label{CV}

Cataclysmic variables are binary stars that consist of a compact object (almost always a white dwarf) and most commonly a relatively normal low mass, main sequence companion.  They are formed in systems initially close enough to participate in a stage of common envelope evolution, resulting in a shrinkage of the orbital separation, and eventually mass transfer events and outbursts.    Cataclysmic variables are somewhat problematic for variable counts as there are several CV types, and the CV phenomenon is known primarily through the longer orbital period, high mass transfer objects, whereas the majority of the systems should be of shorter orbital period with lower mass transfer rates \citep{howe1997} which are less well known.

The number of CVs reported from among the Kepler targets is small \citep[and references therein] {howe2013}, and the Kepler faint limit is not well matched to the quiescent brightness of the most numerous components of the CV population.  A deep survey may well reveal disproportionately more detectable variables of this type than projected from less deep surveys  \citep{howe2001}.  Modeling the CV population is complex, with several distinct stages of evolution, and poorly constrained statistics. 

The approach we have taken is to set a limit to the maximum possible number of CVs.  We assume that all binaries in the galaxy that could evolve to CVs have already done so and remain in the active CV population indefinitely.  The terms to include are the fraction of stars that are binaries and the fraction of binaries that have the right combination of orbital parameters and masses to evolve into CVs.  For the fraction of star systems that are binary or multiple, an observational study of F and G dwarfs by \citet{duqu1991} led to an estimate of 0.57 (not counting substellar companions), whereas \citet{lada2006} suggests that knowledge of the IMF implies a value of 0.33.  For the binary fraction, we use a compromise number of 0.5.  The fraction that evolves to CV has been estimated from Monte Carlo simulations of binary evolution by \citet{howe1997} and  \citet{howe2001} as 0.0038.  The Besan\c{c}on model galaxy does not give explicit binary statistics, so we used the model white dwarf population and  estimate that the probability that each is a member of a CV system to be no greater than (0.5*0.0038).  

We have run a small set of Besan\c{c}on models with the limits set to output all white dwarfs, regardless of apparent brightness, and applying our limit, show these as a dashed line in Figure \ref{GreatCircle}.  There it can be seen that the CV number density upper limit is about an order of magnitude below that of main sequence dwarfs, hence if all potential CVs were active CVs, with detectable outbursts, these would be an important contributor, but would not dominate, the variable star totals.  

The large number of potential CVs suggests that  CV outbursts may be an important contributor to transient object alerts.  Some of these objects will be cataloged prior to flaring, though most will not have been distinguishable as CVs in the quiescent state, and some will appear as blank sky transients.  An estimate of the upper limit to the CV discovery rate can be derived.  The outburst frequency can be estimated in the range 0.1 to 1 per year.  The length of outbursts will typically ensure detection of sufficiently bright events when they occur, unless during the fraction of the year, estimated as 0.5, when the field is not regularly observed.  A simple model gives an upper limit for the CV discovery rate as shown in Figure \ref{Rate-of-Discovery} for an assumed outburst frequency of 0.5 per year.  It is important to note that this rate is several orders of magnitude higher than estimates based directly on CV observations.  However, even if the fraction of CVs that gives detectable outbursts is small, it could dominate the stellar variable discovery rate after an initial survey startup period, and perhaps more importantly could be a major contributor to the blank sky transient rate.  More frequent outbursts would give a steeper decline in the discovery rate. In fact synoptic surveys will contribute to clarifying CV evolution and statistics. 

\section{Dwarf flares}
\label{Flares}

M dwarf flares could be numerous because the stars are so common.    It was noted above that the Kepler variability rms numbers are not very sensitive to rare outliers - the complementary fact is that the rms values do not tell us much about rare, transient events, so the Kepler analysis does not really describe flare frequency.  Furthermore, flares are represented poorly in the Kepler survey, owing to the small number of cool M stars in the Kepler catalog and perhaps also their characteristic ages.  What we do know is that \citet{walk2011} have studied the flare rates in Kepler K and M dwarfs, finding flaring in $\sim1.5$\% of the some 23,000 cool dwarfs observed, with the vast majority having peak flare amplitudes below the likely LSST $5\,\sigma_{cal}$.   Since flaring is thought to be correlated with spotting and rotation, it is likely that most flaring dwarfs will appear in the variable dwarf counts. 

  \citet{kowa2009} have studied the flare rates for M dwarfs, and \citet{hilt2011} have applied the results to the specific question of the frequency of flaring event detections in the LSST survey. While the flare rates, particularly for large flares and in the Galactic plane, are uncertain, and the predictions depend on several assumptions, the numbers are an important guide, as they set a floor to the transient rate.  
  The \cite{hilt2011} predictions for the {\it u}--band are included in Figure \ref{GreatCircle}.

An interesting prediction is the frequency of detectable flares from stars that are below the survey limit.  \citet{hilt2011} estimate this frequency (per epoch per 10-square-degree field) to range from 0.01 at high latitudes to 0.1 at low latitudes, with the frequency in the plane higher but not estimated.   This prediction is for the {\it u}--band, but it is included in Figure \ref{Rate-of-Discovery}, using the low latitude estimate of 0.1, and assuming visits to 600 distinct fields per night.  Although the discovery rate should drop with time, we have no basis for estimating a slope.  Expecting that primarily rare, very strong flares will contribute to this count, the discovery rate may stay rather flat, as shown.

\section{Other variable and transient sources in deep surveys}
\label{Rates}

Galactic stars are only one component of the variable sky, and other types of targets may compete numerically, especially toward the higher latitudes.  We started with the Galactic stars for good reasons: they are important in driving alert totals, and there are good sources of data for statistical descriptions of variability with considerable detail by stellar type.  Beyond that, they serve as a useful model for how to approach the topic of variable detection and alert generation. 

We will mention here three other important sources of variable/transient targets.

\subsection{Quasi Stellar Objects}

QSOs are an important subgroup of AGNs, and for our purposes are defined by the selection criteria of our primary data source, \cite{macl2012}. The amplitude of observed QSO variation depends on the time interval studied, with the occurrence of large amplitude changes increasing with time.  The characteristic timescales, up to hundreds or thousands of days, are similar to survey durations.  We have utilized the cumulative VPDFs from   
\cite{macl2012}.   We describe the cumulative probability as a linear function of $\Delta m$.  \cite{macl2012} present a number of data sets for different observing intervals ranging from 10 to 1000 days.  Data sets for different magnitude cutoffs in different bandpasses are not obviously consistent, so we have brought the different data sets into a single representation for comparison.

\begin{figure}
\includegraphics[width=3.5in]{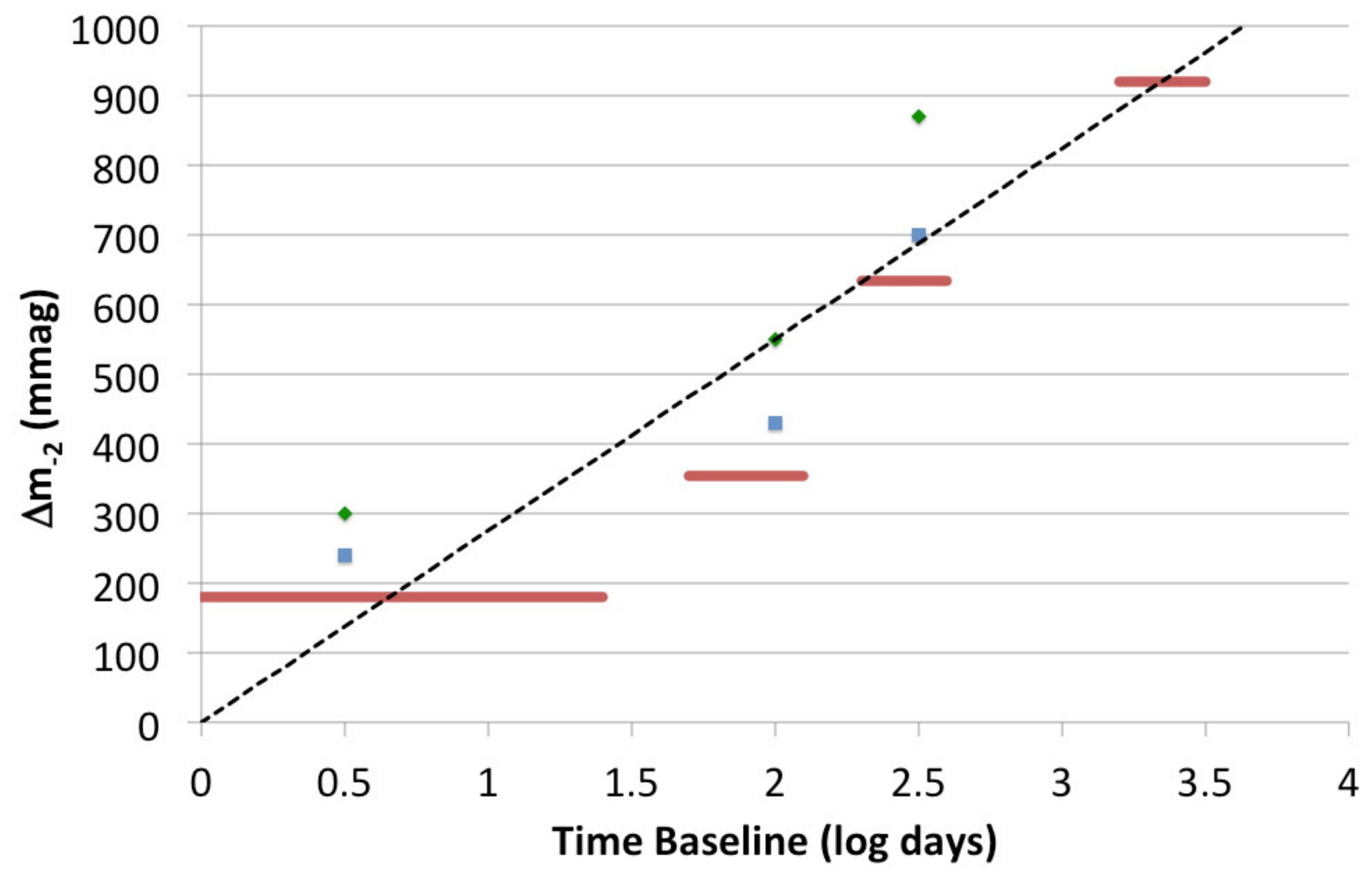}
\caption{ Evaluation of a constant, $\Delta m_{-2}$, that describes the implications of the QSO time variability structure function, based on r-band data of \cite{macl2012}.  The red bars show data from their Figure 2, for QSOs selected as $r < 20$, for several interval ranges.  The symbols represent data from their Figure 18, describing variability of QSOs selected by $u < 22$ (green diamonds) and $u < 24.5$ (blue squares).  The dashed line represents the relation adopted here for computations.  }
 \label{QSO-VPDF-constant}
\end{figure}

We describe the relation analytically as $\log p = -2\Delta m / \Delta m_{-2}(\Delta t)$, where $\Delta m_{-2}$ is the value of the magnitude difference between two measurements corresponding to a cumulative probability $p$ of $\log p = -2$, and is a function of the interval between measurements.  Values for $\Delta m_{-2}$ can be read from the \citet{macl2012} figures for different data sets corresponding to different apparent brightnesses.  Figure \ref{QSO-VPDF-constant} shows our evaluation of its dependence on $\Delta t$.  We see an approximately linear relationship between $\Delta m_{-2}$ and $\log($days$)$, and while there is scatter, there does not appear to be a systematic deviation from linearity over the time scales up to 10 years for different selections of targets.  We adopt the relationship $\Delta m_{-2} = 275*\log($days$)$, shown as a straight line in the figure.

\begin{figure}
\includegraphics[width=3.5in]{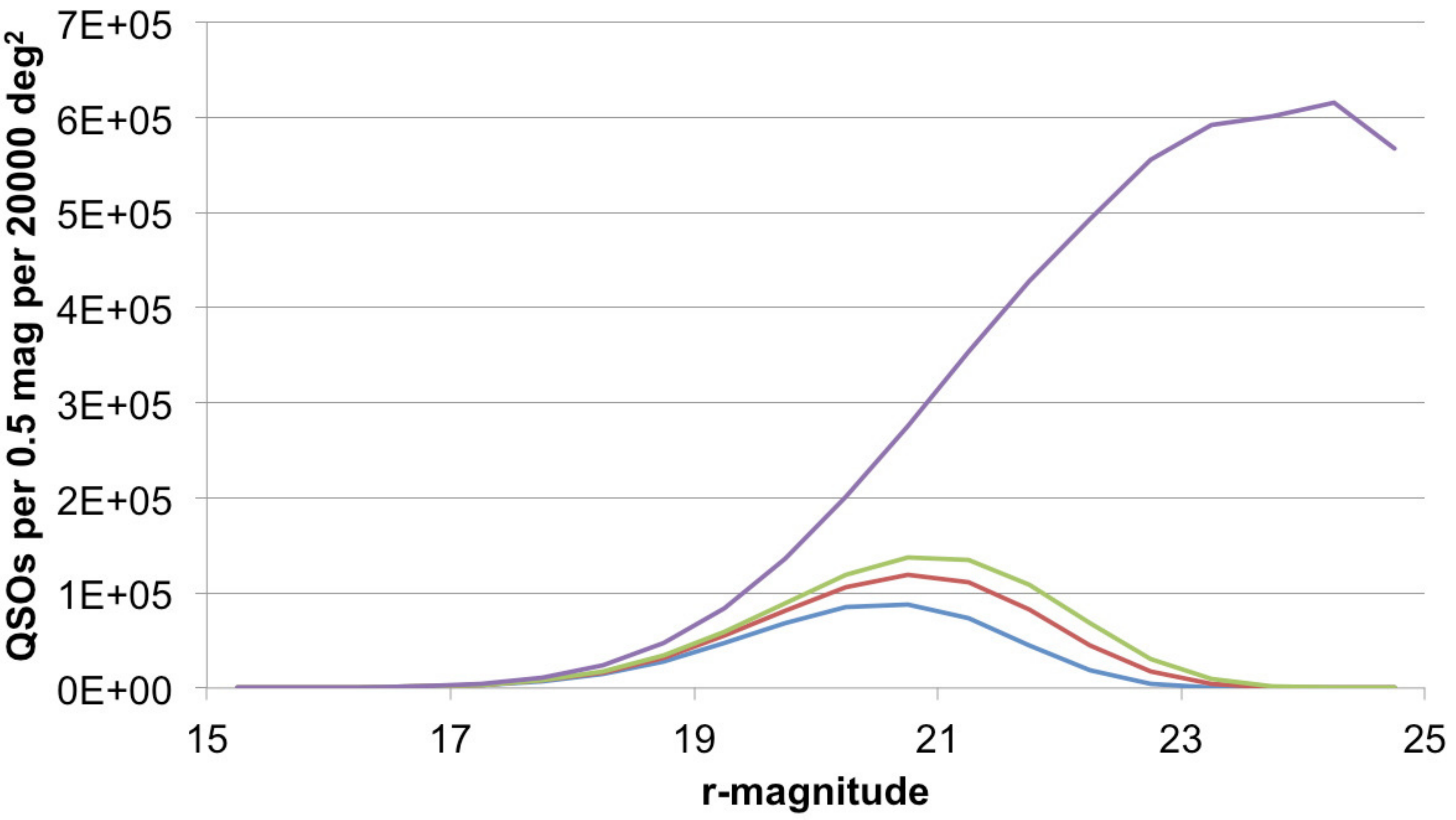}
\caption{Expected number QSOs per 0.5 g-magnitude bin, integrated over 20000 deg$^2$ LSST-accessible sky.  From the bottom, the curves show the counts of QSOs whose variability is expected to be detectable at 5 $\sigma$ in a single-visit after 50, 600 and 3600 days.  The top curve shows the adopted luminosity function.} \label{QSO-fraction}
\end{figure}

The QSO luminosity function is derived from \cite{HRH07}, with a K-correction \citep{pala2013}, and converted to {\it r}--magnitude using a median $g-r$ color relation with redshift.  The adopted relation is shown in Figure \ref{QSO-fraction}.  

With this information, we can integrate over the luminosity function, determining the limiting $5\sigma$ variability detection limit of LSST (single visit), the cumulative fraction of QSOs with expected variability exceeding that value, and summing the total number of detectable variable QSOs at each time step.  The differential with time of that function gives the predicted discovery rate, assuming efficient detection.  
To illustrate this, we have used the techniques described above to evaluate the number of QSOs for which LSST could detect variability with time baselines of 50, 600 and 3600 days.   These distributions are shown in Figure \ref{QSO-fraction}.  

Because most detectable variability is at magnitudes brighter than ~23, the exact trend of the LF at the faint limit is not important. While most detectably variable QSOs will be found early in a survey, the discovery rate falls off more slowly than for stars due to the increase in variability amplitude with time.  The expected detection rate from this model is shown in Figure \ref{Rate-of-Discovery}.  

By simulation, LSST should detect variability in $8\times10^5$ QSOs in a 10-year survey.  If the amplitude of QSO variations is increased by 2$\times$, the number of variable QSOs detectable increases by 21\%.  If the characteristic time constant is increased by 2$\times$, the number decreases by 18\%.

\subsection{Active Galactic Nuclei}

A careful prediction for discovery of AGN should take into account a description of the broad range of variability characteristics.  However, the determination of temporal structure functions for large samples is less satisfactory than for QSOs, and hence the behavior of variability amplitude with survey duration is not well established.  Instead, following the basic logic of the analysis of Galactic stars, we propose to estimate the the number of galaxies, the variable fraction of galaxies,  the number of variable AGNs detectable, and then comment on the temporal sampling. 

\begin{figure}
\includegraphics[width=3.5in]{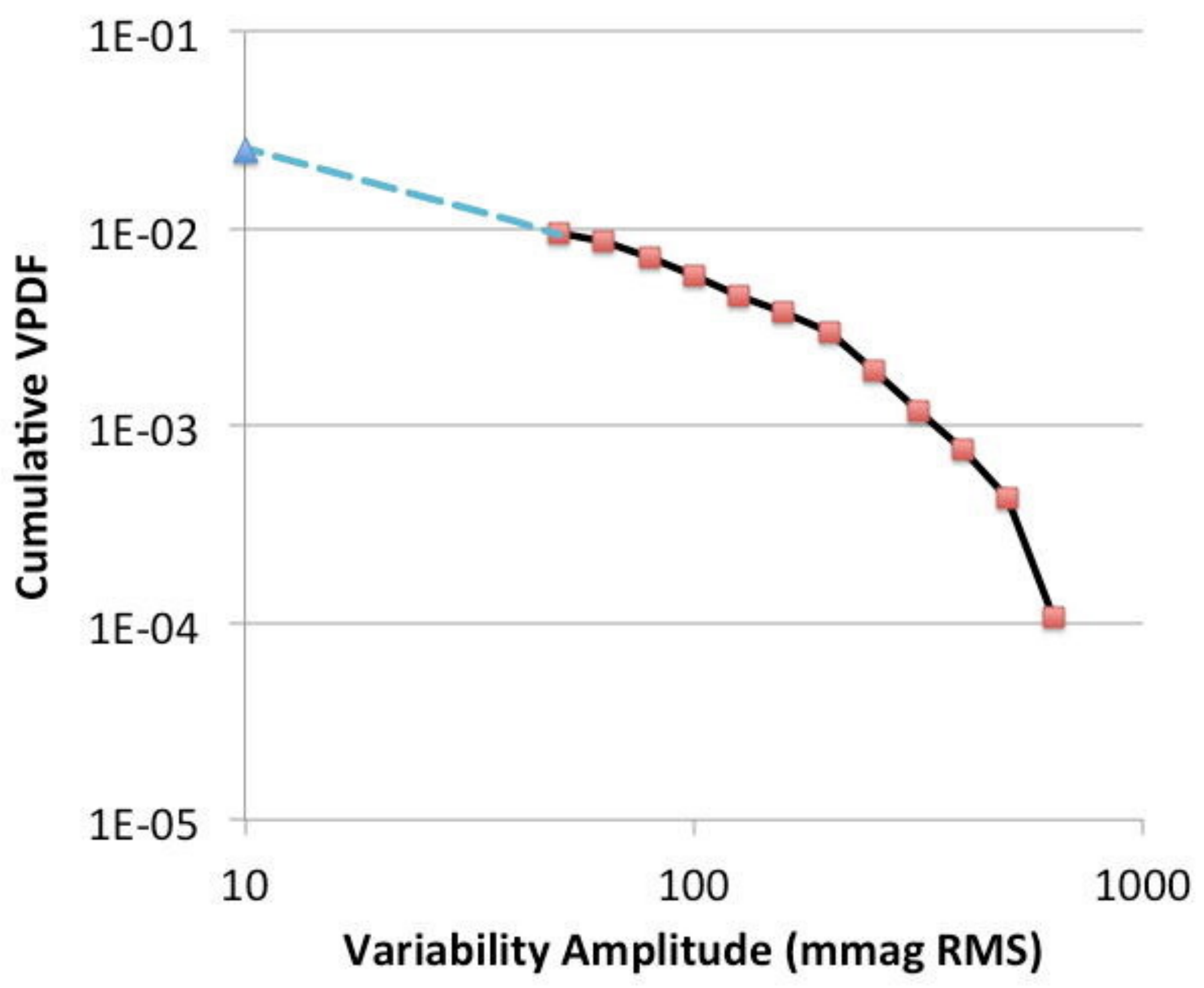}
\caption{Cumulative VPDF for galaxies, showing the data from \citet{kles2012} as squares.  A boundary condition shown as a triangle, explained in the text, implicitly {\it defining} AGNs to have variability $\geq10$ mmag, is the basis for extrapolating the function to smaller amplitudes.    The VPDF is set to zero for variability amplitudes greater than 700 mmag, owing to lack of sufficient data and low probability. }
 \label{AGN-cumulative}
\end{figure}

A VPDF was constructed to satisfy a combination of specific survey data and general constraints.  In surveys, AGNs are identified largely by variability, so the number detected is a function of the photometry and the temporal sampling, and it is not surprising that estimates of the frequency of occurrence vary.   We adopt a recent report \citep{kles2012} that tabulates variability amplitudes for the detected variables among 15,849 cluster galaxies.  The study employed a combination of 2-epoch data (where the brightness difference was reported) and 3-epoch data, for which the $\sigma_{var}$ was reported.  For current purposes we treat 2-epoch differences as representative of target $\sigma_{var}$.  This data is used to construct an observational VPDF that describes the amplitude over the range $\sigma_{var} \sim 50 - 700$ mmag, but leaves the slope for smaller amplitude variability  poorly determined.  Two additional pieces of information are used to establish the slope.  It is assumed that 2.5\% of all galaxies have AGNs \citep{kles2012}, and that all AGNs are variable at the $\sim 10$ mmag level \citep{gask2002}.   The resulting hybrid (observation plus extrapolation) cumulative VPDF is shown in Figure \ref{AGN-cumulative}.  It can be seen that this model converges to $2.5$\% at 10 mmag.  

\begin{figure}
\includegraphics[width=3.5in]{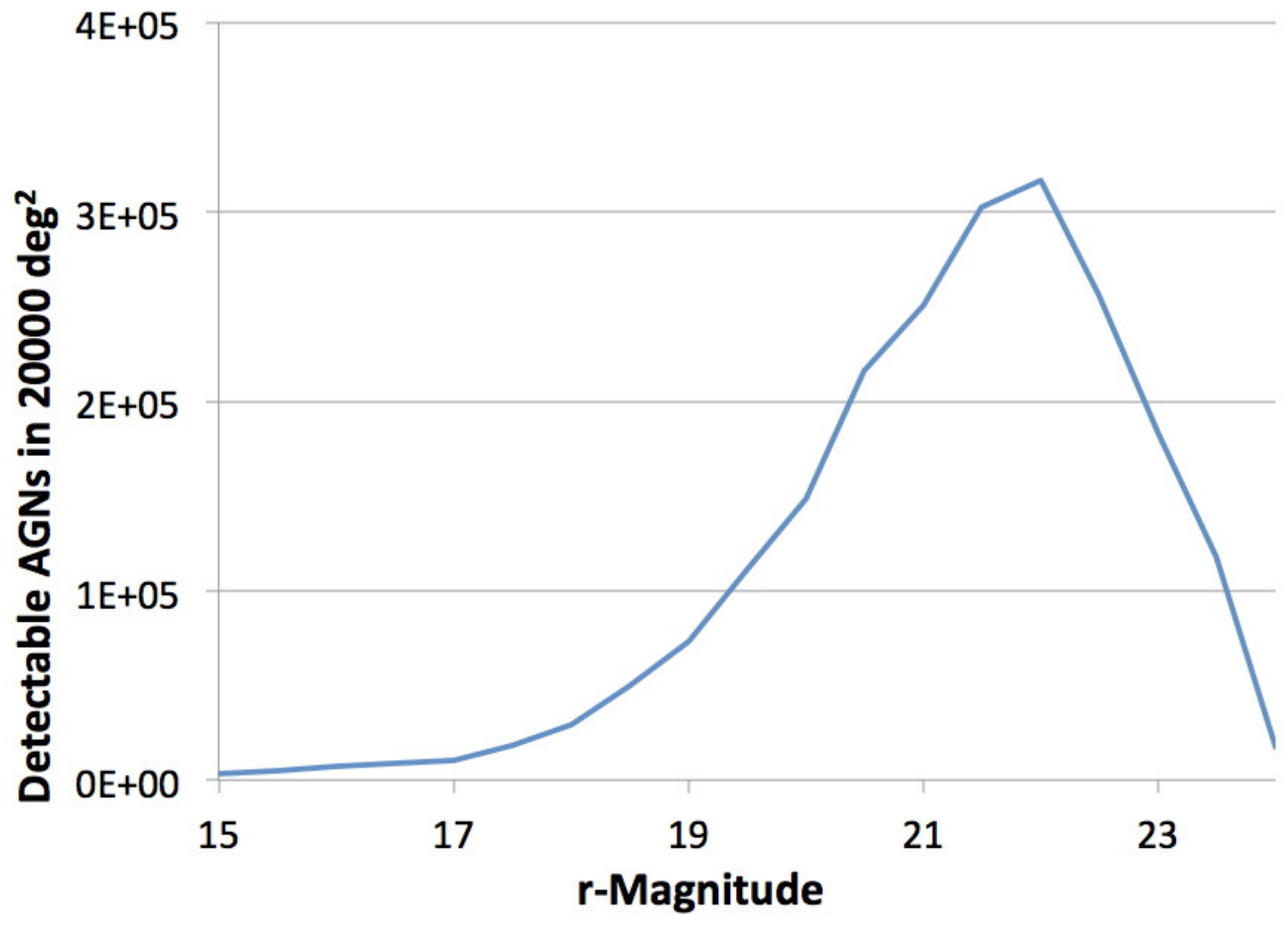}
\caption{Expected number of AGNs per 0.5 r-magnitude bin whose variability can be detected by LSST at the single visit detection threshold, 5 $\sigma_{phot}$, integrated over 20000 deg$^2$ LSST-accessible sky. }
 \label{Variable-AGN}
\end{figure}

In order to apply this VPDF, a distribution function for the number of galaxies vs magnitude is needed.  \cite{metc2001} have compiled a valuable collection of differential density counts.  We employ their distribution function for the {\it R}-band, with a conversion to {\it r}-magnitude by $r = R+ 0.165$ (from \cite{lupt2005}. This distribution predicts $\sim$ 140,000 galaxies per deg$^2$ detectable at the LSST single visit detection threshold.  With a few percent variable, it might be thought that AGNs would dominate the total variable count.  However, going through the calculation with the VPDF and the Galaxy luminosity function, we find that the number of detectably variable AGNs from a pair of epochs will be $\sim150$ per deg$^2$.  The reason is that the Galaxy luminosity function is so strongly weighted toward the faint end that the variability of the overwhelming majority will be undetectable.  The expected distribution with magnitude is shown in Figure \ref{Variable-AGN} for 20000 deg$^2$.

In the case of stars it was a fair assumption that the brightness of most variables would decorrelate between observation epochs.  For galaxies this is not a good assumption, so we also need information about the temporal power in Galaxy variations (e.g. \citep{gask2002}. For present purposes we simply note that the variation is more stochastic than periodic (so $f_{\sigma} \sim0.317$, section \ref{Fsigma}), and the time constant is typically months.  This means that the rate of first-time detection of variability in AGNs will be lower, and the number yet to be detected will decay more slowly than was the case for stars.  For a synoptic survey, the discovery total will be determined more strongly by the duration of the survey rather than by  the number of observations.

For purposes of fielding an estimate, we assume stochastic variation and a time constant of 6 months, so that approximately 1/2 of the yet undetected variables will be discovered in each 6 month interval (neglecting such details as accessibility through the year).  Now the fraction of AGNs observed for the first time to have switched to a detectably different state will be  described as in section \ref{decline}.   As shown in Figure \ref{Rate-of-Discovery}, in this model the discovery rate will begin at $\sim 3000$ per night, and decrease by $50\times$ within 4 years.   It is noted that the difference between the AGN and QSO models may not be signfiicant, as may be seen from the similar trajectories in this figure.

By simulation, the number of variable AGN identified in 10 years should be $\sim 10^6$.   If the amplitude of AGN variations is increased by 2X, the number of variable AGNs detectable increases by 29\%.  The variability structure function for AGN is not well known.  We assumed a single time constant of 6 months.  If a better value were 3 months, the initial detection rate would be approximately doubled, and the decline in detection rate would be faster.  The total detected in a 10-year survey  would not change significantly in this model.

Predictions for AGN detections by GAIA are problematic because of the GAIA optimization for bright point sources, whereby host galaxies may not be identifiable or properly catalogued.  

\subsection{Supernovae}

SNe are a target of LSST key science, but they can also be a source of clutter in the search for certain rarer sources (``One man's trash is another man's treasure", and vice versa).   All of the targets discussed above have the common characteristic that they come from a limited pool and once a member has been discovered, the number remaining to discover is reduced.  This is not the case for SNe, which will be as numerous on the last day of the survey as on the first.   \citet{bail2009} estimate a SNe Ia detection rate for LSST of $1.4\times 10^5$ per year or 380/night, and we adopt this result.  Based on \cite{bulb2012} we multiply this number by 3 to estimate the discovery rate for SNe of all types.

\subsection{Main Belt Asteroids in the LSST survey}
\label{MBA}

The last category of variable targets that will be considered here is the main belt asteroids (MBAs).  A wide and deep survey will detect large numbers.  For LSST, \citet{ivez2009} utilize model orbital element and magnitude distributions to estimate that $N_{MBA} = 5.5\times 10^6$ main belt asteroids will be detectable by LSST.  For comparison with other target types, we partially reverse engineer that simulation, assuming that the asteroids are distributed uniformly in ecliptic longitude, and with an ecliptic latitude dependence from \citet{wolf1978}, we model the distribution of MBA as $F(\beta) = 385 e^{-0.14\beta}$ where F is in objects per square degree and $\beta$ is the latitude in degrees.  Near the ecliptic plane, the number of objects will be in the range 3000 per LSST field, while at ecliptic latitudes greater than about $30^\circ$ the density will drop to a few tens per field.  In the following it will be assumed that the orbits of all of these targets are initially unknown.

Since asteroids move rapidly, this corresponds in some respects to up to thousands of detections per field per observation.  As may be seen in Figure \ref{GreatCircle}, if these cannot be readily identified and tracked, then in the ecliptic at high galactic latitude, asteroids will compete with stars in target numbers.

To add the number of MBAs detected per night to Figure \ref{Rate-of-Discovery}, we reduce the total number of detectable MBAs by $f_{lat}=0.5$ to count just southern hemisphere objects.  We apply a factor of $f_{sky}=0.25$ for the fraction of the sky observed each night, and a factor $f_{opp}=0.5$ to select the fraction that are in a relatively favorable position (with respect to opposition) for detection.  So the MBAs should be observed at a (very uncertain) initial rate of about $N_{MBA}f_{lat}f_{sky}f_{op}=280,000$ per night.  Every subsequent detection on following nights will be a new ``discovery" until the orbit is characterized.  

An adequate analysis of MBAs requires orbital modeling, which is possible, and currently unknown information about the efficiency of orbit characterization from the LSST cadence.  Without detailed justification, we simply assume that the detectable asteroids will be observed with datasets suitable for orbit determination on average once every 2 months, and that the characterization will be successful 50\% of the time.  This model gives the predicted observation rate of objects whose orbit has not yet been characterized as shown in Figure \ref{Rate-of-Discovery}. (A proper simulation of the MBA discovery rate with a distribution of orbits will show that it has a long, low-level tail, due to objects that are not detectable every year.)

From Figure \ref{Rate-of-Discovery}, it appears that MBA discoveries will dominate other variable target counts at the start of the survey by approximately 10$\times$.  MBA targets are of course moving significantly, and to some extent (to be determined) this motion will enable them to be distinguished in individual exposures.  However, even if only 1\% of the MBAs cannot be identified by apparent motion, the unidentified MBAs will still represent a very large contribution of anonymous transients.  Unless MBAs can be immediately identified as moving objects, they will dominate the variable discovery problem at the start of the survey.  The schematic model suggests that a modest success rate at orbit determination will suppress the problem by, initially, about an order of magnitude per year.

\section{Summary and Discussion}

This study was undertaken to determine the scale of the LSST alerts and Broker tasks with respect to the number of variable and transient targets that will be discovered.  The Kepler Q13 data base was used to ``calibrate" variability in a Besan\c{c}on Galactic model and predict the number of detectable variable stars as a function of position on the sky.  An upper limit to the number of cataclysmic variables shows that they should be less numerous than variable dwarfs.  Applying the model to LSST and GAIA predicts that each will detect    $>10^6$ variable stars at high latitude and   $>10^{7}$ at low latitude.  

It is suggested that the alert and Broker tasks will depend mainly on the rate of discovery of {\it new} variables, as on subsequent detections there will already by a near-complete data package and analysis of the target.  The temporal characteristics of the variable stars and the surveys were used to predict the variable star discovery rates for LSST and GAIA, starting at $\simeq  10^5$  per night, but with efficient discovery, dropping by about 1 order of magnitude per year.  For comparison, the rate derived for AGNs begins as $3\times 10^3$ per night, but decays more slowly due to the longer time constant.   QSOs are close to the AGN trend.  Figure \ref{Rate-of-Discovery} suggests that discovery of rare events will become rapidly easier over the interval 2-3 years into the survey.  On the other hand, attempting to push this activity sooner might be an exercise in frustration.

The expected discovery rate for main belt asteroids begins very high, a few$\times 10^5$ per night.   In a deep survey, asteroids present an almost existential problem in the ecliptic, which must be mastered to a high level of success.  They are so numerous that even a small incompleteness could threaten other, difficult, science objectives.  Reducing the discovery rate depends on the success of characterizing each target  - a success rate of 50\% per data set or better appears satisfactory.  

The SN discovery rate is predicted to average $\sim 1100$/night with no decay, and may exceed most other discovery rates a few years into the survey.

Before undertaking this study, we could not determine which of the 3 categories - stars, galaxies, asteroids - might dominate LSST discovery and alert tasks, and Figures 6 and \ref{Rate-of-Discovery} show why it was not obvious, as they all compete for dominance in various parts of the sky and at various times.  

\subsection{What could go wrong and what could be done better?}

This study gives estimated counts for numbers of variable stars. The source sample is large and heterogeneous, but does not represent all infrequent variable types.  However, uncommon targets will represent small corrections to the total counts. 

The Kepler statistics may not fairly represent low amplitude, very slow stellar variation, and its incidence is unknown.  Our statistical representation of Kepler data does not capture rare transient variability.

The population dependence of the stellar variability characteristics has not been studied.  In a simple exercise, it was found that increasing the variability amplitudes by 2X, the number of detected variables in the plane increased by 3.25X, and the number at high latitude by 2.3X. 

The predicted star counts in the Galactic plane are enormous.  Some will be inaccessible owing to blending.  However, variable star discovery may be less limited by confusion than one intuitively expects, since the requirement to have photometrically detectable variability ensures that the variables counted here are magnitudes brighter than the survey faint limit.

The star counts in the plane strongly depend on extinction.  We have used the default Besancon extinction.  Increasing the extinction by 2X (in absorption) results in a decrease in the number of detected variables by a factor of 2X.

The upper limit for CVs may greatly exaggerate their observable numbers.

The study of AGN variability needs long duration sampling of a large number of targets to give a stronger observational basis for the VPDF, and to quantitatively characterize the variability time scales.

The calculations are largely schematic and represent very simplified descriptions.  The assumption of Gaussian statistics is an approximation - it could be improved on the variable side by using more elaborate descriptions of observed variability, and on the instrument side with simulations using actual detection and photometry algorithms, but ultimately probably requires on-sky experience with survey detectors as well.

If false detections are generated by systematic effects such as PSF differences rather than statistical noise, and cannot be suppressed, the variable detection rates suggested here may not be achievable.

Improved dwarf flare statistics are needed, especially for deep observations, over a range of latitudes.

Asteroid discovery/characterization rates and speed of identification could be improved with existing models, Monte Carlo simulations and experience at on-going surveys.

Finally, when the actual discovery rates for various target types have fallen by 1--2 orders of magnitude from the initial rates, second order effects not considered here will enter in and the discovery rates will likely change to a shallower slope or possibly plateau.  Happily, this will be not just  a shortcoming of this analysis, but a welcome discovery space.
 \acknowledgments

We are grateful to our colleagues of the NOAO LSST Science Working Group for discussions, and to participants in Oxford IAU Symposium 285 and the 4th GAIA Alerts Workshop for feedback and encouragement.  David Ciardi provided advice and unpublished data. Annie Robin advised on use of the Besancon models.   Zeljko Ivezi\'{c} and Richard Green gave valuable advice on QSO counts, and Ian McGreer provided the derivation of the {\it r}--magnitude QSO luminosity function from published data.  Patricia Burchat gave valuable insights.

NOAO is operated by the Association of Universities for Research in Astronomy (AURA) under a cooperative agreement with the National Science Foundation.

\vspace{10 mm}

\appendix

The most useful description of the variability fraction proved to be the cumulative VPDF (Section 5) as illustrated in Figure 3 for M0-M2 stars.  The cumulative VPDFs are shown here for all spectral types in Figures 18 - 26, with empirical analytical approximations as described in Section 5 and Table 1.

\vspace{5 mm}
\section{Cumulative Variability Probability Distribution Functions}

\begin{figure}[ht]
\plottwo{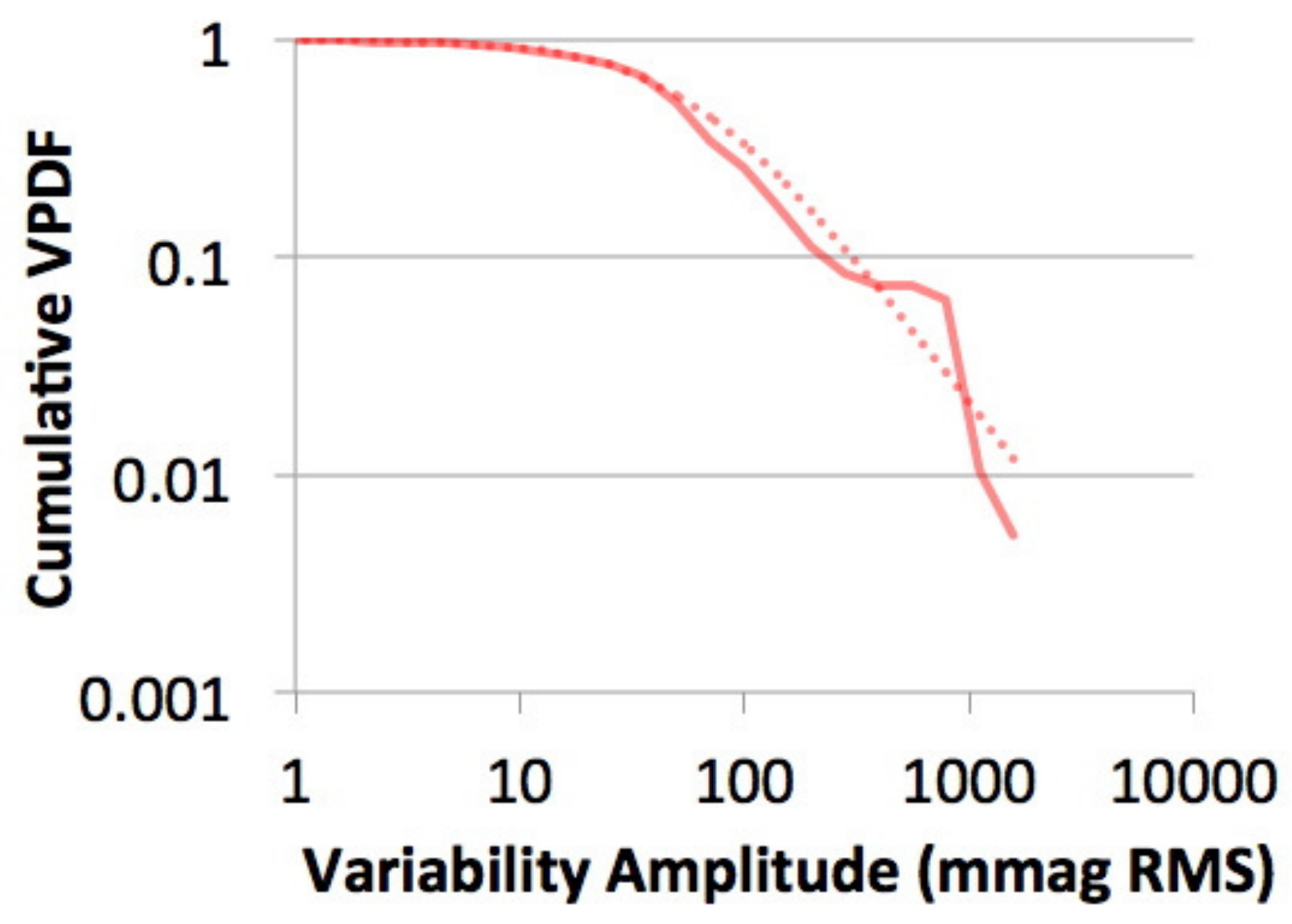}{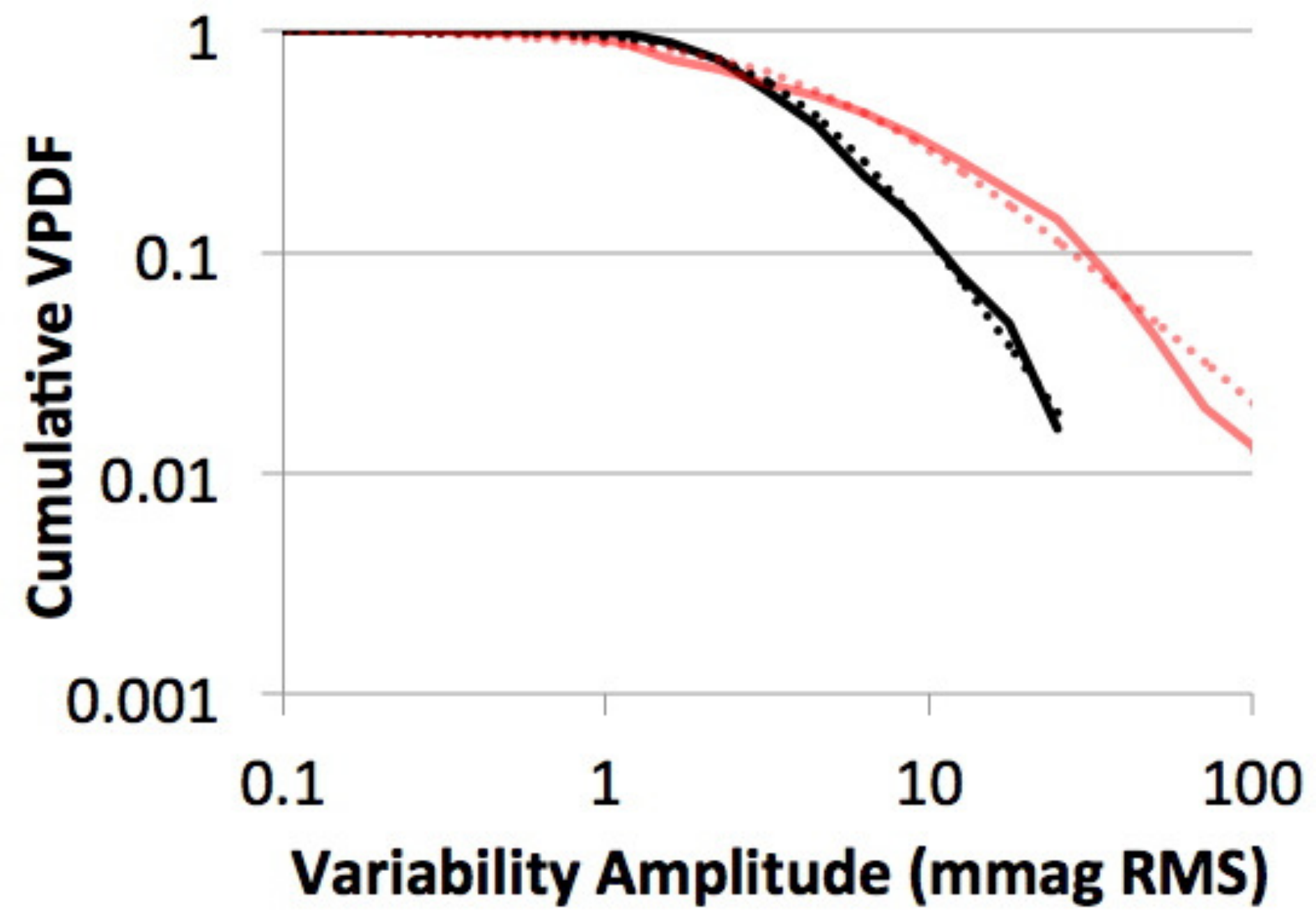}
\caption{  {\it Left}: Cumulative VPDF for Kepler stars M5 and later.  All are assumed to be dwarfs. {\it Right}: Cumulative VPDF for M2-M5 stars.  The red (light) line represents stars of Kepler $\log g \geq 3.5$, and the black (dark) line of $ < 3.5$.  {\it In both figures}, the dotted lines show the empirical fits described in Table \ref{AnalyticalApproximation}}
\end{figure}

\begin{figure}
\plottwo{Cumulative-VPDF-M0-M2-color.eps}{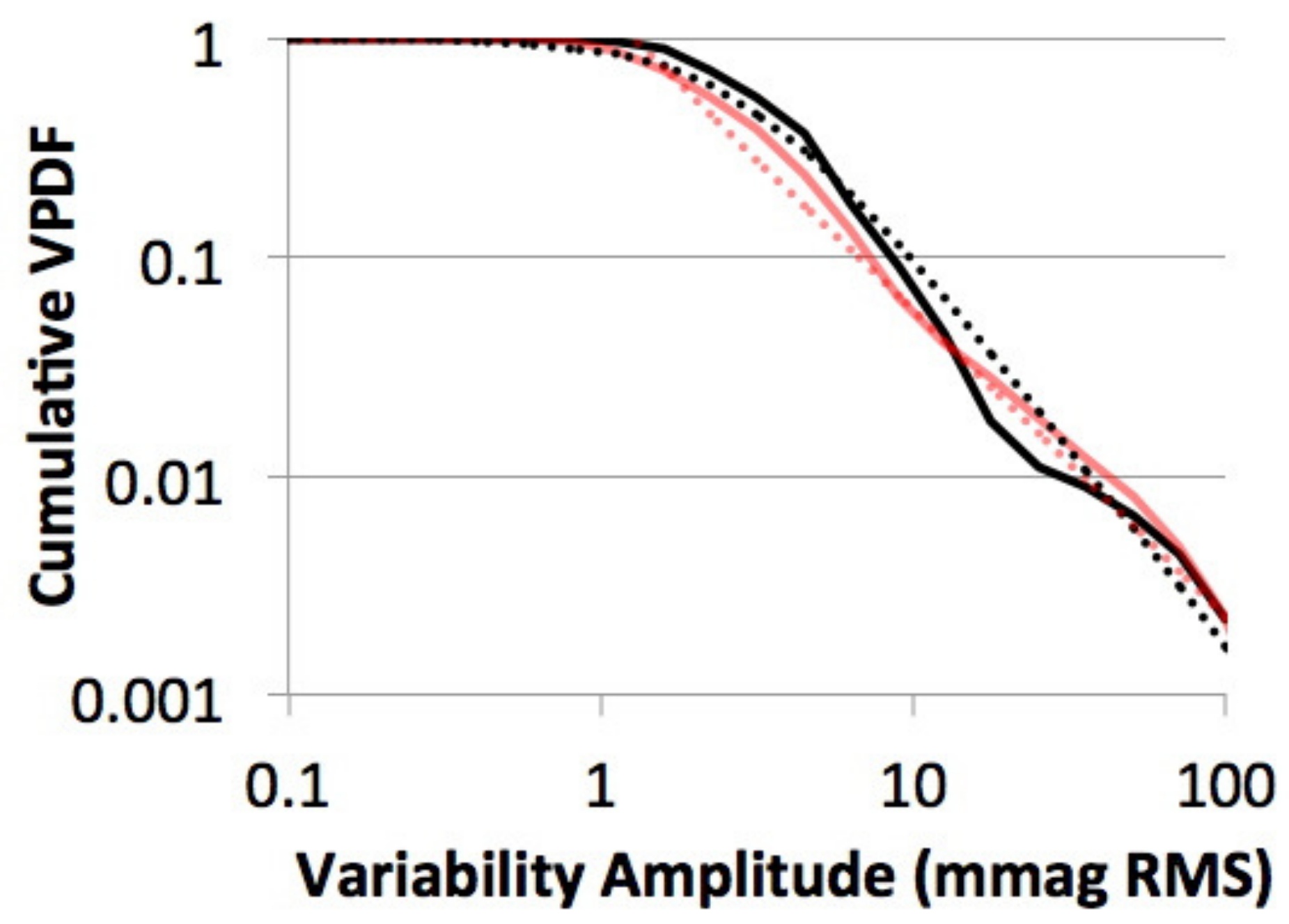}
\caption{  {\it Left}: Cumulative VPDF for Kepler stars M0 - M2 .  {\it Right}: Cumulative VPDF for K5 - M0 stars.  {\it In both figures}: the red (light) line represents stars of Kepler $\log g \geq 3.5$, and the black (dark) line of $ < 3.5$, and the dotted lines show the empirical fits described in Table \ref{AnalyticalApproximation}}
\end{figure}

\begin{figure}
\plottwo{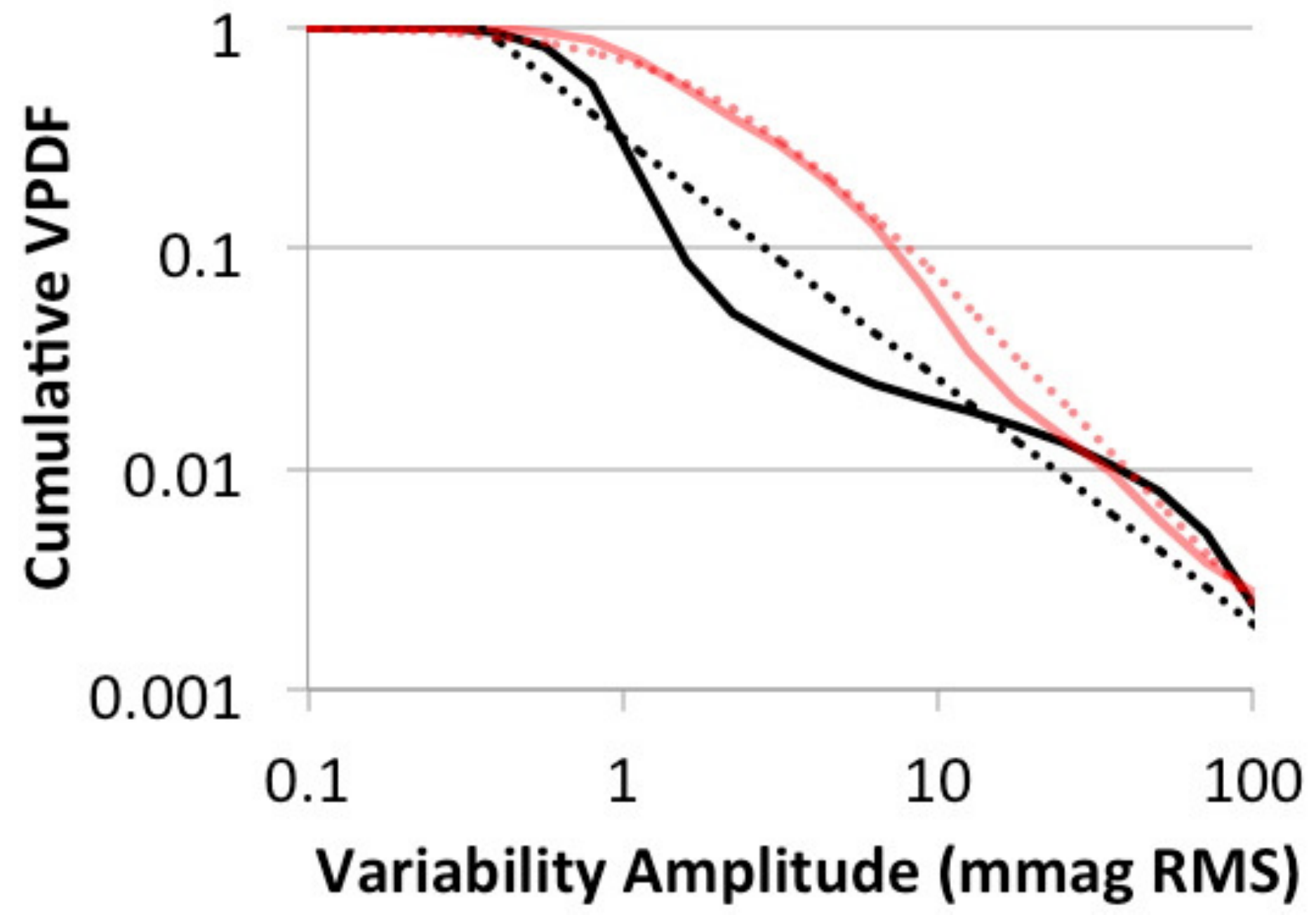}{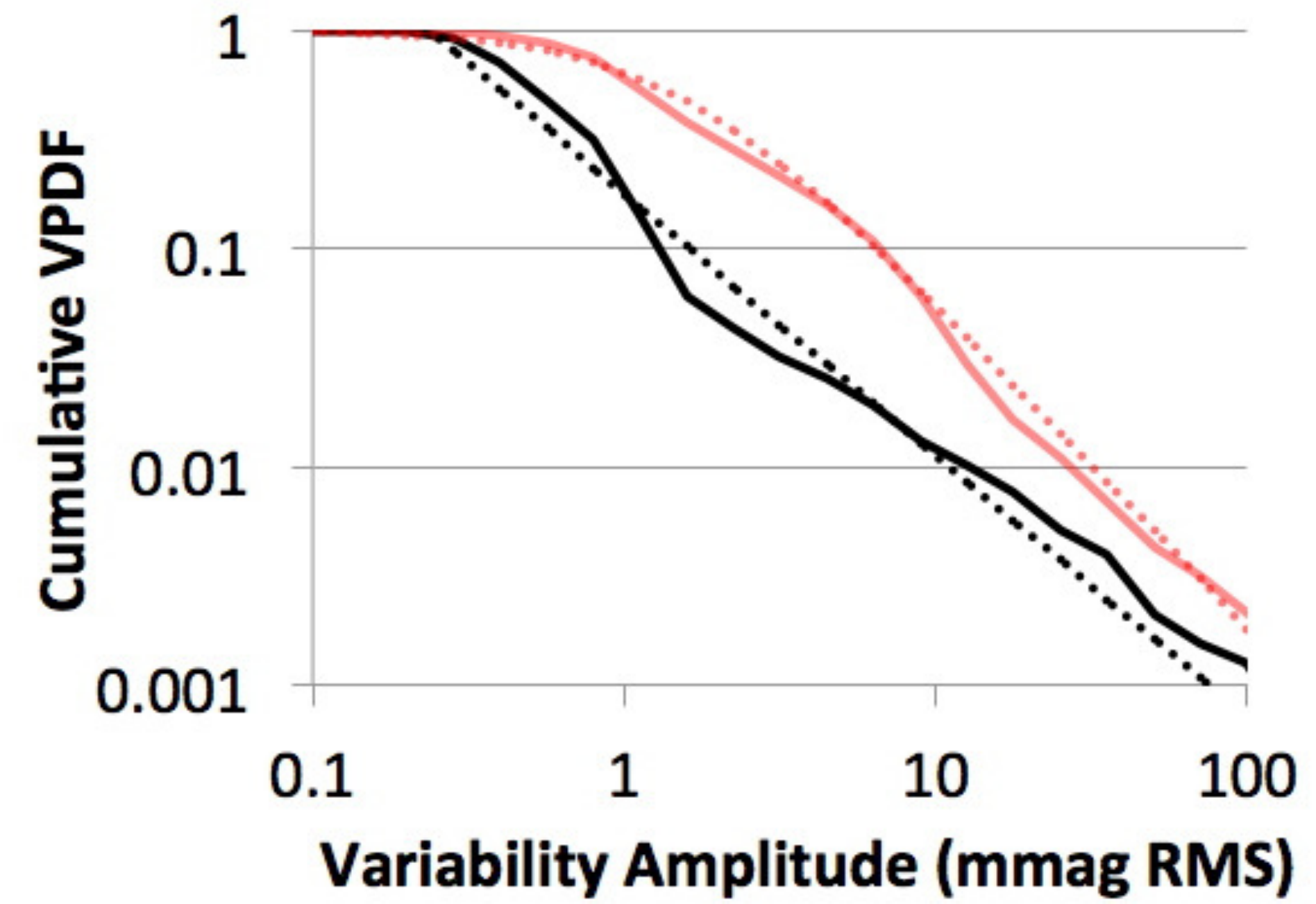}
\caption{   {\it Left}: Cumulative VPDF for Kepler stars K2 - K5.  The red (light) line represents stars of Kepler $\log g \geq 3.75$, and the black (dark) line of $ < 3.75$.  {\it Right}: Cumulative VPDF for K0 - K2 stars. The red (light) line represents stars of Kepler $\log g \geq 4.0$, and the black (dark) line of $ < 4.0$.  {\it In both figures}:  the dotted lines show the empirical fits described in Table \ref{AnalyticalApproximation}}
\end{figure}

\begin{figure}
\plottwo{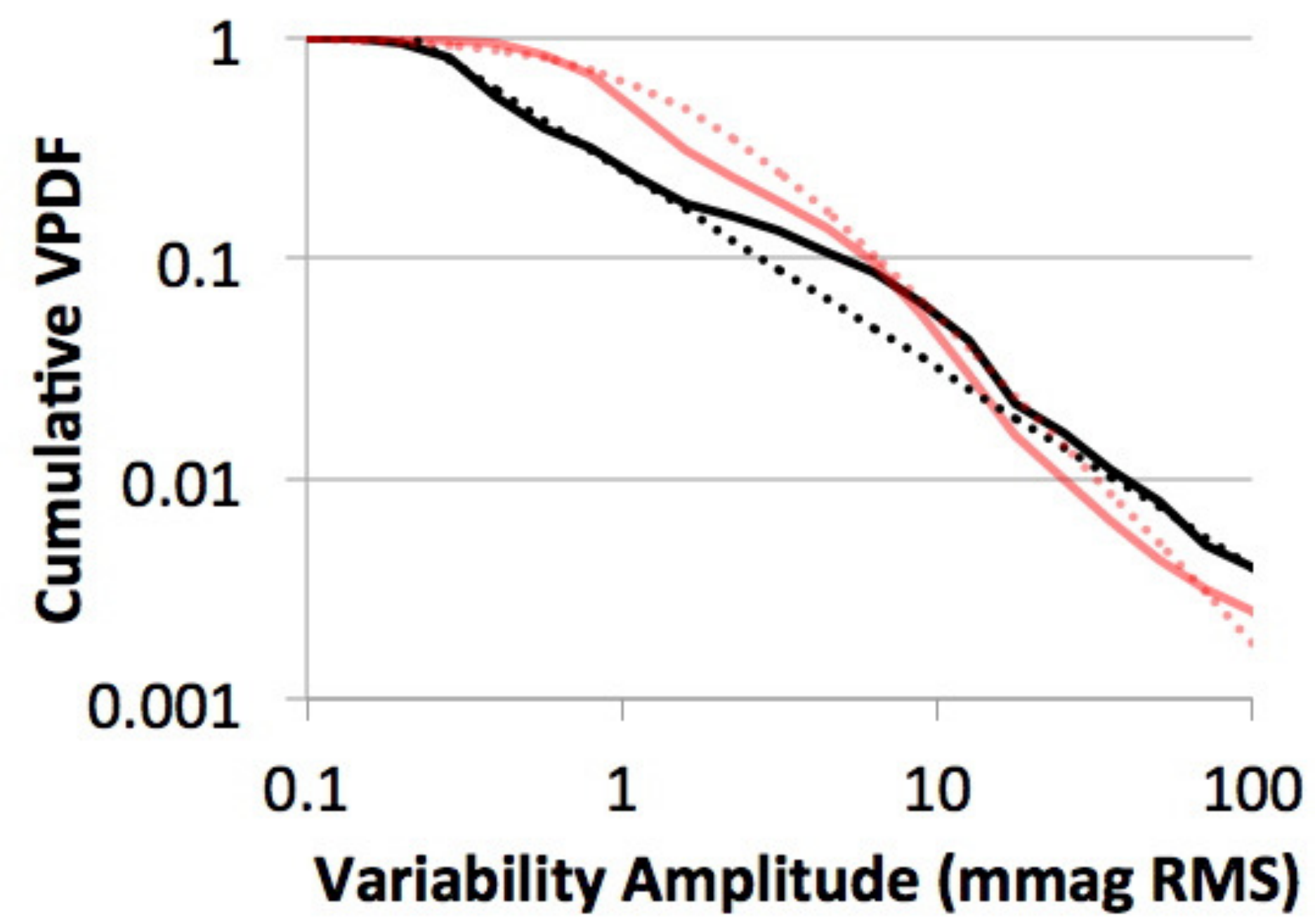}{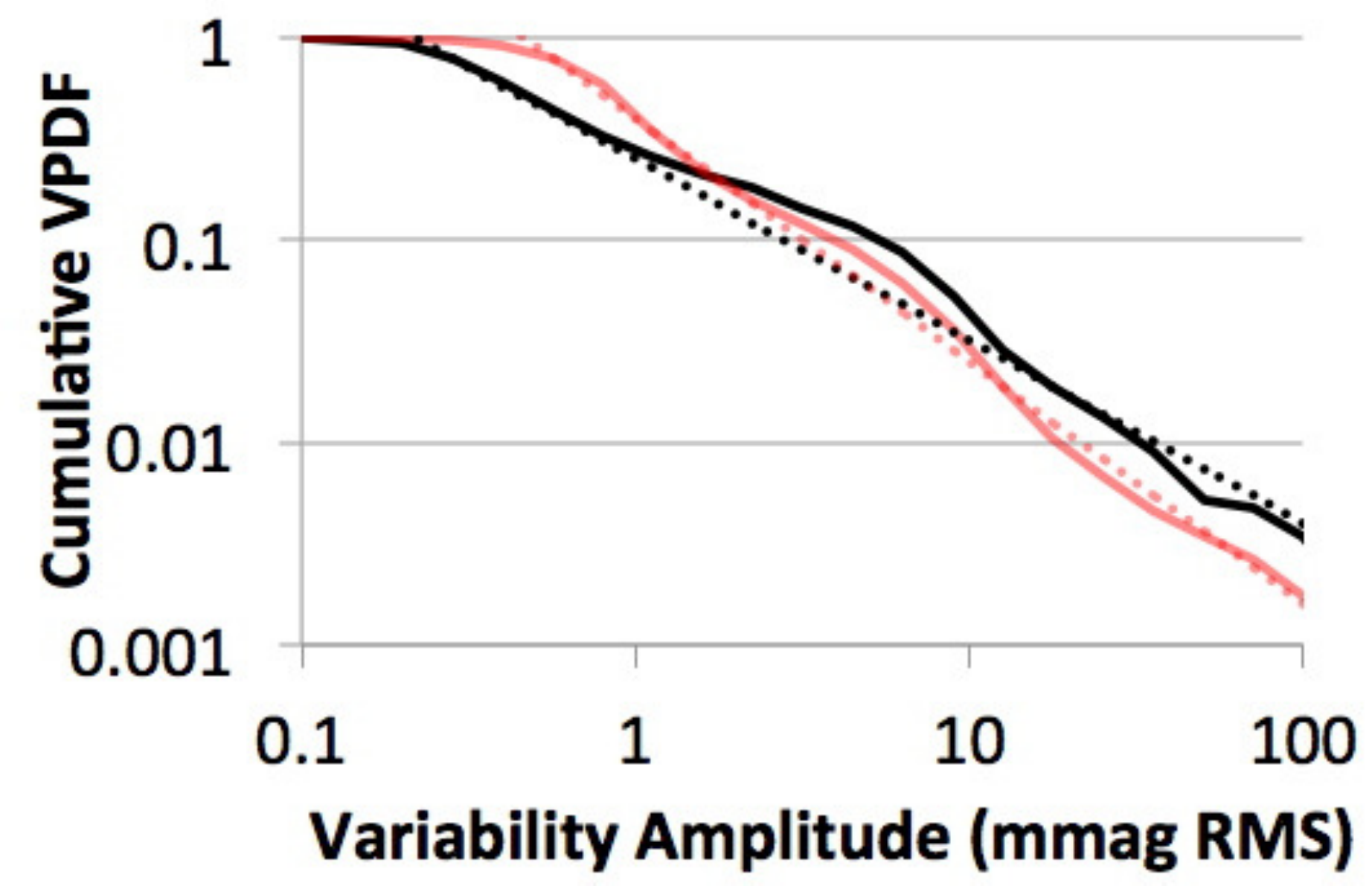}
\caption{   {\it Left}: Cumulative VPDF for Kepler stars G8 - K0.  The red (light) line represents stars of Kepler $\log g \geq 4.1$, and the black (dark) line of $ < 4.1$.  {\it Right}: Cumulative VPDF for G5 - G8 stars. The red (light) line represents stars of Kepler $\log g \geq 4.2$, and the black (dark) line of $ < 4.2$.  {\it In both figures}:  the dotted lines show the empirical fits described in Table \ref{AnalyticalApproximation}}
\end{figure}

\begin{figure}
\plottwo{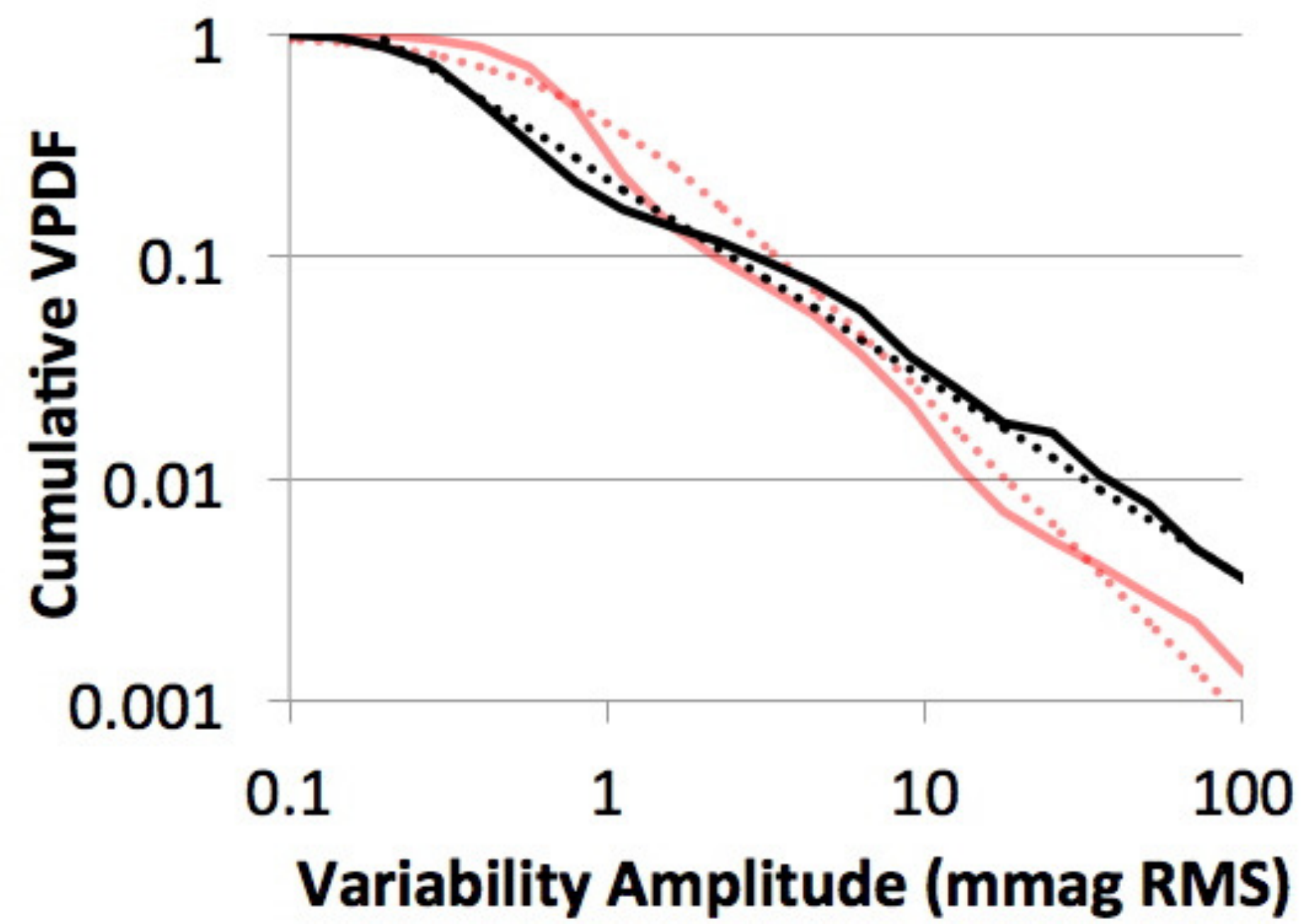}{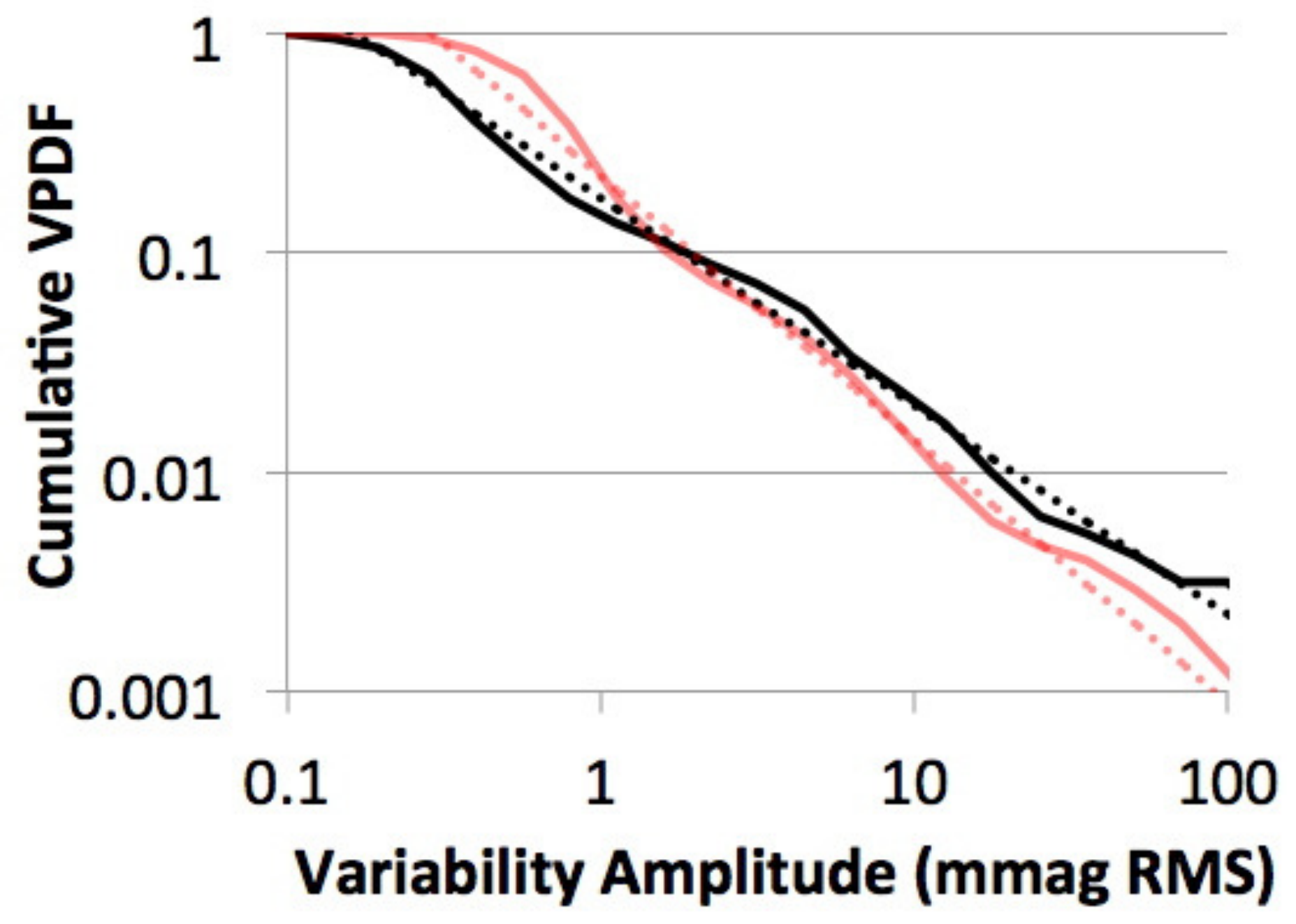}
\caption{   {\it Left}: Cumulative VPDF for Kepler stars G2 - G5.  The red (light) line represents stars of Kepler $\log g \geq 4.3$, and the black (dark) line of $ < 4.3$.  {\it Right}: Cumulative VPDF for G0 - G2 stars. The red (light) line represents stars of Kepler $\log g \geq 4.3$, and the black (dark) line of $ < 4.3$.  {\it In both figures}:  the dotted lines show the empirical fits described in Table \ref{AnalyticalApproximation}}
\end{figure}

\begin{figure}
\plottwo{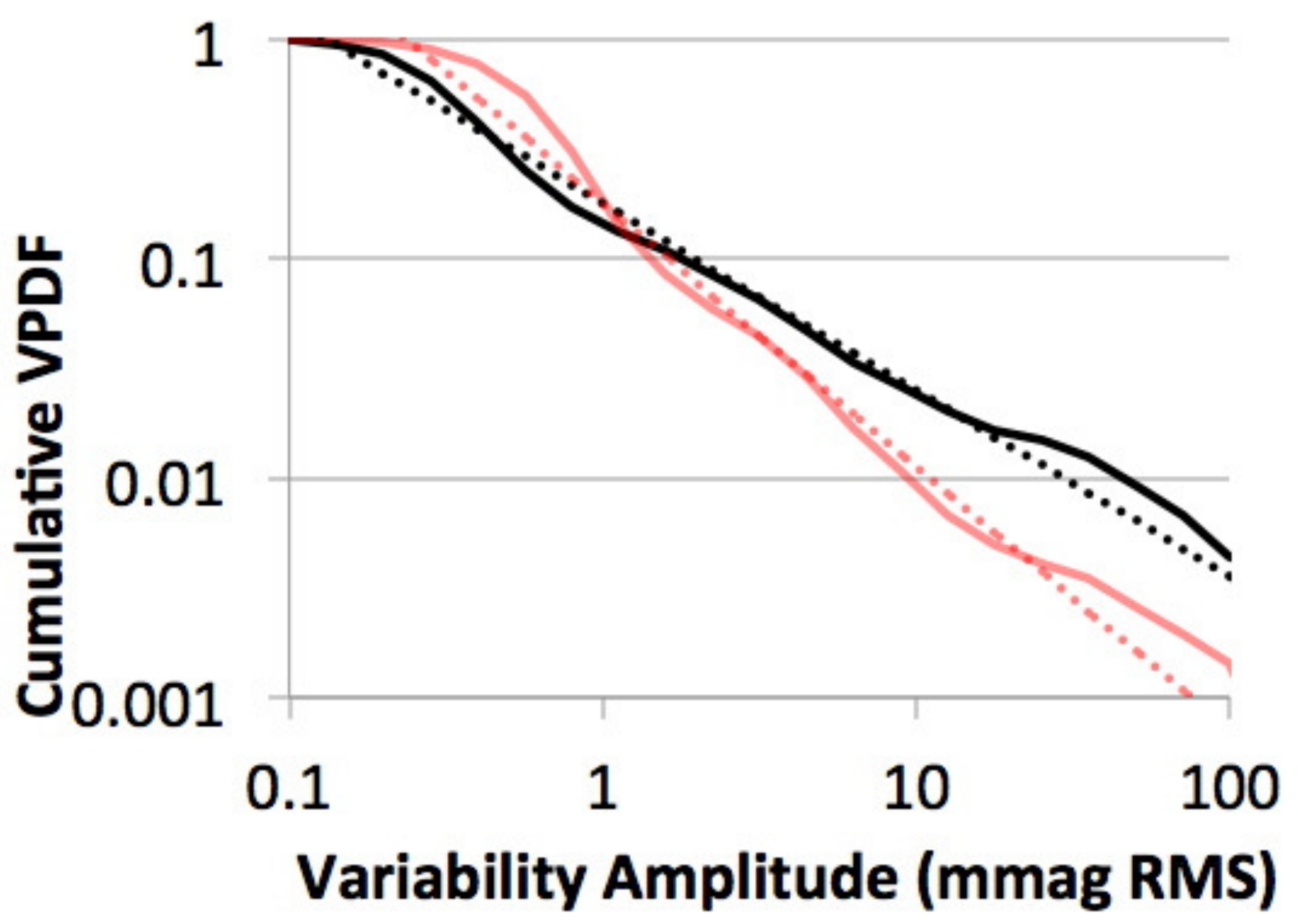}{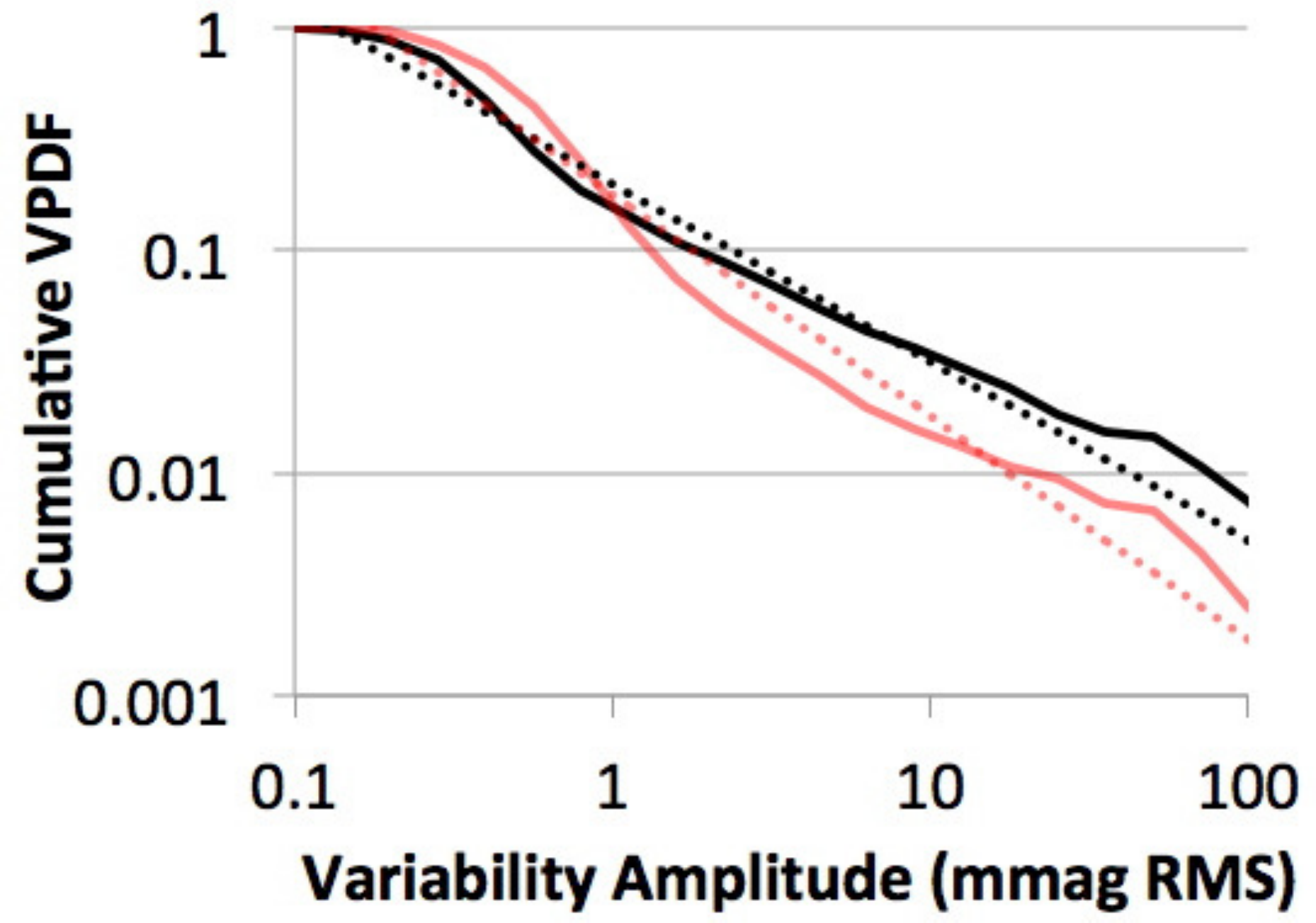}
\caption{  {\it Left}: Cumulative VPDF for Kepler stars F8 - G0.  The red (light) line represents stars of Kepler $\log g \geq 4.3$, and the black (dark) line of $ < 4.3$.  {\it Right}: Cumulative VPDF for F5 - F8 stars. The red (light) line represents stars of Kepler $\log g \geq 4.25$, and the black (dark) line of $ < 4.25$.  {\it In both figures}:  the dotted lines show the empirical fits described in Table \ref{AnalyticalApproximation}}
\end{figure}

\begin{figure}
\plottwo{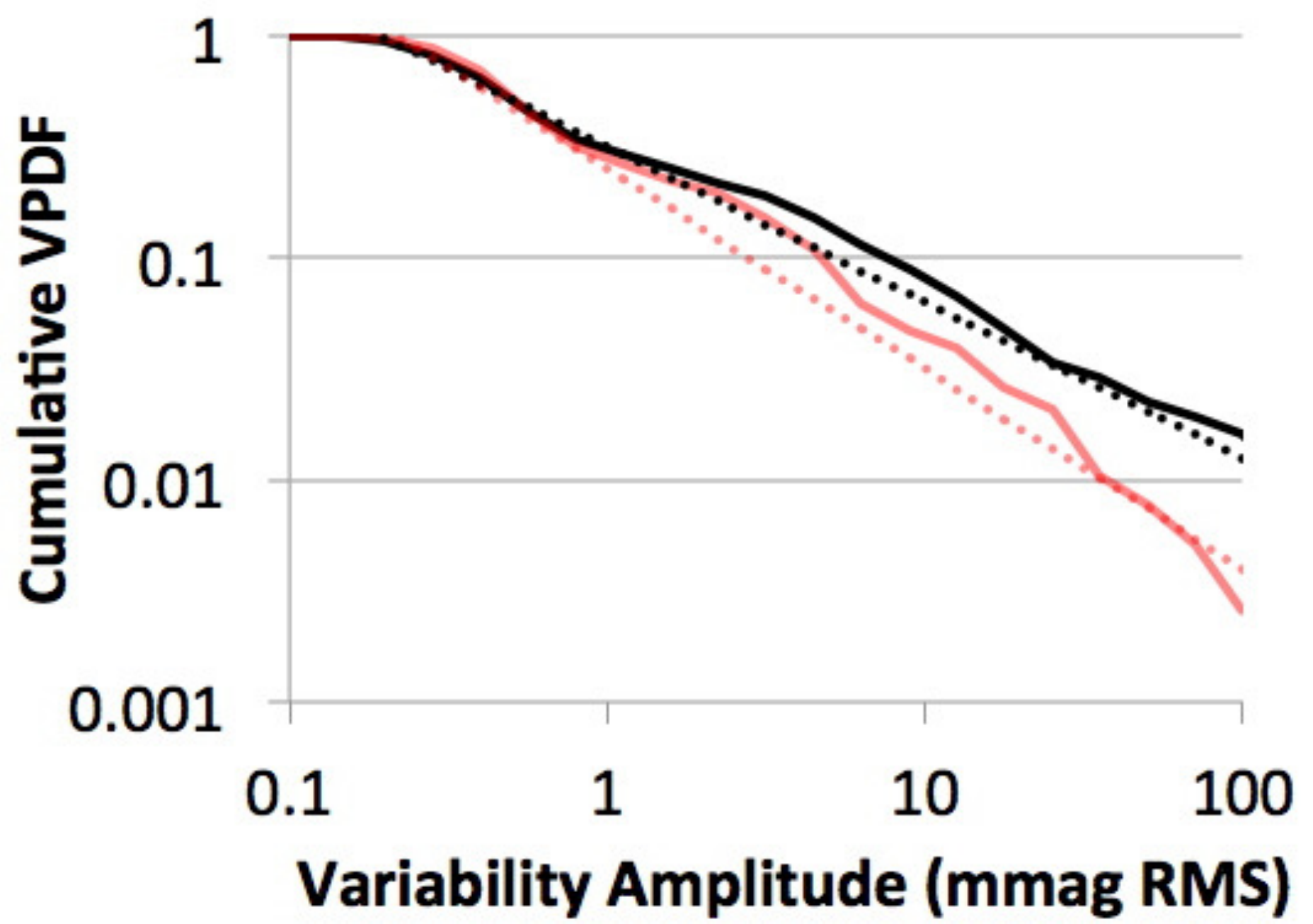}{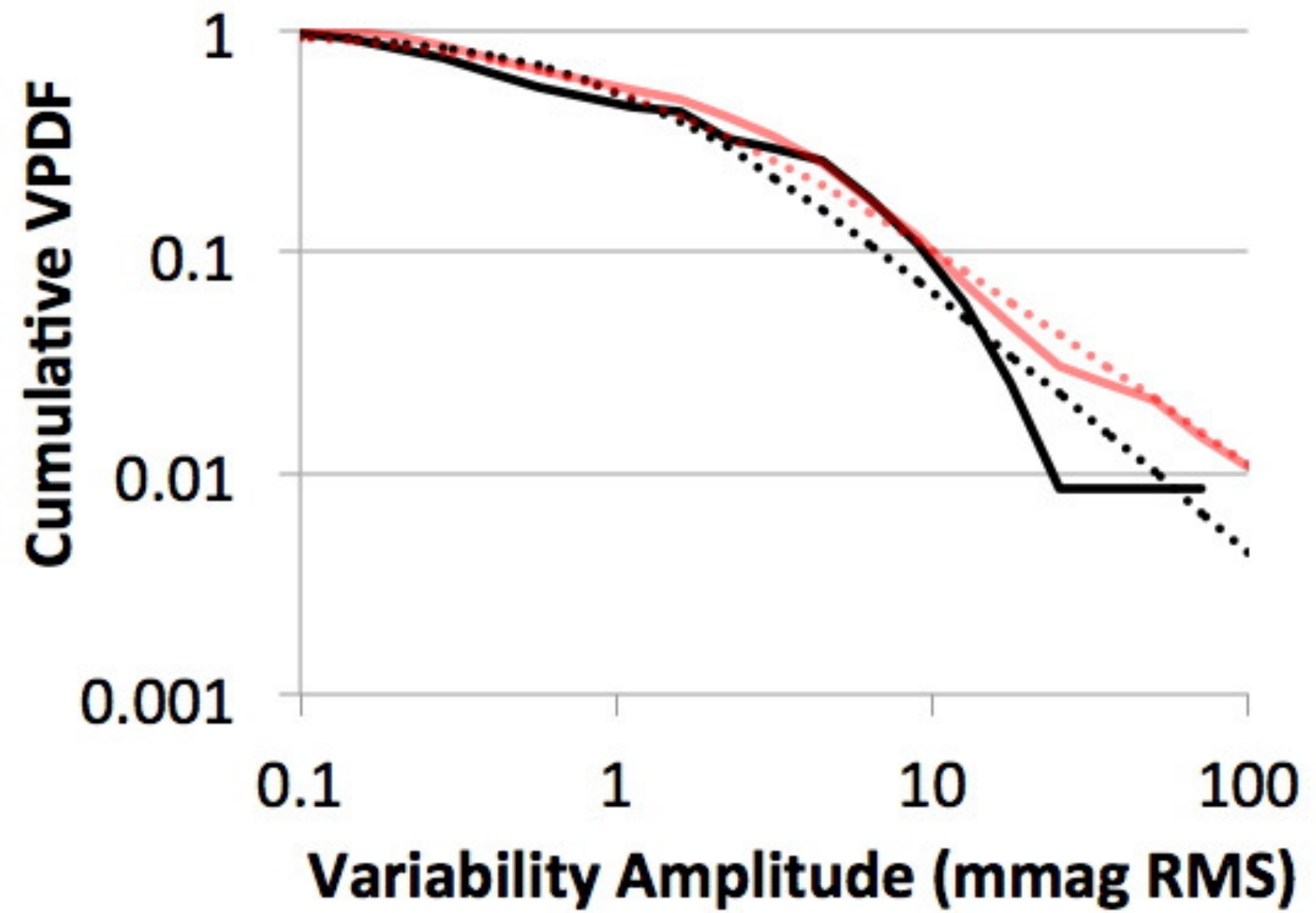}
\caption{   {\it Left}: Cumulative VPDF for Kepler stars F2 - F5.  The red (light) line represents stars of Kepler $\log g \geq 4.3$, and the black (dark) line of $ < 4.3$.  {\it Right}: Cumulative VPDF for F0 - F2 stars. The red (light) line represents stars of Kepler $\log g \geq 3.7$, and the black (dark) line of $ < 3.7$.  {\it In both figures}:  the dotted lines show the empirical fits described in Table \ref{AnalyticalApproximation}}
\end{figure}

\begin{figure}
\plottwo{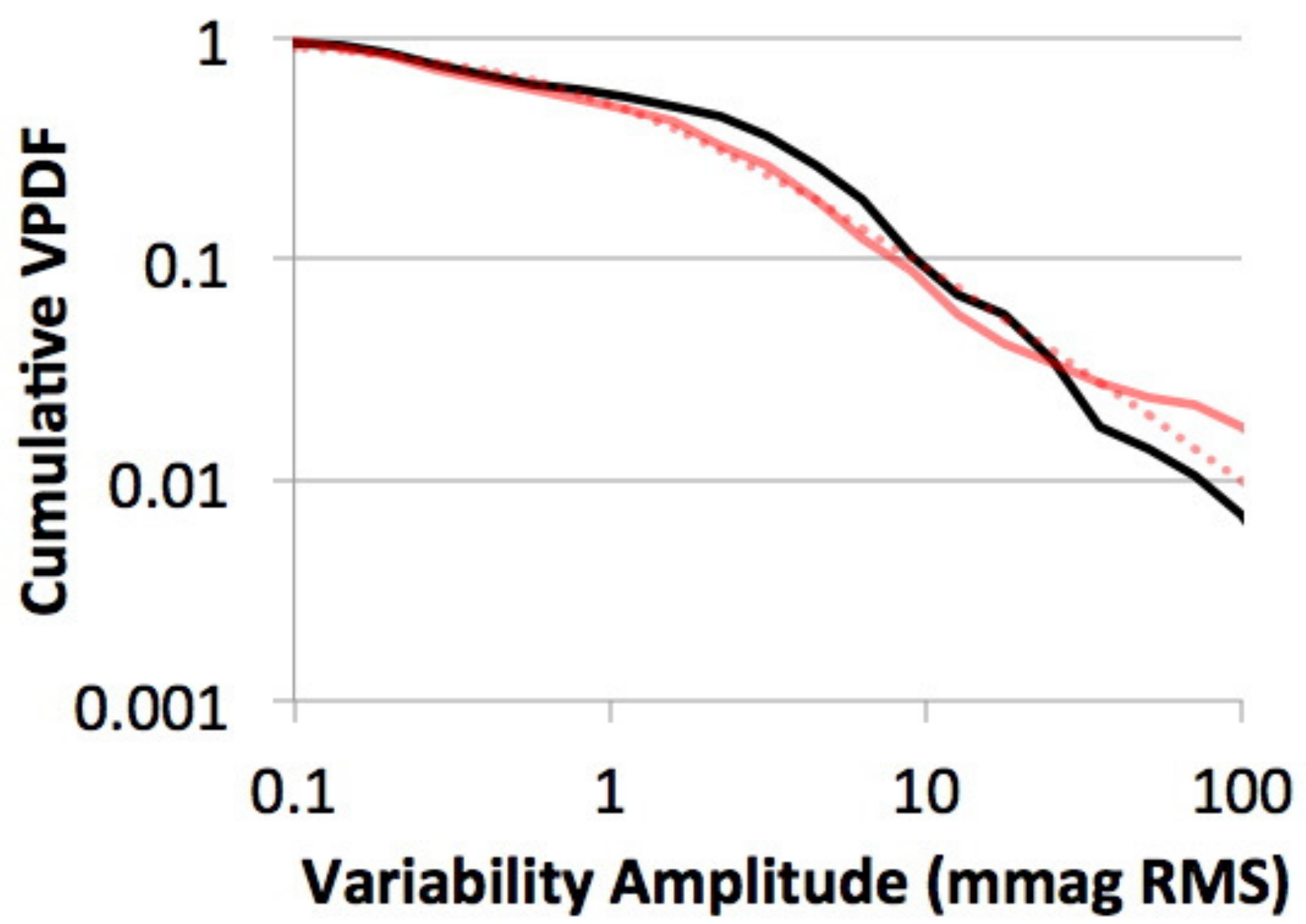}{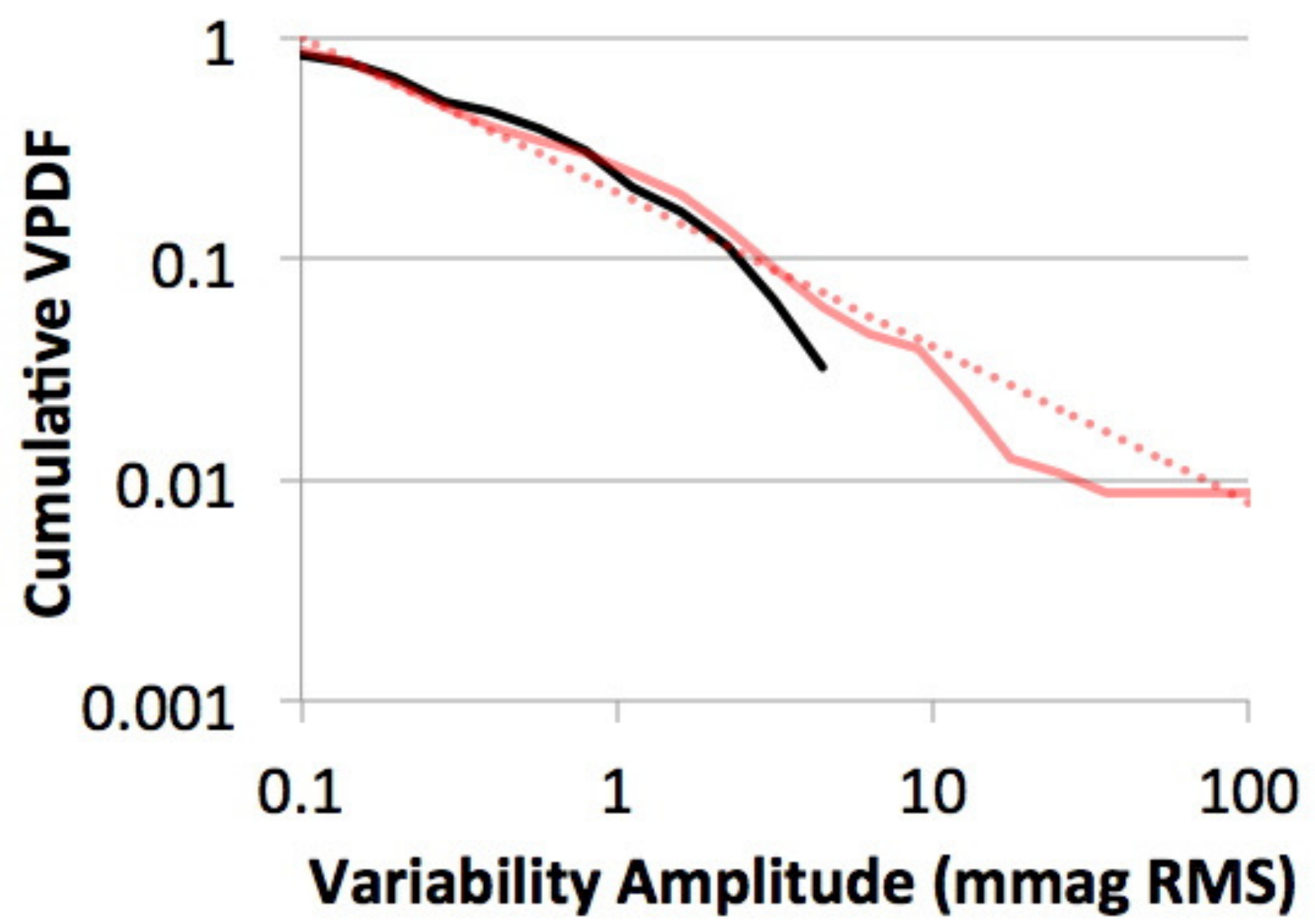}
\caption{  {\it Left}: Cumulative VPDF for Kepler stars A5 - F0.  The red (light) line represents stars of Kepler $\log g \geq 3.7$, and the black (dark) line of $ < 3.7$.  {\it Right}: Cumulative VPDF for A2 - A5 stars. The red (light) line represents stars of Kepler $\log g \geq 3.7$, and the black (dark) line of $ < 3.7$.  {\it In both figures}:  there is no significant difference between $\log g$ groups, and the dotted lines show the empirical fits described in Table \ref{AnalyticalApproximation}}
\end{figure}

\begin{figure}
\plottwo{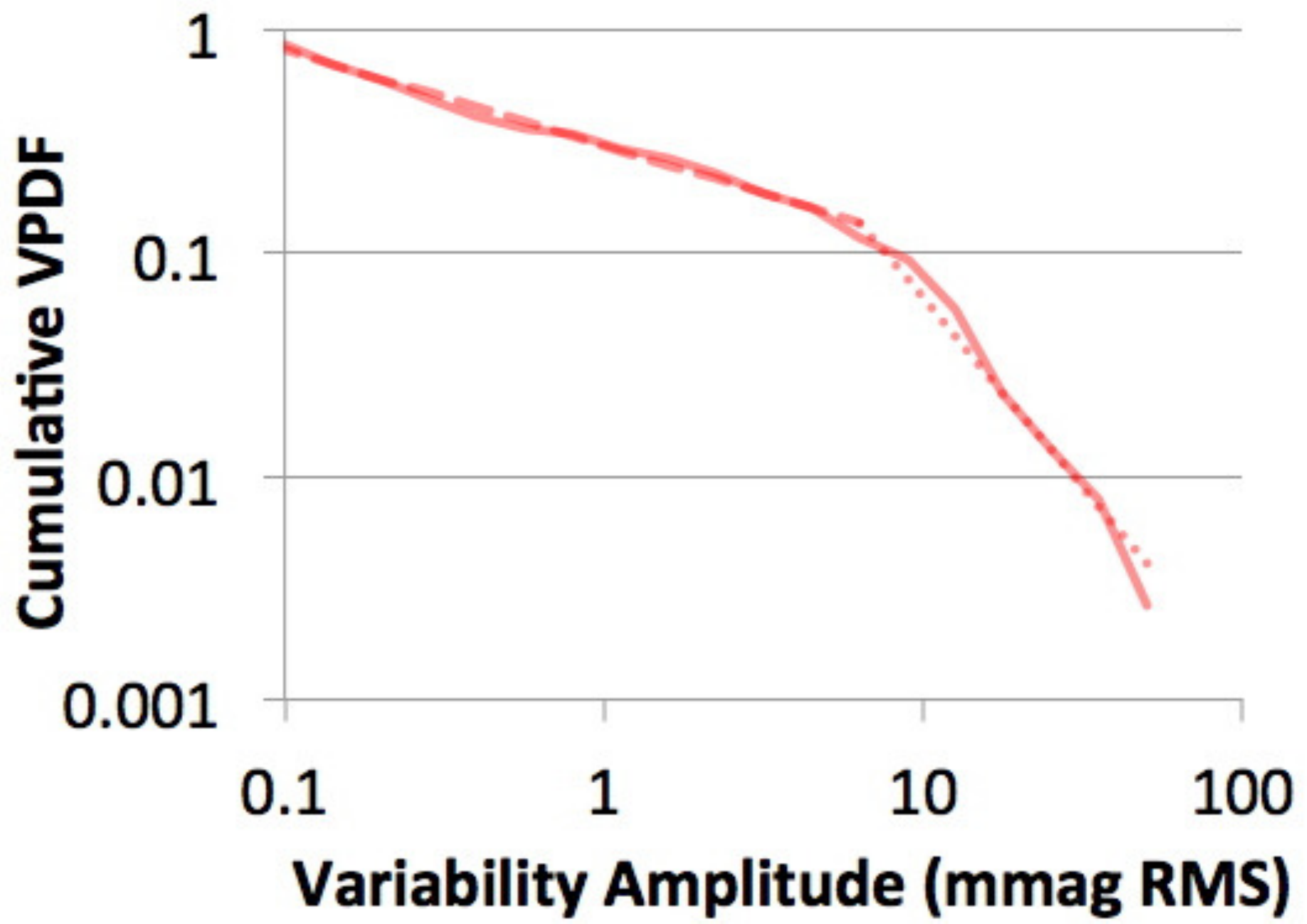}{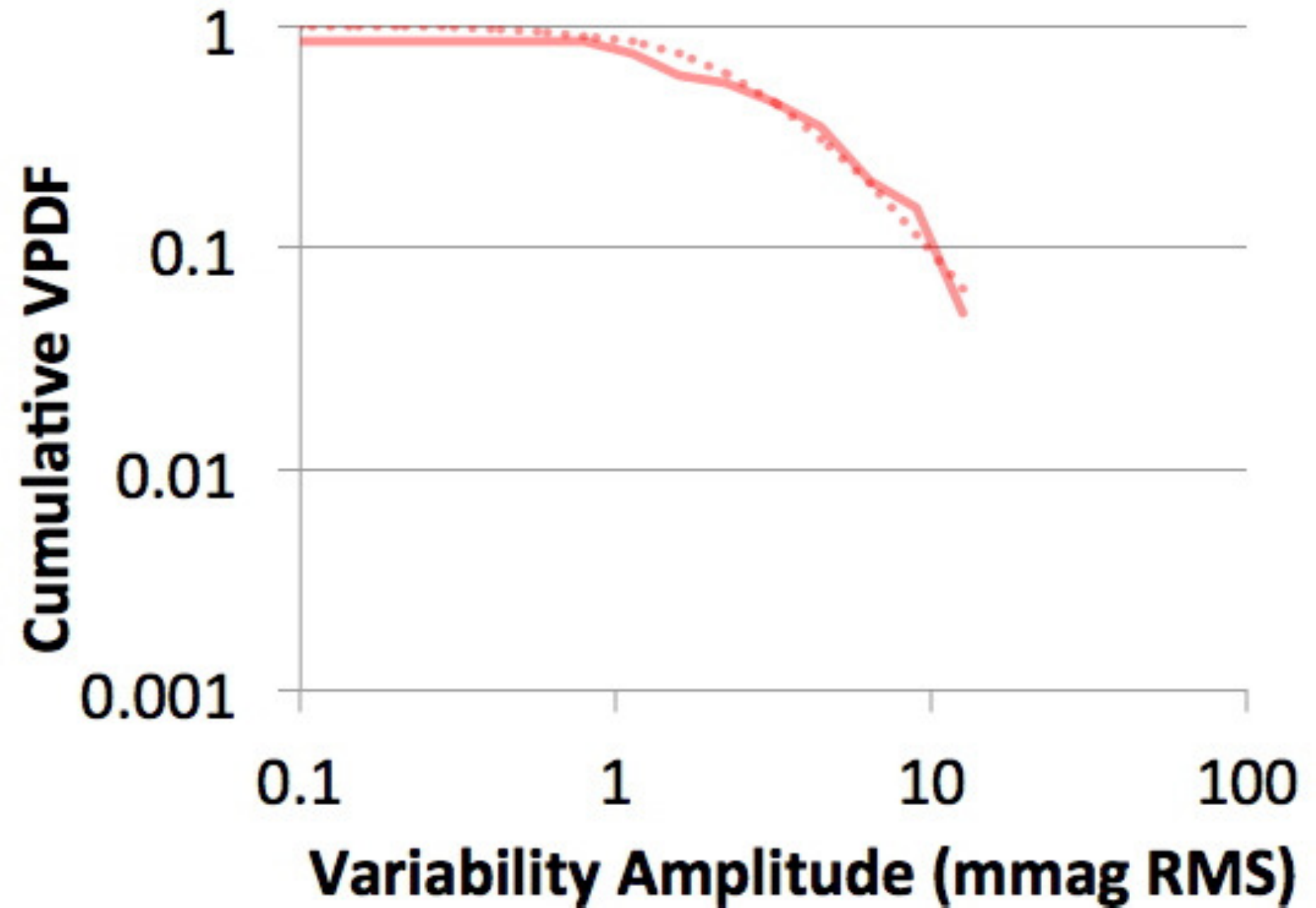}
\caption{   {\it Left}: Cumulative VPDF for Kepler stars B9 - A2.    {\it Right}: Cumulative VPDF for B5 - B8 stars.  {\it In both figures}:  there are insufficient data to distinguish different $\log g$ groups.  The dashed and dotted lines show the empirical fits described in Table \ref{AnalyticalApproximation}}
\end{figure}

\clearpage

\end{document}